# The Control of the Elementary Quantum Systems Radiation Using Metamaterials and Nanometaparticles

V.V. Klimov

*Lebedev Physical Institute, Russian Academy of Sciences, Leninsky prosp. 53, Moscow, 119991, Russia*

Translated by A.V. Sharonova

*The most important direction in the development of fundamental and applied physics is the study of the properties of optical systems at nanoscale in order to create optical and quantum computers, biosensors, single photon sources for quantum informatics, devices for DNA sequencing, sensors of various fields, etc. In all these cases, nanoscale light sources - dye molecules, quantum dots (epitaxial or colloidal), color centers in crystals, and nanocontacts in metals - are of key importance. In the nanoenvironment, the characteristics of these elementary quantum systems - pumping rates, radiative and non-radiative decay rates, local density of states, lifetimes, level shifts - experience changes that can be used intentionally to create nanoscale light sources with desired properties. This review presents an analysis of actual theoretical and experimental works in the field of elementary quantum systems radiation control using plasmonic and dielectric nanostructures, metamaterials, and nanoparticles made from metamaterials.*

**Content**

# 1. Introduction

At present, the most important direction in the development of fundamental and applied science is the study of the properties of optical systems at nanoscale with the aim of creating nanolasers and spasers [1,2], nanoantennas [3-7], metasurfaces [8,9,10], sensors [11,12,13], optical information processing systems [14,15], devices based on quantum optics and topological photonics [16,17,18]. In all these cases, a nanoscale light source - dye molecules, epitaxial or colloidal quantum dots, color centers in crystals, or nanoscale contacts in metals - is of key importance. Pump rates, radiative and non-radiative decay rates (Purcell effect [19]), lifetime, level shifts, directional patterns - all these characteristics of elementary quantum systems (EQS) in the nanoenvironment will experience changes that can be purposefully used to create nanoscale radiation sources with required characteristics. These characteristics include the efficient use of pumping, high fluorescence intensity, small losses, radiation pattern, and directivity, enabling one to detect and redirect photons effectively. On the other hand, the creation of the element base of quantum computers requires a long lifetime of the EQS excited states. It is also possible to do this if there are no states of light where the EQS can decay [20-23].

Currently, the control of spontaneous emission based on the Purcell effect is a key element for many applications, including single-photon sources [24], integrated quantum optics [25,26], nanolasers [1,27] active metamaterials [28], biosensors based on enhanced fluorescence [29], ultrafast modulators of light-emitting diodes [30], and nonlinear photonic structures [31].

The urgency of this direction is especially associated with the success in the field of nanotechnology and bottom-up and top-down nanofabrication. These technologies make it possible to manufacture nanoelements and assemble them in nanostructures of various shapes and sizes from various materials. In fact, everything that is required can be made, the main thing is to develop the physical principles of operation and the design of the optical nanodevice. Here, a precise description and optimization play a key role, and are impossible without a deep understanding of optical processes at nanoscale.

This review will analyze both classical and the most modern works in the field of radiation control of elementary quantum systems (EQS) with making use of plasmonic and dielectric nanostructures, metamaterials, and nanoparticles from metamaterials (nanometaparticles). A typical geometry of the considered systems consists of a control element (nanoantenna) and an EQS (Fig. 1).

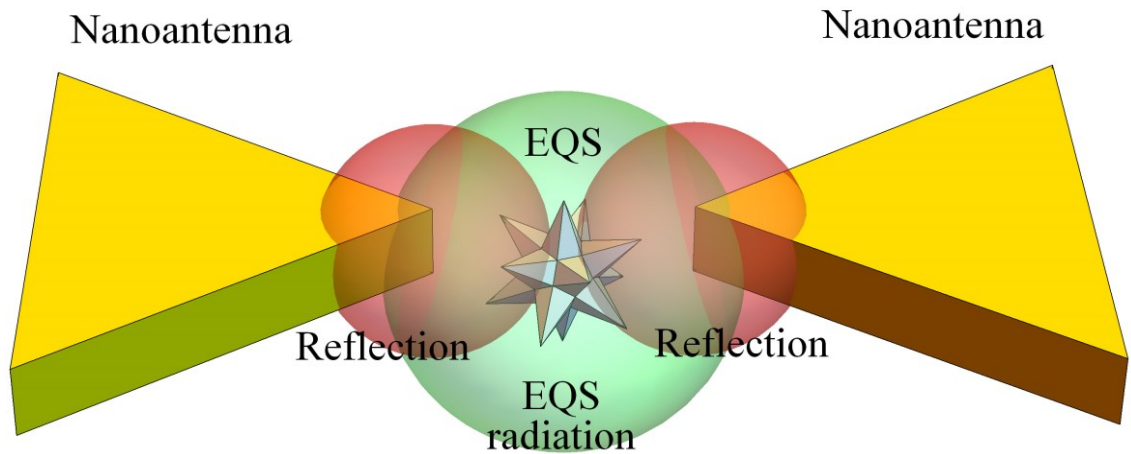

Figure: 1. A typical scheme of EQS radiation control using a nanoantenna. EQS radiation is reflected from the antenna and has an additional effect on it. If the reflected field and the radiation are in phase, then the constructive interference takes place, and the EQS emits more effectively and, conversely, if the reflected field and the radiation are out of phase, then the destructive interference takes place, and the EQS emits weaker.

Figure 2 shows one of the realizations of a directional nanoscale light source based on an electrically excited nanoparticle inside a nanocontact [32].

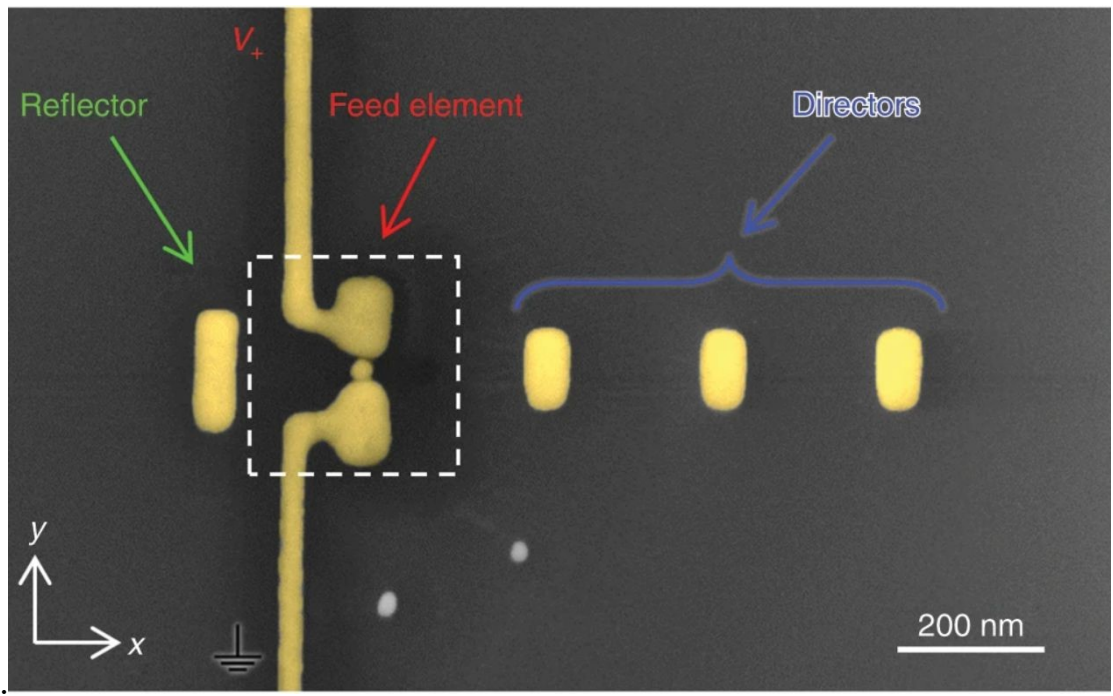

Fig. 2. A nanoscale light source based on an electrically-driven Yagi-Uda nanoantenna [32].

The plan for the rest of the review is as follows. Section 2 will present general approaches to the description of spontaneous emission and fluorescence of the EQS in a nanoenvironment. Here, basic definitions will be introduced, and fundamental formulas presented, that are required to understand the radiation processes of atoms and molecules. Particular attention will be paid to works devoted to the fundamental relations between radiative and non-radiative decay channels of excited states in the presence of nanoparticles. This is very important, since in an experiment only one of these quantities can often be measured. In spectroscopy, along with allowed ("dipole") transitions, forbidden ("multipole") transitions are also of great importance. For them, works on the Purcell effect will also be analyzed.

The effect of plasmonic nanoparticles and nanoantennas on the EQS radiation will be considered in Section 3. In such systems, it is possible to realize resonance effects at subwavelength scales. In plasmonic systems, losses are large, and therefore, at present, much attention is paid to methods for controlling the emission of an EQS using dielectric nanoparticles and nanostructures. This direction will be analyzed in Section 4. Section 5 will be devoted to EQS near nanometaparticles and metamaterials. The processes of influence of the environment on the emission of acoustic and elastic waves will be described in Section 6. In conclusion (Section 7), we will talk about the practical importance, relevance and prospects of this area, paying special attention to its applications and experimental achievements.

## 2. General Approaches to the Description of EQS Radiation

### 2.1 Historical Background

An excited quantum system is capable of emitting in a spontaneous and stimulated manner, as predicted by Einstein's theory of equilibrium radiation [34] and rigorously shown in Dirac's quantum theory of radiation [35]. According to Dirac's theory, the probability of a photon emitting into a given mode is proportional to (1 + N), where the first term corresponds to spontaneous, and the second - to stimulated emission, N is the number of radiation quanta in the considered oscillation mode. The total decay rate $\Gamma_{21} = 1/\tau_{21}$ of the spontaneous emission of the excited state "2" of the quantum system to the lower state "1" is determined by the Fermi's golden rule [36]:

$$\Gamma_{21} = \frac{1}{\tau_{21}} = \frac{2\pi}{\hbar}|H_{21}|^2 \rho_0(\omega), \qquad (1)$$

where $H_{21}$ is the matrix element of the interaction Hamiltonian connecting states "2" and "1", and

$$\rho_0(\omega)d\omega = 2\frac{Vd^3k}{(2\pi)^3} = \frac{V\omega^2 d\omega}{\pi^2 c^3} \qquad (2)$$

is the mode density of final states at the frequency of transition in free space (Rayleigh-Jeans law), V stands for quantization volume. In the case of the dipole interaction of EQS with the electric field $\hat{H} = -\hat{\mathbf{d}}\hat{\mathbf{E}}$, the expression (1) can be represented as

$$\Gamma_{21} = \frac{4\pi^2 \omega |\mathbf{d}_{21}|^2}{3\hbar V}\rho_0(\omega) = \frac{4|\mathbf{d}_{21}|^2 \omega^3}{3\hbar c^3}, \qquad (3)$$

where $\mathbf{d}_{21}$ is the matrix element of the EQS dipole moment operator.

For a long time, only free atoms and molecules had been studied, and therefore it had been taken for granted that the rate of spontaneous emission is constant. However, in 1946 Purcell [19] drew an attention to the fact that the resonator system selects a limited number of higher quality modes of oscillations and that it is possible to increase the emission rate of an excited quantum system by the so-called Purcell factor $F_P$, by placing it in a cavity with natural frequency tuned to the transition frequency of the atom:

$$\Gamma_{21} = \frac{4\pi^2 \omega |\mathbf{d}_{21}|^2}{3\hbar V}\rho_0(\omega) F_P, \quad F_P = \frac{3\lambda^3 Q}{8\pi^2 V_c}, \qquad (4)$$

where $Q$ is the quality-factor of the cavity, and $V_c$ is its volume.

Expression (4) is rather qualitative than quantitative, since it does not take into account a number of parameters of the electromagnetic field and properties of the resonator. In particular, Purcell's formula does not take into account the frequency detuning between the resonator mode and the frequency of the EQS radiation. A generalization of Purcell's formula to this case was made by Bunkin and Oraevsky [37,38], who instead of (4) proposed a formula that is valid for an arbitrary detuning of the cavity oscillations frequency $\omega_c$ with respect to the quantum transition frequency $\omega$ of the atom:

$$\Gamma_{21} = \frac{4|\mathbf{d}_{21}|^2 \omega^3}{3\hbar c^3} F_P, F_P = \frac{3\lambda^3}{8\pi^2 V_c}\frac{\omega_c^2/Q}{(\omega-\omega_c)^2 + \omega_c^2/Q^2} \qquad (5)$$

However, the generalized formula (5) does not allow the quantitative description of the spontaneous emission processes depending on the position of the atom, the orientation of its dipole moment, frequency, and the spatial structure of the resonant mode. Moreover, as we will see below, the rate of spontaneous emission of an atom, molecule, or quantum dot can change significantly in non-resonant cases. Thus, formulas (4), (5) cannot serve as a quantitative basis for controlling of the processes of EQS radiation. Therefore, in the following sections, a more accurate description of the influence of the nanoenvironment on the rate of spontaneous emission will be presented. In this case, we, as a rule, will assume that the excited state decays exponentially, that is, that the interaction of the atom with the field is rather weak and is described by the Fermi's golden rule (1).

Note again that the Purcell factor can be less than 1, that is, the decay rate can decrease in the absence of the required modes significantly [20-23]. Such a decrease in the decay rate and, accordingly, an increase in the EQS lifetime is extremely important for the practical implementation of quantum computers.

## 2.2. Description of the EQS Radiation Process in the Nanoenvironment

To describe the radiation processes of an EQS located in free space near an arbitrary body, one can use a clear classical approach [39-44], EQS are approximated by a nonrelativistic oscillator consisting of a resting charge $-e$ and an oscillating charge $e$ located at the distance $\delta \mathbf{r}$ from the resting charge. If the oscillator is in an arbitrary nanoenvironment at a point $\mathbf{r}'$, it is affected by an additional reflected electric field $\mathbf{E}^{(R)}(\mathbf{r})$ in comparison with the case of free space (see Fig.1), and the equation of motion of the oscillating charge takes the form:

$$m\ddot{\mathbf{d}} + m\gamma_0 \dot{\mathbf{d}} + m\omega_0^2 \mathbf{d} = e^2 \mathbf{E}^{(R)}(\mathbf{r}',t), \qquad (6)$$

where $\mathbf{d}_0 = e\delta\mathbf{r}$ is the electric dipole moment of the oscillator, $\gamma_0 = \frac{2e^2}{3c^3}\frac{\omega_0^2}{m}$ is the total linewidth in vacuum, $m$ is the mass of the moving charge, $\omega_0$ is the frequency of oscillations in vacuum. To find the reflected field $\mathbf{E}^{(R)}(\mathbf{r}',t)$, it is necessary to solve the complete system of Maxwell's equations, where the source of the field is the dipole moment of the oscillator.

If the reflected field $\mathbf{E}^{(R)}(\mathbf{r})$ and the EQS radiation in free space are in phase, then the constructive interference occurs, and the EQS emits more effectively, and vice versa, if the reflected field and the EQS radiation are out of phase, then the destructive interference takes place, and the EQS emits weaker. Figure 3 shows the interference effects arising from the interaction of the EQS with the mirror.

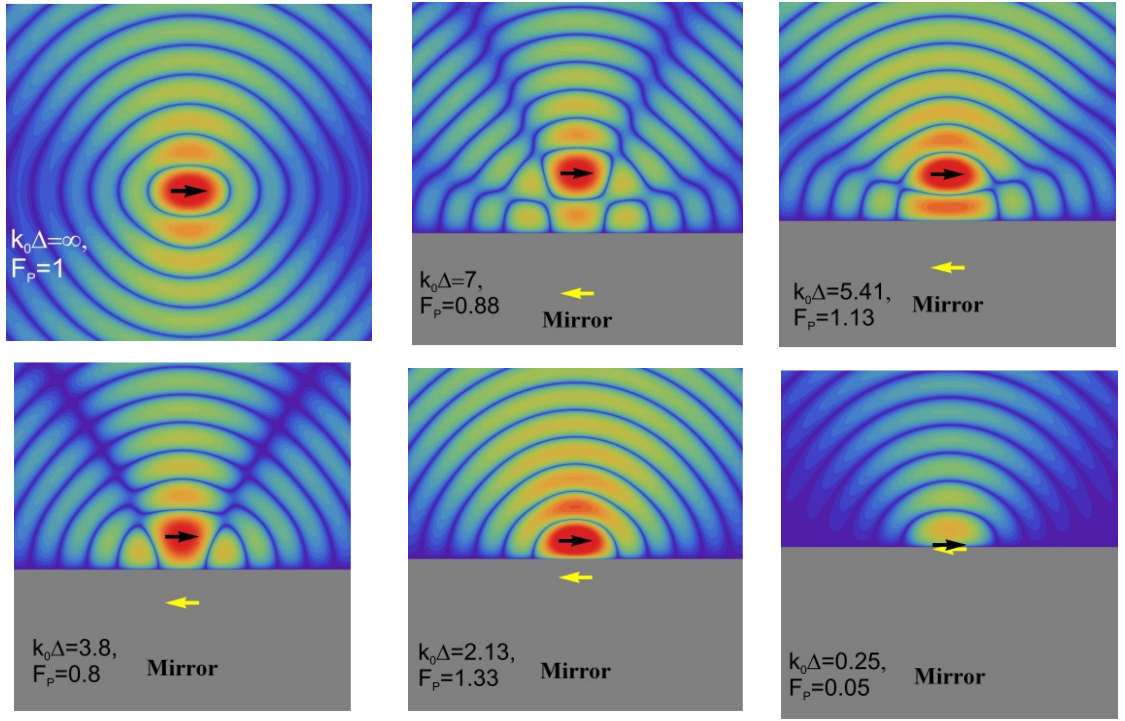

Fig. 3. An illustration of the interference nature of the Purcell effect. In Figures 3 (a) - (f), the spatial distribution of the imaginary part of the total field Im(**d**•**E**) for a dipole oriented parallel to the surface is shown by a pseudocolor. a) The dipole in free space, b) - f) the dipole is located at the distance $k_0\Delta$ = 7, 5.41, 3.8, 2.13, and 0.25 from the mirror, respectively. Corresponding Purcell factor $F_P$ (9),(33) is shown on each figure. The arrows show the real source and its image.

From Fig. 3, it can be seen that as the EQS with the horizontal orientation of the dipole moment approaches the mirror, the total radiated power, oscillating, decreases to zero. In [45], a visual analysis of interference effects is presented by the example of the radiation from a waveguide defect in a plane photonic crystal.

Assuming that all quantities are proportional to $\exp(-i\omega t)$ from (6), we obtain the dispersion equation for an atomic oscillator in an arbitrary environment [39-44]:

$$\omega^2 + i\omega\gamma_0 - \omega_0^2 + \frac{e^2}{md_0^2}\mathbf{d}_0\mathbf{E}^{(R)}(\mathbf{r}',\omega) = 0 \qquad (7)$$

In the case of small corrections to the frequency of free oscillations and smallness of the oscillator linewidth in comparison with the nanocavity linewidth (weak coupling regime), the solution of this equation for the EQS oscillation frequency can be written in the form [39-44]:

$$\omega = \omega_0 - \frac{i}{2}\gamma_0 - \frac{e^2}{2m\omega_0}\frac{\mathbf{d}_0 \mathbf{E}^{(R)}(\mathbf{r}',\omega_0)}{d_0^2} \qquad (8)$$

Separating the real and imaginary parts of (8), we obtain an expression for the rate of spontaneous emission $\Gamma$ (Purcell factor) in the weak coupling regime

$$\frac{\Gamma}{\gamma_0} = F_P = 1 + \frac{3}{2}\operatorname{Im}\frac{\mathbf{d}_0 \mathbf{E}^{(R)}(\mathbf{r}')}{d_0^2 k_0^3} \qquad (9)$$

Hereinafter, $k_0 = \omega_0/c$ denotes the wave vector of free space.

The electric field of the dipole oscillator can be expressed in terms of the Green's function $\ddot{G}(\mathbf{r},\mathbf{r}',\omega)$ of Maxwell's equations in the presence of an environment with a dielectric constant $\varepsilon(\mathbf{r},\omega)$

$$\mathbf{E}(\mathbf{r},\omega) = \ddot{G}(\mathbf{r},\mathbf{r}',\omega)\mathbf{d}_0 = \left(\ddot{G}^{(0)}(\mathbf{r}-\mathbf{r}',\omega) + \ddot{G}^{(R)}(\mathbf{r},\mathbf{r}',\omega)\right)\mathbf{d}_0$$
$$\mathbf{E}^{(R)}(\mathbf{r},\omega) = \ddot{G}^{(R)}(\mathbf{r},\mathbf{r}',\omega)\mathbf{d}_0 \qquad (10)$$

$$\nabla \times \left(\nabla \times \ddot{G}(\mathbf{r},\mathbf{r}',\omega)\right) - k_0^2 \varepsilon(\mathbf{r},\omega)\ddot{G}(\mathbf{r},\mathbf{r}',\omega) = 4\pi k_0^2 \ddot{1}\delta(\mathbf{r}-\mathbf{r}') \qquad (11)$$

In (11), $G_{ij}^{(0)}(\mathbf{r},\omega)$ stands for the Green's function of the free space [46]:

$$\ddot{G}_{ij}^{(0)}(\mathbf{r},\omega) = \left[k_0^2 \frac{(\delta_{ij} - \mathbf{n}_i \mathbf{n}_j)}{r} + \frac{(3\mathbf{n}_i \mathbf{n}_j - \delta_{ij})}{r^3}(1 - ik_0 r)\right] e^{ik_0 r}, \qquad (12)$$

$\mathbf{n} = \mathbf{r}/r$ is the unit vector in the direction from the EQS to the observation point. When (10) is substituted into (9), the expression for the rate of spontaneous emission of the EQS takes on the symmetric form:

$$\frac{\Gamma}{\gamma_0} = F_P = \frac{3}{2}\operatorname{Im}\frac{\mathbf{d}_0 \ddot{G}(\mathbf{r}',\mathbf{r}',\omega)\mathbf{d}_0}{d_0^2 k_0^3} = 1 + \frac{3}{2}\operatorname{Im}\frac{\mathbf{d}_0 \ddot{G}^{(R)}(\mathbf{r}',\mathbf{r}',\omega)\mathbf{d}_0}{d_0^2 k_0^3} \qquad (13)$$

It is important to note that these expressions describe the total rate of the spontaneous emission, including the energy loss in matter in the case of a complex permittivity. However, in practice it is also necessary to know the radiative decay rate $\Gamma^{rad}$, that is, the rate of the EQS energy radiation into the surrounding space [44]. In the classical case, this part can also be found by calculating the energy flux at infinity, where there are no bodies anymore:

$$\frac{\Gamma^{rad}}{\gamma_0} = F_{P,rad} = \frac{\int \mathrm{Re}\left(\left(\mathbf{E}^{(0)} + \mathbf{E}^{(R)}\right) \times \left(\mathbf{H}^{(0)} + \mathbf{H}^{(R)}\right)\right)_{r \to \infty} d\Omega}{\int \mathrm{Re}\left(\mathbf{E}^{(0)} \times \mathbf{H}^{(0)}\right) d\Omega}. \tag{14}$$

In (14), $\mathbf{E}^{(0)}$, $\mathbf{H}^{(0)}$ are the electric and magnetic fields of the dipole $\mathbf{d}_0$ in free space, and $\mathbf{E}^{(R)}$, $\mathbf{H}^{(R)}$ are the reflected fields, and the integration is carried out over the angle $d\Omega$, where the radiation propagates.

The difference between the total and radiative decay rates determines the rate of absorption by the environment of the energy emitted by the EQS (non-radiative decay rate, $\Gamma^{nonrad}$):

$$\frac{\Gamma^{nonrad}}{\gamma_0} = F_{P,nonrad} = \frac{\Gamma - \Gamma^{rad}}{\gamma_0} = F_P - F_{P,rad} \tag{15}$$

The expressions obtained within the classical picture describe the picture of spontaneous emission completely.

A universal quantum-mechanical description of the EQS spontaneous emission in the presence of arbitrary objects was proposed in [47,48]. For the total transition rate from the state *i* to the state *f* in the weak coupling regime, the following expression holds:

$$\frac{\Gamma_{fi}}{\gamma^0} = \frac{3}{2} \mathrm{Im} \frac{\mathbf{d}^{fi} \overleftrightarrow{G}(\mathbf{r},\mathbf{r},\omega_0) \mathbf{d}^{fi}}{\left|\mathbf{d}^{fi}\right|^2 k^3}, \tag{16}$$

where $\mathbf{d}^{fi}$ is the matrix element of the dipole moment operator between the states *i* and *f*.

Expression (16) coincides with the classical expression (13) exactly, if the dipole moment of the classical oscillator $\mathbf{d}_0$ and the dipole moment of the transition $\mathbf{d}^{fi}$ have the same orientations. This coincidence is extremely remarkable since it allows one to replace specific complex quantum-mechanical calculations by finding the classical Green's function of the electromagnetic field.

In quantum mechanics, in contrast to the classical situation, the initial state can decay into several final ones. In this case, in (16) it is necessary to sum over all final states.

The above expressions for the EQS radiation rate are valid for an individual emitter with a fixed orientation of the dipole moment. Modern technologies make it possible to implement such a regime of EQS radiation control and to provide a greater flexibility in the creation of nanooptical devices. Often, however, it is difficult and unnecessary to ensure a fixed orientation of the dipole moment, and therefore, to describe experiments with a non-fixed orientation of the dipole moment, it is necessary to average (16) over its orientations, and as a result the expression for the averaged EQS radiation rate $\langle \Gamma \rangle$ will be as follows:

$$\langle \Gamma \rangle = \frac{2|\mathbf{d}|^2}{3\hbar} \operatorname{Im} \operatorname{Tr} \ddot{G}(\mathbf{r},\mathbf{r},\omega_0) \qquad (17)$$

Taking into account the definitions (3) or (4), (17) can be written in the form:

$$\langle \Gamma \rangle = \frac{4\pi^2 \omega |\mathbf{d}|^2}{3\hbar} \rho_L(\mathbf{r},\mathbf{r},\omega_0), \qquad (18)$$

where

$$\rho_L(\mathbf{r},\mathbf{r},\omega_0) = \frac{\operatorname{Im} \operatorname{Tr} \ddot{G}(\mathbf{r},\mathbf{r},\omega_0)}{2\pi^2 \omega} \qquad (19)$$

stands for the so-called local density of states [49,50]. It is advisable to use expression (18) when real averaging over the orientations of the dipole moments is required or for a qualitative (integral) estimation of the spontaneous emission rate from the results of numerical simulation. In general, one should use (13) or (16). Note that a direct generalization of (18) to the case of transitions with a higher multipolarity (see section 2.4) is nontrivial and therefore impractical.

2.3. Strong Coupling of EQS to Nanoenvironment

The conditions of applicability of the weak-coupling regime considered in the previous section are usually satisfied, and to describe the processes of spontaneous emission one can use (8) and (9).

If a resonator or a nanoantenna interacting with the EQS has a very narrow linewidth and is located close to the EQS, then a strong coupling regime may arise, when the replacement of the frequency ω by ω₀ in (7) is incorrect, and to find the spectrum of the system it is necessary to find all solutions of (7). In the case of resonant interaction between the EQS and the environment, instead of (7) we will have [51-54]:

$$\omega^2 + i\omega\gamma_0 - \omega_0^2 + \omega_0 \left\{ \frac{g^2}{\omega - \Omega_{res}} - \frac{g^{*2}}{\omega + \Omega_{res}^*} + \eta \right\} = 0, \quad (20)$$

where $\Omega_{res} = \omega_{res} - i\Gamma_{res}/2$ is a complex quantity characterizing the frequency and decay rate of the resonant mode, and $g$ is a parameter characterizing the coupling strength and depending on the geometry of the system and the distance of the EQS from the resonator. The case $g^2/(\gamma_0 \Gamma_{res}) \ll 1$ corresponds to the case of weak coupling (8). In addition to the purely resonant terms, the nonresonant term

$$\eta = \frac{3}{8} \frac{\varepsilon - 1}{\varepsilon + 1} \frac{\gamma_0}{(k\Delta)^3} \quad (21)$$

is also taken into account in the expression (20). This term is large in the region of strong coupling, that is, in the case when the EQS is located near (at a distance $\Delta \ll \lambda$) of the nanocavity or nanoantenna, and corresponds to a purely electrostatic interaction without retardation, that is, to the frequency shift of the EQS. If the EQS is located at the distance $r$ from the center of the spherical resonator, the expression for the parameter of the coupling strength with the resonant TM mode of the n$^{th}$ order has the form [51]:

$$g^2 = \gamma_0 \Gamma_{res} \frac{3}{4} n(n+1)(2n+1) \left( \frac{h_n^{(1)}(\omega_{res} r/c)}{(\omega_{res} r/c)} \right)^2, \quad (22)$$

where $h_n^{(1)}(z)$ is the spherical Hankel function.

In the case of a strong coupling regime, the spectrum becomes significantly more complicated, doublets (Rabi splitting) and triplets appear in it, and the very concept of the decay rate of the EQS spontaneous emission becomes inapplicable [51-54]. Fig. 4 shows how the frequency shifts and the relative linewidths (decay rates) change with distance in the EQS + spherical resonator with a permittivity $\varepsilon = 6$ (TM$_{901}$ resonance) [51].

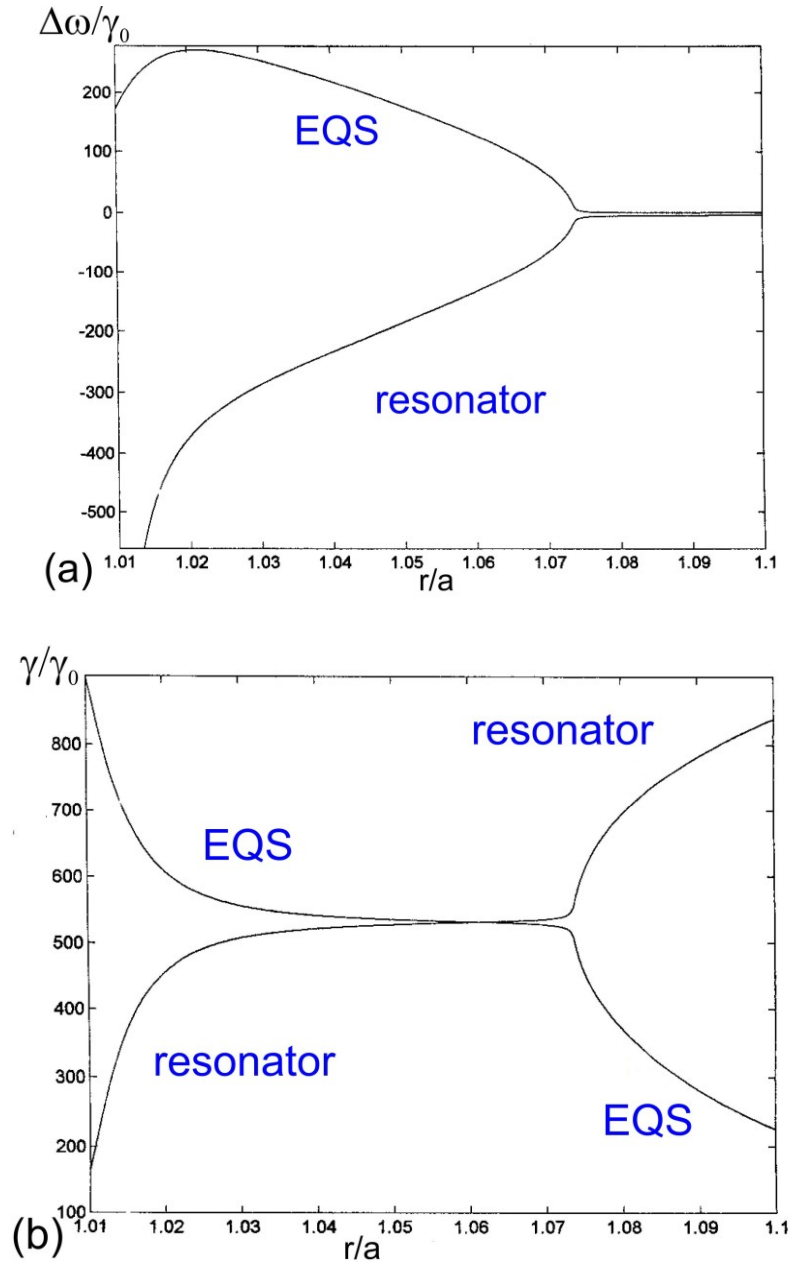

Fig. 4. Frequency shifts (a) and relative linewidths (b) of the system EQS + spherical resonator of the radius $a$ ($\varepsilon = 6$, TM$_{901}$ resonant mode, $\omega_{res}a/c = 5.548733$,

$\omega_0/\Gamma_{res} = 10^4$) as a function of the position ($r/a$) of the radially oriented EQS ($\omega_0/\gamma_0 = 10^7$) in the case of the EQS frequency slightly higher than the exact resonance frequency [51].

It can be seen from this figure that, at large distances from the resonator, $r/a > 1.1$ ($g/\sqrt{\gamma_0 \Gamma_{res}} < 1$), the decay rate of the radially oriented EQS increases at the approach to the nanocavity surface (the Purcell effect [19]). This review is devoted to this effect. At the distance $r/a \approx 1.07$ ($g/\sqrt{\gamma_0 \Gamma_{res}} \approx 10$), the weak coupling regime is completely replaced by the strong coupling regime. In this regime splitting of the initially identical frequencies of the EQS and the resonator appears (Rabi splitting, Fig. 4a) appears. In Fig.4b one can see the significant changes in the linewidths of the EQS and resonator: the EQS linewidth increases at the approach to the resonator surface, while the linewidth corresponding to the resonator decreases. In the case when the EQS transition frequency is slightly less than the exact resonance frequency, the situation is reversed: the EQS linewidth begins to decrease as it approaches the resonator surface, while the resonator linewidth becomes larger [51].

The realization of the strong coupling regime of epitaxial InAs quantum dots to microcavities based on photonic crystals was demonstrated experimentally in [55-57]. In [58], the regime of strong coupling of SiV⁻ centers to a linear nanoresonator based on a perforated diamond nanowaveguide with a triangular cross-section was demonstrated. A review of works on the study of the strong interaction regime in semiconductors is given in [59].

Recently, the regime of strong coupling of TDBC dye molecules to a plasmonic nanoparticle was demonstrated in [60]. The use of TDBC dye molecules is very beneficial for achieving the strong coupling regime, since excitons in J-aggregates formed by these molecules have much larger transition dipole moments than the total dipole moments of the constituent molecules. Fig.5 shows the geometry of the experimental sample and the experimental extinction spectra obtained on it. Figure 5b shows clearly the formation of the dispersion curves anti-crossing, indicating a transition to the strong interaction mode (see Fig. 4a). [61] provides an overview of the current situation in the study of the strong coupling regime.

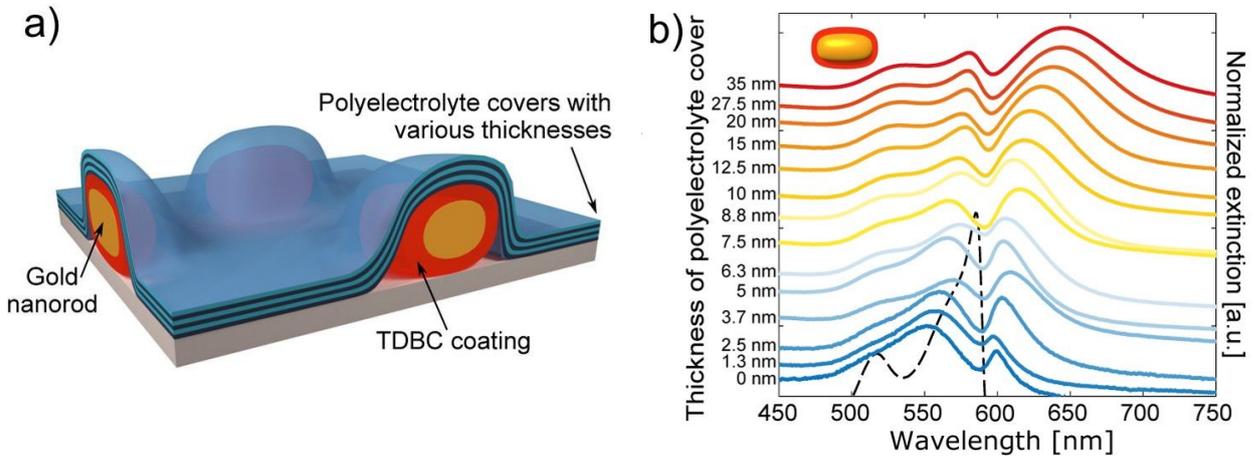

Fig. 5. a) Geometry of an experimental sample demonstrating the strong coupling of the TDBC dye molecules to the gold nanorod of the size of 25 nm × 57 nm. b) Extinction spectra of such a system at different thicknesses of the polyelectrolyte layer, regulating the position of the plasmon resonance and, consequently, the coupling strength. The dotted line shows the TDBC spectrum without nanoparticles [60].

With a further increase in the coupling strength, that is, when the decay rates (linewidths) and frequency shifts become comparable to the fundamental frequency, that is, under the condition that $g/\sqrt{\gamma_0 \Gamma_{res}} \gg 1, g/\omega_0 \sim 1$, the ultrastrong coupling regime occurs. In principle, such a regime cannot be achieved on the basis of systems with single EQS, when it can be shown [62] that $g/\omega_0 \sim \alpha^{3/2}$, where $\alpha \approx 1/137$ is the fine structure constant. Therefore, to demonstrate the ultrastrong coupling regime, systems consisting of multiple EQS should be used. Quite recently, it was demonstrated in [63] that the regime of deep ultrastrong coupling of localized plasmons and photons (g/ω₀ ~ 0.55) was achieved at room temperature. A review of recent works on the ultrastrong coupling regime is given in [64, 65].

In the rest of the review, the weak coupling regime conditions $g/\sqrt{\gamma_0 \Gamma_{res}} \ll 1, g/\omega_0 \ll 1$ will be considered satisfied.

2.4. Forbidden Transitions

Above, we have considered the description of spontaneous emission of electric dipole (allowed) transitions. However, in applications (spectroscopy, laser physics, quantum computing, etc.), forbidden magnetic-dipole, MD, and electric quadrupole, EQ, transitions are also of great importance. These transitions are also influenced by the nanoenvironment. Moreover, the influence of the nanoenvironment on these transitions is often more significant than on allowed ones [66]. The theory of spontaneous emission for forbidden transitions was presented in [40, 67] for emission of a molecule near a planar interface and in [68 - 75] for boundaries of an arbitrary shape.

The theory of spontaneous emission for arbitrary forbidden transitions is constructed analogously to the case of allowed transitions [68]. The general expression for the total decay rate is:

$$\frac{\Gamma}{\gamma_0} = \frac{\operatorname{Im} \int d^3\mathbf{r} \int d^3\mathbf{r}' \mathbf{J}^*(\mathbf{r},\omega) \ddot{\mathbf{G}}(\mathbf{r},\mathbf{r}',\omega) \mathbf{J}(\mathbf{r}',\omega)}{\operatorname{Im} \int d^3\mathbf{r} \int d^3\mathbf{r}' \mathbf{J}^*(\mathbf{r},\omega) \ddot{\mathbf{G}}^{(0)}(\mathbf{r},\mathbf{r}',\omega) \mathbf{J}(\mathbf{r}',\omega)}, \qquad (23)$$

where $J_j(\mathbf{r},\omega)$ is the classical current corresponding to the transition, and the Green's functions $G$ and $G^{(0)}$ are defined in (11) and (12).

For quadrupole *EQ* transitions

$$J_i^{EQ}(\mathbf{r}) = \frac{i\omega}{3!} \sum_{i,j} Q_{ij} \nabla_j \delta(\mathbf{r}-\mathbf{r}'); Q_{ij} = \int d\mathbf{r} \rho(\mathbf{r})\left(3 r_i r_j - \mathbf{r}^2 \delta_{ij}\right) \qquad (24)$$

we have

$$\frac{\Gamma^{EQ}}{\gamma_0^{VQ}} = 5 \frac{\operatorname{Im} \lim_{\mathbf{r}\to\mathbf{r}'} Q_{ij}^* Q_{kl} \nabla_j \nabla_l' G_{ik}(\mathbf{r},\mathbf{r}';\omega)}{k^5 \sum_{ij} |D_{ij}|^2} \qquad (25)$$

For magnetic-dipole *MD* transitions

$$\mathbf{J}^{MD} = -c[\mathbf{m}_0 \nabla]\delta(\mathbf{r}-\mathbf{r}'); \mathbf{m}_0 = \frac{1}{2c}\int d\mathbf{r}[\mathbf{r}\mathbf{j}(\mathbf{r})] \qquad (26)$$

for the rate of spontaneous emission, respectively, we have:

$$\frac{\Gamma^{MD}}{\gamma_0^{MD}} = \frac{3}{2}\frac{\operatorname{Im}\mathbf{m}_0^* \ddot{\mathbf{G}}^{MD} \mathbf{m}_0}{m_0^2 k_0^5} \qquad (27)$$

where $G^{MD}$ – Green's functions of the magnetic dipole

$$k_0^2 G_{ij}^{MD}(\mathbf{r},\mathbf{r}',\omega) = -\varepsilon_{ikl}\varepsilon_{jmn}\nabla_k \nabla'_m G_{ln}(\mathbf{r},\mathbf{r}';\omega) \qquad (28)$$

If there are several multipole current components (for example, for chiral molecules when both the electric and magnetic dipole moments of the transition are nonzero), then there is the interference between different decay channels [69-73].

## 2.5. Exact Solutions to the Problem of EQS Radiation

As shown in the previous section, the rate of spontaneous emission of any quantum systems in an arbitrary environment is determined by the classical Green's function of the system (11) and (12). However, since it is often difficult to obtain specific results, exact analytical solutions are of great importance.

An exact analytical description of the spontaneous emission rate is known for 3 geometries:

• EQS near flat interfaces [40],
• EQS near spherical interfaces [42,43,74,76,77,78],
• EQS near cylindrical interfaces [79,80].

The geometries of these problems are shown in Fig. 6.

The simplest is the description of the emission of a molecule near a spherical particle with a dielectric constant $\varepsilon_1$, located in a medium with the dielectric constant $\varepsilon_2$. In the case of a radially oriented dipole, only TM modes of the particle are excited, and the expression for the spontaneous decay rate has the form [42,43,74]:

$$\frac{\Gamma_{rad}^{ED}}{\gamma_0} = 1 - \frac{3}{2}\operatorname{Re}\sum_{n=0}^{\infty} n(n+1)(2n+1) q_n \left(\frac{h_n^{(1)}(\sqrt{\varepsilon_2}k_0 r)}{k_0\sqrt{\varepsilon_2}r}\right)^2,$$

$$q_n = \frac{\varepsilon_1 \left[z_2 j_n(z_2)\right]' j_n(z_1) - \varepsilon_2 \left[z_1 j_n(z_1)\right]' j_n(z_2)}{\varepsilon_1 \left[z_2 h_n^{(1)}(z_2)\right]' j_n^{(1)}(z_1) - \varepsilon_2 \left[z_1 j_n^{(1)}(z_1)\right]' h_n^{(1)}(z_2)}, \qquad (29)$$

where $j_n$ and $h_n$ are the spherical Bessel and Hankel function, $z_{1,2} = \sqrt{\varepsilon_{1,2}} k_0 a$ and $q_n$ are Mie reflection coefficients. In the case of the tangential orientation of the EQS dipole moment, the expression is a little more complicated, since in this case both TE and TM modes are excited [42,43,74]. A generalization to the case of EQS radiation in the presence of several spherical particles can be found in [81].

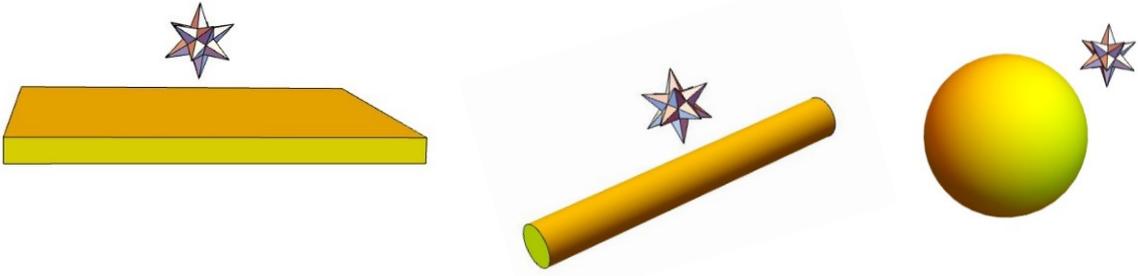

Figure: 6. Geometries of the problem of the spontaneous emission of the EQS, where there is an exact analytical description: a) lamellar (layered) structures, b) a circular cylinder, c) spherical nanoenvironment. The star denotes EQS.

In the case of an EQS near a planar system, the total decay rate is completely determined by the reflection coefficient of plane waves from this surface [40,41]. For a dipole located at the distance $d$ from the surface, we have the following expressions:

$$\frac{\Gamma_\perp}{\gamma_0} = 1 + \frac{3}{2k_0^3} \text{Re} \int_0^\infty r_{12}^p e^{2i\beta_1 d} \frac{q^3}{\beta_1} dq \qquad (30)$$

$$\frac{\Gamma_\parallel}{\gamma_0} = 1 - \frac{3}{4k_0^3} \text{Re} \int_0^\infty \left[\beta_1^2 r_{12}^p - k_0^2 r_{12}^s\right] e^{2i\beta_1 d} \frac{q}{\beta_1} dq \qquad (31)$$

In (30) and (31)

$$r_{ij}^p = \frac{\varepsilon_j \beta_i - \varepsilon_i \beta_j}{\varepsilon_j \beta_i - \varepsilon_i \beta_j}, \quad r_{ij}^s = \frac{\beta_i - \beta_j}{\beta_i + \beta_j} \qquad (32)$$

are usual Fresnel reflection coefficients for p and s polarized waves, $\beta_i = \sqrt{\varepsilon_i k_0^2 - q^2}$, the origin is at the interface.

In the case of an EQS located in vacuum at the distance $d$ from an ideally conducting plane ("mirror"), instead of (27) and (28) we have:

$$\frac{\Gamma_\perp}{\gamma_0} = \left(1 + 3\left(\frac{\sin(\tau)}{\tau^3} - \frac{\cos(\tau)}{\tau^2}\right)\right)_{\tau=2k_0 d}$$
$$\frac{\Gamma_\parallel}{\gamma_0} = \left(1 - \frac{3}{2}\left(\frac{\sin(\tau)}{\tau} + \frac{\cos(\tau)}{\tau^2} - \frac{\sin(\tau)}{\tau^3}\right)\right)_{\tau=2k_0 d}$$

(33)

Note that the expressions (33) can be obtained by the method of images, suggesting that the field reflected by the mirror is equal to the field of an image dipole, located at the mirror-image point relative to the real dipole, and the orientation of the fictitious dipole coincides with the orientation of the dipole perpendicular to the surface and is antiparallel to it in the case of its tangential orientation (see Fig. 3). In the limit $d \to 0$ from (33), it is easy to find that $\Gamma_\perp / \gamma_0 = 2, \Gamma_\parallel / \gamma_0 = 0$, and this can also be explained by the method of images readily.

In the case of a EQS near a cylindrical object, one can also find exact expressions for the rate of spontaneous emission, but they are cumbersome to review and are given in [79,80].

2.6 Quasistatic Case

In the case of nanobodies or nanoparticles, to describe the EQS spontaneous emission, it is often possible to use the perturbation theory with respect to the small parameter

$$ka = 2\pi a / \lambda \ll 1,$$ (34),

where $a$ is the characteristic size of the nanobody and $\lambda$ is the radiation wavelength. In this case, the expression for the reflected field in (7) can be represented as a series in powers of the wave vector $k$ [80,82]:

$$\frac{\mathbf{d}_0 \mathbf{E}^{(R)}(\mathbf{r},\mathbf{r},\omega_0)}{d_0^2} = a_1(\mathbf{r}) + b_1(\mathbf{r})k + c_1(\mathbf{r})k^2 + id_1(\mathbf{r})k^3 + ..., \quad (35),$$

where $a_1, b_1, c_1, d_1$ are the coefficients that can be found from the solution of the corresponding quasistatic problems. The first three terms describe the nearfield effects, while the radiation effects arise from the 4th term. Substituting the expansion (35) into (9) for the total rate of the EQS spontaneous emission near the nanobody, we obtain

$$\frac{\Gamma}{\gamma_0} = \frac{3}{2}\mathrm{Im}\left(\frac{a_1(\mathbf{r})}{k^3} + \frac{b_1(\mathbf{r})}{k^2} + \frac{c_1(\mathbf{r})}{k}\right) + 1 + \frac{3}{2}\mathrm{Re}(d_1(\mathbf{r})) + .... \quad (36).$$

In (36), the first term is nonzero only for absorbing media and describes nonradiative losses, while the second and third terms are nonzero even in the absence of absorption. These terms describe radiative losses. Thus, to find the nonradiative and radiative losses in the first approximation, it is sufficient to find $a_1(\mathbf{r})$ and $d_1(\mathbf{r})$, respectively.

To find the principal term with $a_1$, it suffices to solve the quasistatic problem with a dipole source [82]:

$$\begin{aligned} \mathbf{E}(\mathbf{r}) &= -\nabla\varphi(\mathbf{r}) \\ \nabla(\varepsilon(\mathbf{r})\nabla\varphi)(\mathbf{r}) &= 4\pi(\mathbf{d}_0\nabla')\delta^{(3)}(\mathbf{r}-\mathbf{r}')), \\ a_1 &= \frac{\mathbf{d}_0\mathbf{E}(\mathbf{r}=\mathbf{r}')}{d_0^2} \end{aligned} \quad (37)$$

where $\nabla, \nabla'$ mean the gradient in the coordinates $\mathbf{r}$ and $\mathbf{r}'$ (the atom position), respectively. If the EQS is located very close to the surface (at a distance $\Delta \ll \lambda$), then the expressions for the nonradiative decay rate have a universal form [82, 104 Chapter 6]:

$$\frac{\Gamma^{nonrad}}{\gamma_0} = \begin{cases} \dfrac{3}{16(k\Delta)^3}\mathrm{Im}\dfrac{\varepsilon-1}{\varepsilon+1}, & \text{radial dipole orientation} \\[2mm] \dfrac{3}{32(k\Delta)^3}\mathrm{Im}\dfrac{\varepsilon-1}{\varepsilon+1}, & \text{tangential dipole orientation} \end{cases} \quad (38)$$

It is difficult to find the 4th term $d_1(\mathbf{r})$ by direct calculation. However, physical considerations allow us to understand that it describes the dipole radiation of the entire system, that is, the radiative decay rate will be described by the expression [44,82,85]:

$$\frac{\Gamma^{rad}}{\gamma_0} = \frac{|\mathbf{d}_{total}|^2}{|\mathbf{d}_0|^2} = \frac{|\mathbf{d}_0 + \delta\mathbf{d}|^2}{|\mathbf{d}_0|^2}, \qquad (39),$$

where $\mathbf{d}_{total}$ is the total dipole moment of the EQS + nanobody system, which can be found from the asymptotics of the solution (37)

$$\varphi(\mathbf{r}) \xrightarrow[r\to\infty]{} \frac{\mathbf{d}_{total} \cdot \mathbf{r}}{r^3}. \qquad (40).$$

As a result, to find changes in the rate of EQS radiation in the presence of any nanoobject of the sizes that are small in comparison with the radiation wavelength, it is sufficient to solve the quasistatic problem (37) on the EQS near this nanoobject. The class of such problems is much wider than in the case of objects of arbitrary sizes. If the quasistatic condition (34) is applicable, one can find not only solutions for a sphere, plane, and cylinder, but also for a spheroid, ellipsoid, and a cluster of two spherical or spheroidal nanoparticles [80,82-88].

2.7. Mode Expansions of the Green's Function and Their Relationship with the Purcell Formula

The interaction of the EQS radiation with the nanoenvironment depends fundamentally on the resonance modes of the latter, and the development of this direction has begun with the Purcell's work [19] on the resonant interaction of the EQS with the environment. Therefore, for a deeper understanding of the physics of the EQS interaction with the environment and, consequently, for effective control of its radiation, a fundamental comprehension of the environment resonant properties is necessary. Here, however, there are several unresolved problems.

In ordinary closed cavities, the eigenfrequencies are real, the eigenfunctions are orthogonal [89, 90] and localized in the cavity volume. This allows us to expand any field in the resonator in terms of them and describe the process of excitation of the resonator [89,90].

In the case of open resonators - and this case is the subject of this review - such an expansion is not always possible, since, due to radiation, the eigenfrequencies are complex numbers $\Omega_n = \omega_n + i\Gamma_n/2$, leading to eigenmode fields increase at infinity.

$$E_n(\mathbf{r}) \sim \frac{\exp(i\Omega_n r/c)}{r} = \frac{\exp(i\omega_n r/c)\exp(\Gamma_n r/2c)}{r} \xrightarrow[r\to\infty]{} \infty \quad (41)$$

L.A. Weinstein [90] has been the first to draw attention to this fact. The eigenmodes of open systems are also known as decay states, leaky modes or quasimodes. The main problem in the description and in computer simulation of quasimodes or modes of open resonators is to determine how they can be normalized. Since these modes increase at infinity, the usual normalization procedures do not work, and non-standard solutions are needed.

Several approaches to the normalization of quasimodes were proposed in [91-97]. In [97], it was proposed to normalize the eigenfunctions that are solutions of Maxwell's equations without sources

$$\nabla \times (\nabla \times \mathbf{E}_n) - k_{0,n}^2 \varepsilon(\mathbf{r})\mathbf{E}_n = 0 \quad (42)$$

using the relation

$$\int_V d\mathbf{r}\varepsilon(\mathbf{r})\mathbf{E}_n^2(\mathbf{r}) - \lim_{k\to k_n}\frac{\int_{S_v} dS\left(\mathbf{E}_n \cdot \frac{\partial \mathbf{E}(k)}{\partial s} - \mathbf{E}(k) \cdot \frac{\partial \mathbf{E}_n}{\partial s}\right)}{k_n^2 - k^2} = 1, \quad (43)$$

where each term tends to infinity, but their sum remains finite. With this normalization, the expression for the Green's function in an arbitrary nanoenvironment can be expressed in the form [97]:

$$G_{ij}(\mathbf{r},\mathbf{r}';\omega) = \frac{\omega}{2}\sum_n \frac{E_i^{(n)}(\mathbf{r})E_j^{(n)}(\mathbf{r}')}{\omega - \omega_n} \quad (44)$$

If the radiation occurs near the frequency of one high-Q eigenmode, then the resonant approximation of the Green's function

$$G_{ij}^{(n)}(\mathbf{r},\mathbf{r}';\omega) \approx \frac{\omega_n}{2} \frac{E_i^{(n)}(\mathbf{r}) E_j^{(n)}(\mathbf{r}')}{\omega - \omega_n} \tag{45}$$

can be used.

Substituting (45) into (16) we obtain

$$\frac{\Gamma}{\gamma_0} = \frac{\lambda^3 3(\mathbf{nE}^{(n)})^2}{32\pi^3} \frac{\omega \Gamma_{res}^{(n)}}{(\omega - \omega_{res}^{(n)})^2 + \Gamma_{res}^{(n)2}}, \tag{46}$$

which coincides with the Purcell-Bunkin-Oraevsky formula (5) if one put

$$V_c(\mathbf{r}) = \frac{4\pi}{(\mathbf{nE}^{(n)}(\mathbf{r}))^2}; \quad \Gamma_{res}^{(n)} = \frac{\omega_{res}^{(n)}}{Q} \tag{47}$$

The Purcell's formula in this form appears in many works [93,95,96,91,92].

However, as it has been already mentioned, the determination of the mode volume remains a rather arbitrary procedure, requiring a verification each time.

To control radiation, one must be able not only to amplify, but also to suppress it. This is important, when all the EQS energy needs to be transferred to the resonator modes to excite certain nonlinear processes. In this case, of course, we are not talking about quenching of fluorescence, that, in our case, is reduced to the fact that the excitation energy of the EQS heats the nanoparticle or nanoantenna. This process is well-studied and described completely using the nonradiative part of the spontaneous decay rate (15), (38). We are also not talking about placing the EQS in conditions where there are no radiation modes (in the band gap of a photonic crystal [20-23]). It would be much more interesting to create conditions under which the energy of the excited EQS excites a dielectric particle in free space without radiation loss effectively. This is a nontrivial problem associated with the nonradiating modes existing in the resonator - pseudomodes, that were described theoretically only recently in [98].

2.8. Using the Reciprocity Theorem to Calculate the Purcell Factor

In some symmetric cases or cases where only one eigenmode plays the main role, the Purcell factor can be calculated using the Lorentz reciprocity theorem [89]. In particular, the radiation of an atom near a single-mode waveguide can be reduced to the problem of finding the mode of this waveguide [99,100]. The reciprocity theorem can also be effective for to accurate computation of the Purcell factor averaged over all dipole orientations [101].

## 2.9. Fluorescence Intensity and Other Observed Parameters of EQS radiation

In addition to the spontaneous emission rate, several other EQS radiation characteristics are also important for applications. In particular, the directivity factor $D$ of the EQS radiation, describing the properties of the field in the far zone, is of particular importance:

$$D = \frac{4\pi p(\theta,\varphi)}{P_{rad}}, \tag{48}$$

where $p(\theta,\varphi)$ is the angular distribution of the radiation power, and $P_{rad}$ is the total radiation power obtained by integrating $p(\theta,\varphi)$ over the angles $\theta,\varphi$ of the spherical coordinate system. This characteristic is essential for efficient collection of photons and construction of detectors.

Another important characteristic of the EQS is its radiation efficiency determined by the ratio of the radiated power $P_{rad}$ to the power supplied to the EQS:

$$\eta = \frac{P_{rad}}{P_{rad} + P_{loss}} = \frac{P_{rad}}{P_{rad} + P_{nonrad} + P_{loss}^0} \tag{49}$$

In (49), $P_{loss}$ and $P_{loss}^0$ stand for all losses and internal losses of the EQS in vacuum, respectively. If the internal EQS losses are described by the quantum (internal) efficiency:

$$\eta_q = \frac{P^0}{P^0 + P_{loss}^0}, \tag{50}$$

where $P^0$ stand for the radiation power of the EQS in vacuum, then the expression for the total radiative efficiency of the EQS (49) takes the form [102]:

$$\eta = \frac{P_{rad}/P^0}{P_{rad}/P^0 + P_{nonrad}/P^0 + (1-\eta_q)/\eta_q} = \frac{F_{P,rad}}{F_{P,rad} + F_{P,nonrad} + (1-\eta_q)/\eta_q}, \quad (51)$$

where $F_{P,rad}, F_{P,nonrad}$ are the radiative and non-radiative Purcell factors (13)-(15).

Above, we have presented a theory of the influence of the nanoenvironment on the radiation properties of a single optical transition. However, in practice, more often it is necessary to deal with more complex processes, for example, with fluorescence, when several different transitions are involved, associated with both absorption of radiation and its emission (Fig. 7). In the case of fluorescence, it is assumed that the excitation of the upper level $S_2$ of the EQS occurs at the frequency $\omega_2$, and the fluorescence occurs at the frequency $\omega_1$ from the level $S_1$, where the system falls during the vibrational relaxation of the $S_2$ level at the rate $K$. It has been already shown above that the linewidths (the decay rates) $\Gamma_1$, $\Gamma_2$, depend on the nanoenvironment substantially. The pumping field $E_{ext}$ also depends significantly on the nanoenvironment. Thus, in the case of an EQS in a nanoenvironment, all quantities characterizing fluorescence depend on the position of the EQS relative to the nanoenvironment.

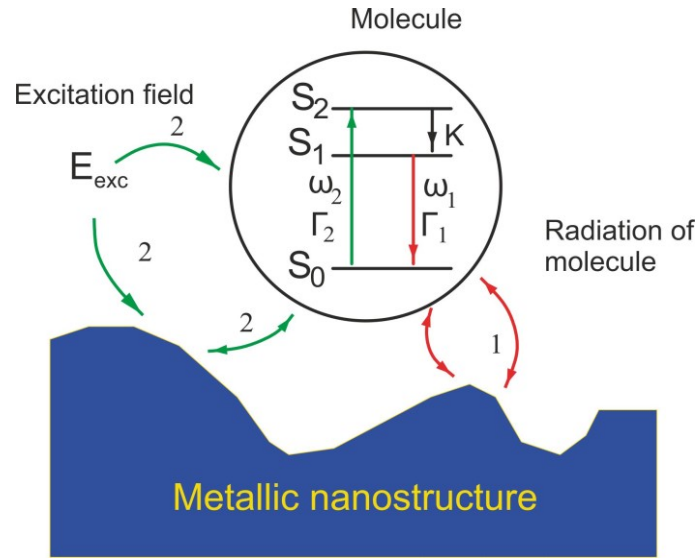

Fig. 7. The scheme of parameters characterizing fluorescence in the presence of nanoobjects is adapted from [103]. Excitation of the upper level $S_2$ of the EQS occurs at the frequency $\omega_2$, and the fluorescence occurs at the frequency $\omega_1$ from the level $S_1$, where the system falls during the vibrational relaxation of the $S_2$ level at the rate K.

As a result of solving the generalized Bloch equations, the formula for the fluorescence intensity can be written in the form [103]:

$$I_{fl}(\mathbf{r}) = \frac{\hbar\omega_1(\mathbf{r})}{\dfrac{1}{\tilde{\Gamma}_1(\mathbf{r})} + \dfrac{2}{K} + \dfrac{1}{\Gamma_{pump}(\mathbf{r})}\left(1 + \dfrac{\Gamma_2(\mathbf{r})}{K}\right)} \eta_1(\mathbf{r}),  \qquad (52)$$

where $\tilde{\Gamma}_1(\mathbf{r}) = \Gamma_1(\mathbf{r}) + \Gamma_{1,nonopt} = \Gamma_{1,rad}(\mathbf{r}) + \Gamma_{1,nonrad}(\mathbf{r}) + \Gamma_{1,nonopt}$ is the total rate of optical and non-optical transitions from the $S_1$ level to the $S_0$ level, $1/K \sim 10^{-10} \div 10^{-12} s$ is the vibrational relaxation time of the $S_2$ level, and $\Gamma_1(\mathbf{r})$ is the rate of optical spontaneous transitions from the $S_1$ level to the $S_0$ level. The pumping rate $\Gamma_{pump}(\mathbf{r}) = 4|\Omega(\mathbf{r})|^2/\Gamma_2$ is determined by the Rabi frequency $\Omega(\mathbf{r}) = \mathbf{dE}_{ext}(\mathbf{r})/\hbar$, that in its turn depends on the EQS transition dipole moment and the external electric field $\mathbf{E}_{ext}(\mathbf{r})$. The factor $\eta_1(\mathbf{r}) = \Gamma_{1,rad}(\mathbf{r})/\tilde{\Gamma}_1(\mathbf{r})$ determines the radiation efficiency of the fluorescence, that is, the probability of the real photon emission during the process.

It can be seen from (52) that the spatial dependence of the fluorescence intensity on the position of the molecule is determined by the corresponding linewidths and frequency shifts, that can be found using the methods described above. In addition, the intensity of fluorescence is determined by the value of the local Rabi frequency (local electric field), which also undergoes significant changes in the presence of nanobodies.

In the case of a strong external field (the case of saturation, $\Gamma_{pump}(\mathbf{r}) \gg \Gamma_1, K$), the expression (52) is simplified greatly:

$$I_{fl}(\mathbf{r}) = \frac{\hbar\omega_1(\mathbf{r})\Gamma_{1,rad}(\mathbf{r})}{1 + 2\tilde{\Gamma}_1/K} \qquad (53)$$

It is important to note that at not large nonradiative losses $\left((\Gamma_{1,nonrad}(\mathbf{r}) + \Gamma_{1,nonoptical})/K \leq 1\right)$, an increase in the rate of radiative transitions near nanobodies (for example, near plasmon resonances in nanoparticles) leads to a proportional increase in the fluorescence intensity. This effect makes it possible to develop new types of nanosensors. The case of a strong external field is especially important, since it allows one to exclude from consideration the spatial structure of the exciting field.

In the case of small pumping fields $\Gamma_{pump}(\mathbf{r}) \ll \Gamma_1, K$, instead of (52), respectively, we have

$$I_{fl}(\mathbf{r}) = \frac{\hbar\omega_1(\mathbf{r})\Gamma_{pump}(\mathbf{r})}{\left(1 + \frac{\Gamma_2(\mathbf{r})}{K}\right)} \eta_{1q}(\mathbf{r}) \qquad (54)$$

In this case, the intensity is determined by a nonlinear combination of factors varying in space significantly, that, in its turn, can be found again using the methods outlined above.

The combination of the regimes of strong and weak exciting fields in the experiment enables an unambiguous determination of both the dependence of the transition linewidth on the position of the EQS (Purcell factor) and the spatial structure of the local pumping field.

## 3. EQS Radiation near Plasmonic Nanoparticles and Nanostructures

Even though theoretical approaches stated above are based on the fundamental principles of quantum and classical electrodynamics, the experimental verification of specific theoretical results is necessary, since the radiation process is extremely complicated, and moreover, the use of the concept of permittivity becomes inapplicable on nanoscale sooner or later.
A systematic theoretical and experimental study of the effect of the nanoenvironment on the EQS radiation has been initiated by considering plasmonic nanostructures with a negative permittivity in the optical or near-IR frequency range. Negative permittivity allows surface plasmons [104], causing an increase in local fields, small mode volumes and, as a result, a significant enhancement of spontaneous emission processes and the fluorescence intensity increase. In this frequency range, even noble metals (gold and silver) have significant losses, reducing the proportion of radiative processes during EQS emission. However, despite this, there are good reasons, both technological and physical, for controlling the processes of EQS emission using plasmon materials [105].

Experiments with individual EQSs near macroscopic bodies are extremely complicated and currently the most reliable experimental data are available only for the EQS lifetime near a partially reflecting flat surface [106-108] (see Fig.8a). In these experiments, fluorescence was studied in samples with different widths of the dielectric layer separating the $Eu^{3+}$ ion from the mirror. The measurement results are shown in Fig.8b.

Figure 8b shows how the lifetime oscillations of $Eu^{3+}$ ions occur, due to the interference of $Eu3^+$ ion radiation with the reflection from the metal film. Another feature of Fig. 8b is that the molecules lifetime tends to zero as approaching the metal surface (see also Fig.3). This phenomenon is well-known and called fluorescence quenching. Here, an extremely important aspect of the description and control of the spontaneous emission of the EQS manifests itself, consisting of the presence of two main channels for the decay of the excited state: radiative and nonradiative. This is often forgotten, leading to incorrect conclusions. In this case, at small distances from the surface, the dominance of the nonradiative channel manifests itself, and this is well described by the expression (38).

The spherical geometry is studied theoretically better than others [44,42,43,74,76-78]. The first experiments verifying the correctness of the theoretical description of the emission of Nile Blue dye molecules near the plasmon (Au) nanosphere, were carried out in [109,110] (Fig.9). Figure 9b shows again a good agreement between the theory and the experiment down to exceedingly small (a few nm) distances from the molecule to the surface.

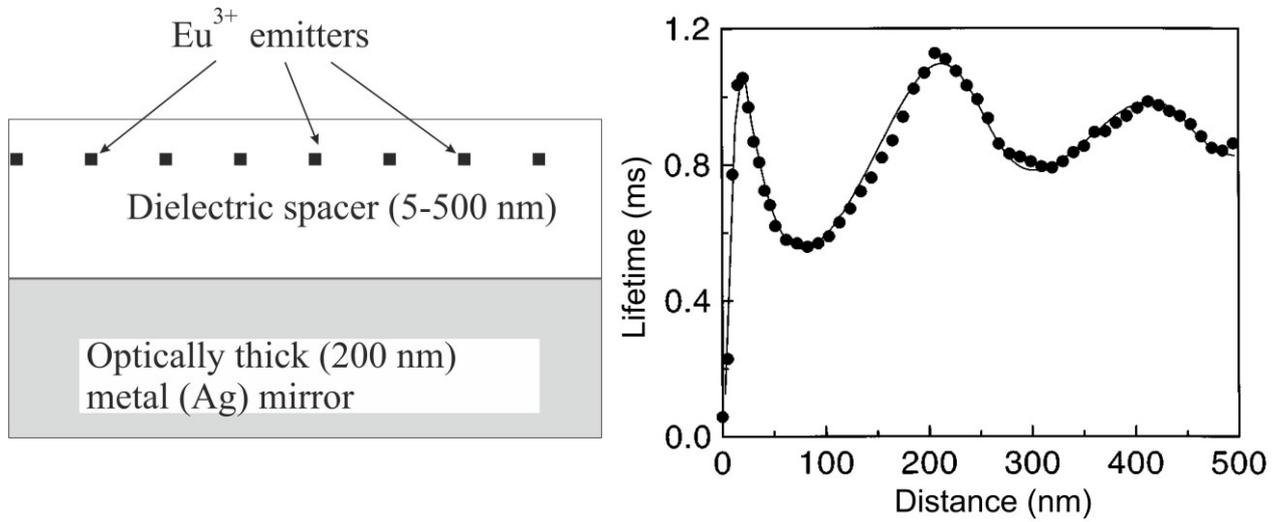

Fig. 8. a) Geometry of the problem of the emission of $Eu^{3+}$ ions near an Ag mirror; b) the lifetime $\tau = 1/\Gamma$ of the excited $Eu^{3+}$ ion as a function of the distance to the mirror [108] (dots are experimental data, solid curve is the theory, see (30) and (31)).

The study of the emission of CdSe quantum dots in the presence of a silver nanowire was carried out in [111] (Fig. 10). This work demonstrates that the spontaneous emission rate increases by 10-15 times (Fig.10b). In fact, not infinite nanowires, but elongated spheroids were studied in [111]. An analytical theory of spontaneous emission near spheroids and triaxial ellipsoids was constructed in [83-85]. The ellipsoid geometry [85] is important since both pumping and radiative rates can be enhanced due to the presence of plasmon resonances with different eigenfrequencies. For more symmetric nanoparticles, there is no such effect.

In [93,112,113], the methods of calculating the spontaneous emission rate of the EQS near plasmonic nanoparticles were proposed basing on the extraction of resonance modes or quasimodes and their numerical calculation (see 2.7). Fig. 11 shows the dependence of the total and nonradiative rates on the radiation wavelength of the dipole located near the nanocylinder. Calculating the field distribution of only one quasimode allows one to find a very good description of the spontaneous emission decay rate.

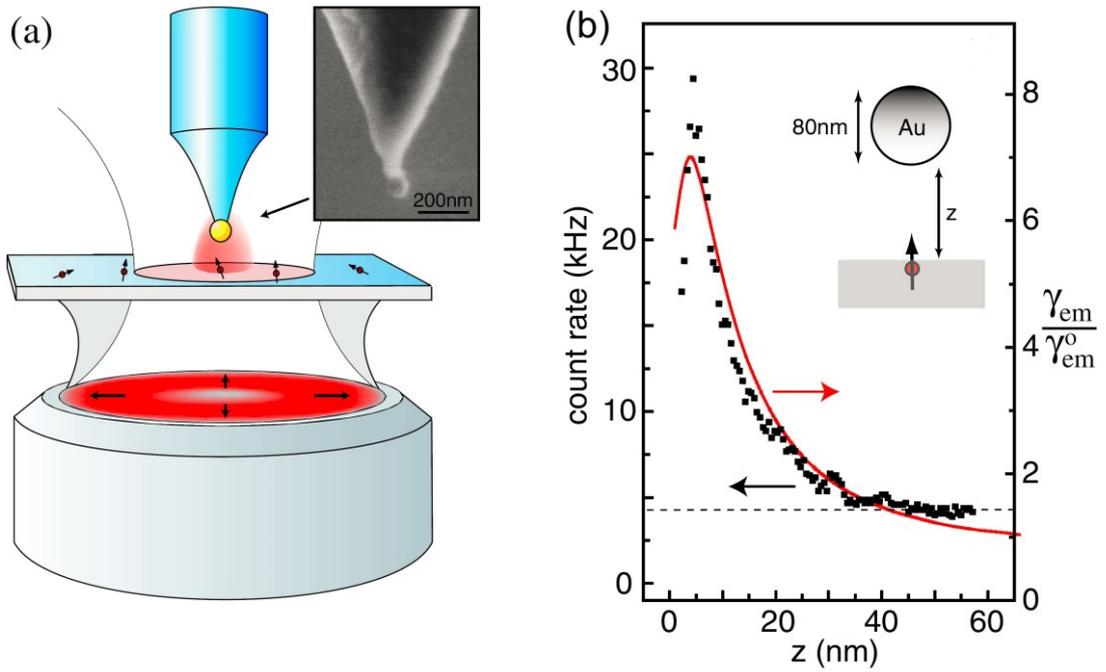

Fig. 9. a) The geometry of the experiment; b) the radiative decay rate of the spontaneous emission of Nile Blue dye molecules as a function of their distance from the surface of the sphere [109,110].

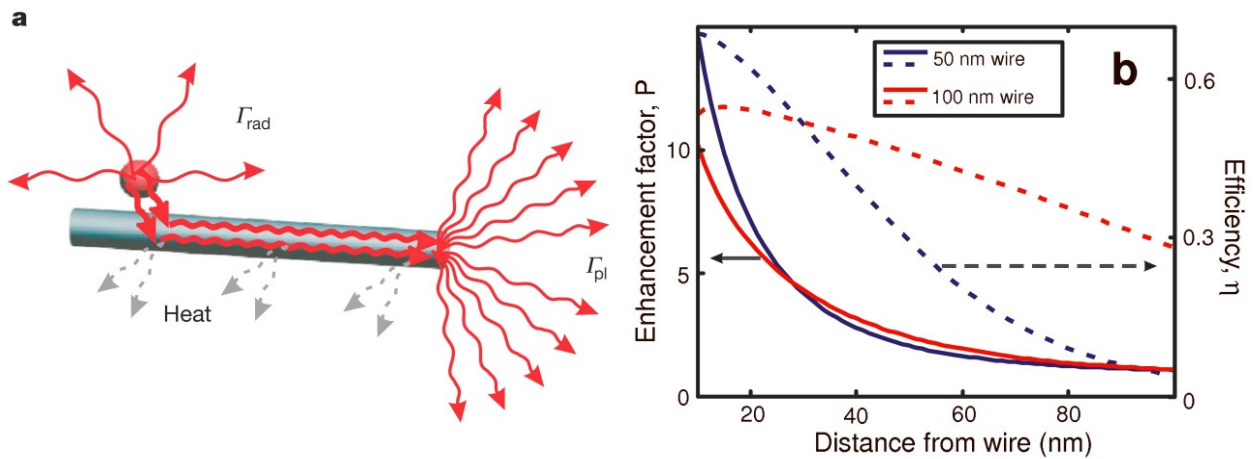

Fig. 10. a) The geometry of the experiment to control the emission of a CdSe quantum dot using a silver nanowire; b) the Purcell factor and the radiation efficiency of the CdSe quantum dot emission [111].

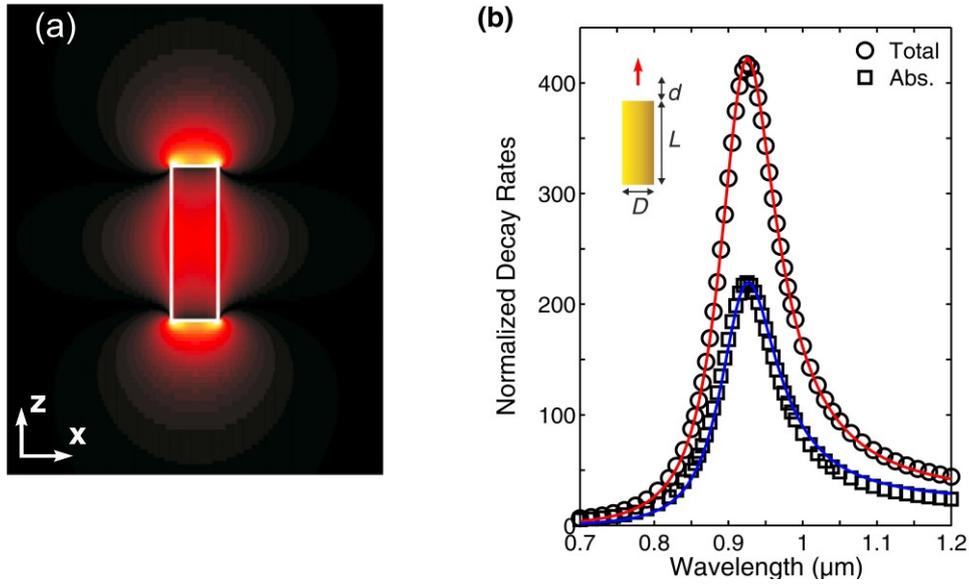

Fig. 11. The spontaneous emission near a gold cylinder 30 nm in diameter and 100 nm in length in a medium with permittivity $n=1.5$. (a) Distribution of $|E_z|$ in a quasimode with a complex frequency of $2\pi c/\omega = 920 + 47i$ nm. (b) The relative decay rate of the spontaneous emission of a dipole located on the cylinder axis (red arrow) at the distance $d = 10$ nm. The circles and the squares show the results of the full vector calculation of the total and nonradiative decay rates. The solid lines are the results of an approximate calculation taking into account only one quasimode [93].

From Fig. 11b it is seen that, indeed, the results of an approximate calculation using only one quasimode give particularly good agreement with the results of the exact numerical solution of Maxwell's equation. However, in each specific case, the question arises whether this method is more convenient than the direct numerical calculation.

To control the emission of the EQS and efficiently detect single photons, the use of more complex than single nanoparticles systems - so-called optical nanoantennas – is of great interest [3-7]. The optical nanoantennas are direct analogs of radio-frequency antennas [114-116], allowing one to control of EQS emission effectively as well. Two or more nanoparticles of various shapes are usually used as optical nanoantennas. In [117-120], an exhaustive analytical description of the theory of radiation from an EQS in the presence of a nanoantenna of two spherical or spheroidal nanoparticles was given.

An experimental study of the radiation of terrylene molecules in a thin layer of p-terfinil near a cluster of 2 nanospheres was carried out in [121] (Fig. 12a), where a significant increase in both total and radiative decay rates was demonstrated, as well as good agreement with theoretical calculations. In [122], the emission of CdSe / CdS quantum dots located in the gap between a gold film and a gold nanodisk 20 nm thick and 1.4 to 2.1 μm in diameter was studied theoretically and experimentally (see Fig. 12b). This geometry is often referred to as a patch antenna. The theoretical results and experimental measurements of the total spontaneous decay rate are in good agreement and indicate that the Purcell factor reaches values of several tens. The radiation pattern was also measured, and the measurement results were also in good agreement with the theoretical predictions.

In [123], a similar patch nanoantenna was studied, but instead of a gold disk, a silver nanocube was used, and instead of quantum dots - ruthenium-based dye molecules (ruthenium metal complex dye, Ru dye) with a long internal lifetime $\tau_0 = 1/\gamma_{0,sp} = 600 \pm 50$ ns (Fig. 13). As in [122], in [123] only the total decay rate was measured. The increase in the radiative decay rate was determined on the basis of indirect reasoning and turned out to be of the order of $10^3$. These works demonstrated an increase in both total and radiative decay rates by more than an order of magnitude at a gap thickness of about 11 nm and good agreement between theory and experiment.

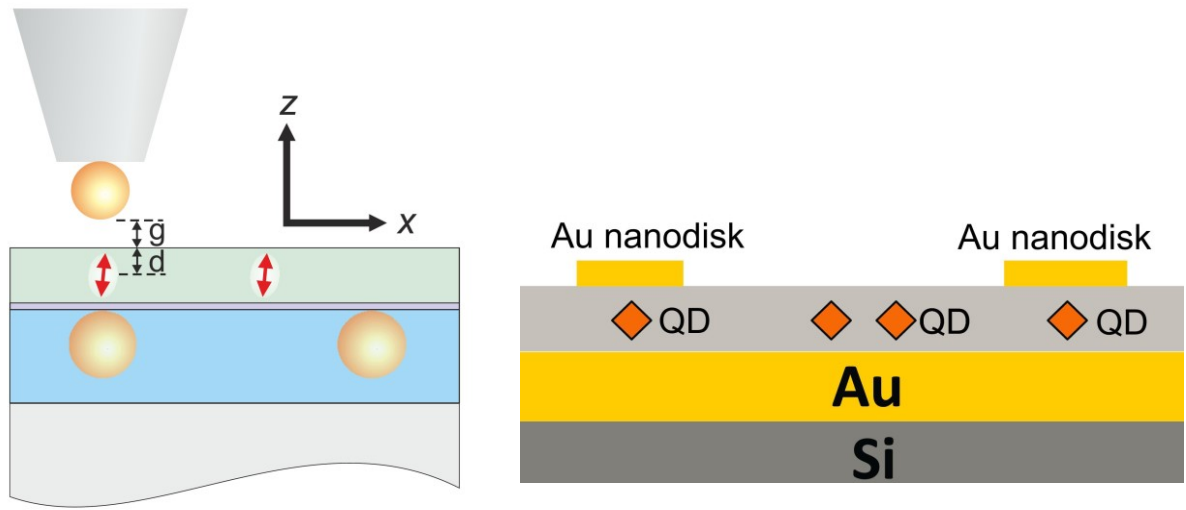

Fig.12 a) Geometry of the problem of radiation of terrylene molecules in a thin layer of p-terfinil near a cluster of 2 nanospheres [121]; b) geometry of the problem of emission of CdSe / CdS quantum dots located in the gap between a gold film and a gold nanodisc 20 nm thick and 1.4 to 2.1 μm in diameter [122].

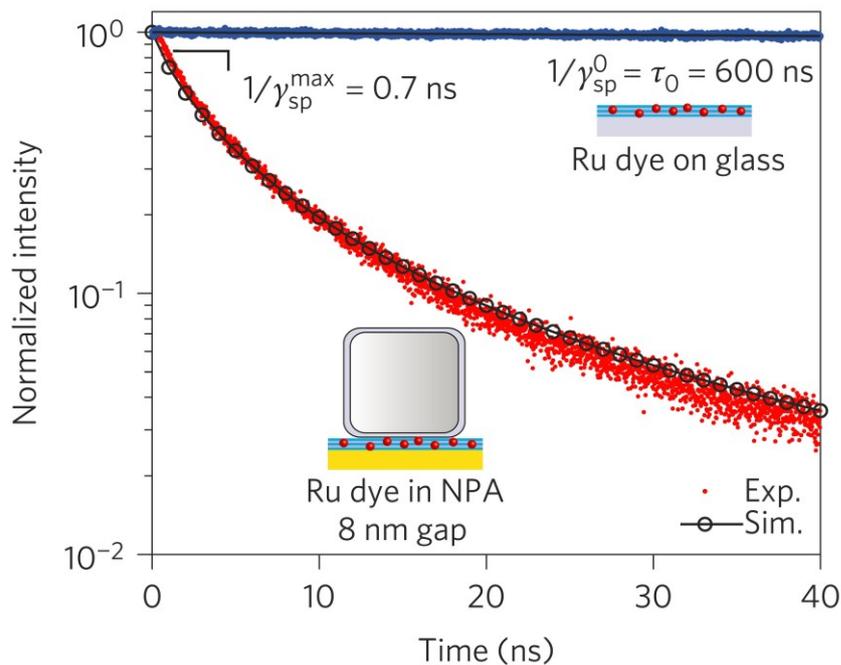

Fig.13. The geometry of the problem and the decay curve for ruthenium-based dye molecules located under a silver cube [123].

A theoretical and experimental study of the fluorescence of CdSe/CdS/ZnS colloidal quantum dots in the presence of a patch nanoantenna based on a silver triangular nanoprism was carried out in [124,125]. These studies demonstrated an increase in both total and radiative decay rates by more than an order of magnitude.

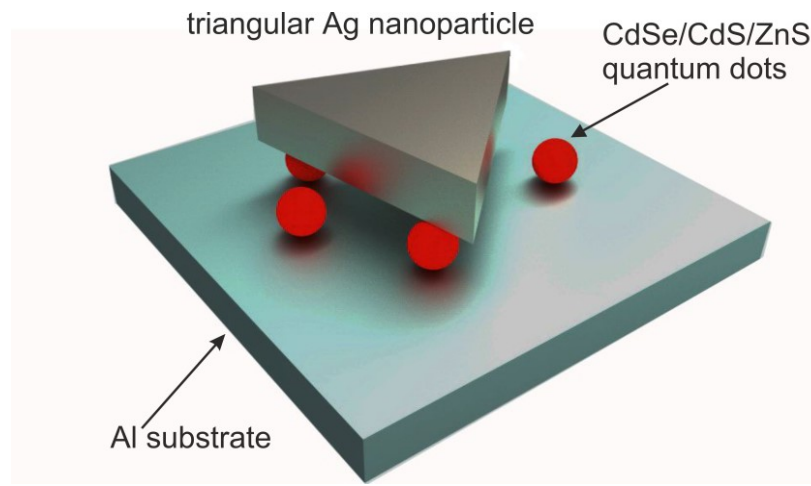

Fig. 14. Geometry of the problem of fluorescence of CdSe/CdS/ZnS colloidal quantum dots in the presence of a patch nanoantenna based on a silver triangular nanoprism [124,125].

An important geometry of optical nanoantennas is two-element nano-antennas, so-called bow-tie antennas (Fig.1). The synthesis of high-quality plasmonic bow-tie nanoantennas was first demonstrated in [126], where nano-antennas were etched from a single gold monocrystal using a focused ion beam. The use of a single monocrystal of gold made it possible not only to improve the geometry of the nanoantenna significantly (compare two images in the upper right corner of Fig.15a), but also to increase its photoluminescence substantially (by 2 orders of magnitude) (Fig. 15b), indicating a considerable enhancement in local fields and, hence, the figure of merit of plasmon oscillations in the nanoantenna.

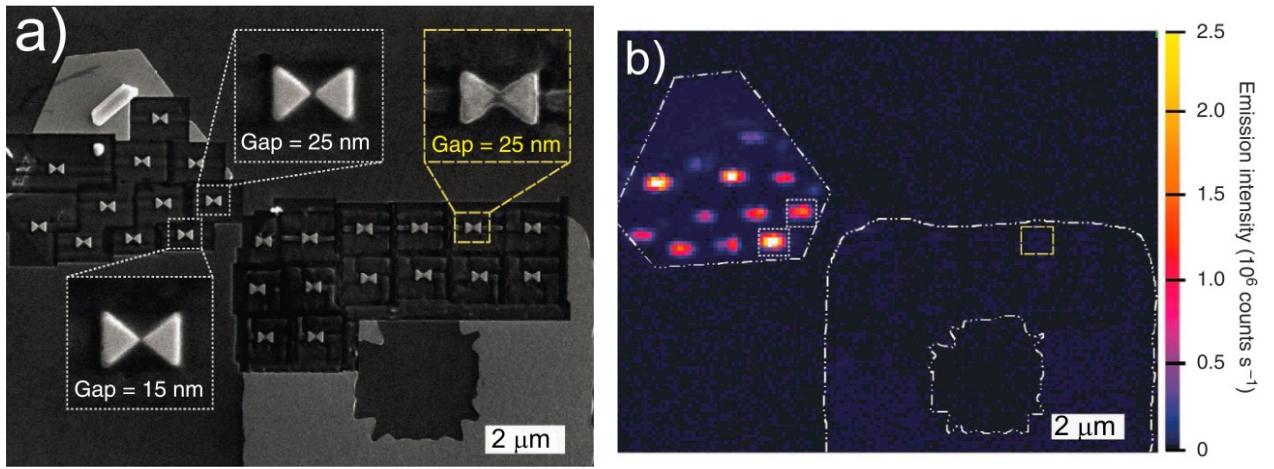

Fig. 15. a) A SEM-image of gold bow-tie nanoantennas based on a single monocrystal (upper left corner) and a polycrystalline film (upper right corner); b) image of a two-photon photoluminescence (TPPL) map of the sample shown in Fig.15a [126].

In [127,128], the influence of a gold bow-tie nanoantenna on the emission of a single InGaAs quantum dot located at a distance of 25 nm from it was studied (Fig. 16a). The theoretical calculation showed (Fig.16b) that the radiative decay rate should increase by an order of magnitude, while the non-radiative decay rate should be of an order of magnitude lower. However, in the experiment, no change in the total decay rate was recorded.

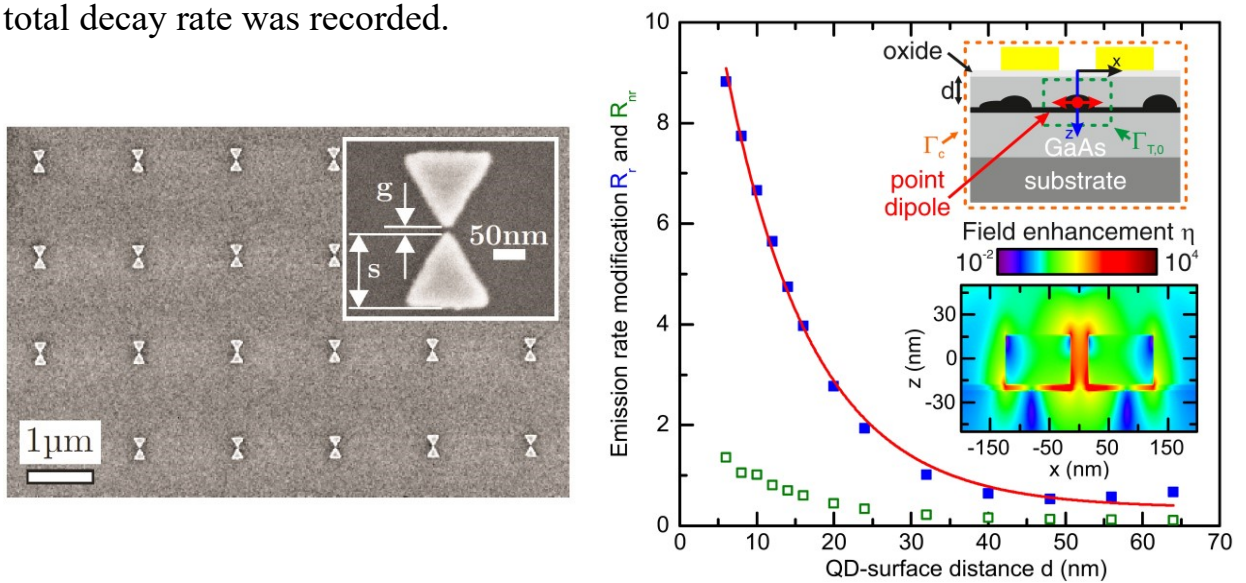

Fig. 16. a) Geometry of plasmonic bow-tie nanoantennas and b) theoretical calculation of the radiative (blue squares) and nonradiative (green squares) Purcell factor [127,128],

EQS radiation near a trimer from a cylinder and two triangular prisms (Fig. 17) was studied theoretically in [129], where an increase in the radiative decay rate by more than 2 orders of magnitude was demonstrated. In this case, the radiation efficiency (49) of an EQS with a poor (~ 1%) internal quantum efficiency (50) increases up to 30% due to the increase in the radiative decay rate (see Section 2.9)! [129]. A remarkable feature of this system is the fact that the presence of several plasmon resonances makes it possible both to excite the EQS efficiently and to increase the radiative Purcell factor significantly.

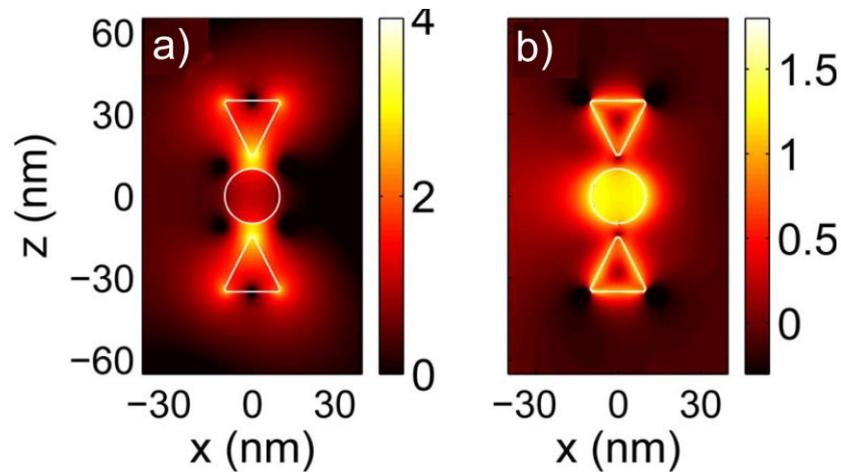

Fig. 17. Plasmonic modes of a gold nanowire trimer [129]. Distribution of the near electric field intensity, normalized to that of the incident field (log10 scale, p polarization) at the plasmonic resonances $\lambda = 390$ nm (a) and $\lambda = 334$ nm (b).

The effect of a more complex plasmonic nanoantenna on the emission of NV centers in nanodiamonds of the size of 100 nm, containing on average 400 NV centers per crystal, was studied in [130,131]. As an antenna in [130], a set of concentric $TiO_2$ ridges was used, with the periods selected in such a manner that plasmons excited by the NV center radiation scattered resonantly on the ridges and the upward directed radiation occurred (Fig. 18). In this case, the rate of spontaneous emission is increased by 20 times approximately. In a similar work [131], non-concentric ridges were used, leading to a shift of radiation direction from the axis of the system by 17.3° and that is convenient from the experimental point of view. In both cases, a high efficiency of collection of photons was achieved.

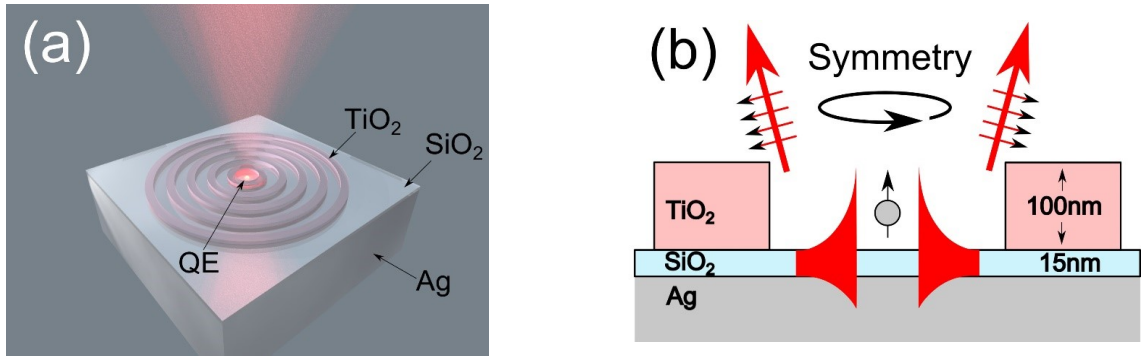

Fig. 18. Geometry of the problem of a hybrid plasmonic bullseye antenna effect on the NV center emission in a nanodiamond [130].

Much of the work on EQS radiation control using plasmonic nanostructures is based on an analogy with conventional radio frequency antennas, a classic example is the Yagi-Uda antenna [114-116], which consists of a feed, a reflector and several directors (Fig. 19). The reflector suppresses the backward radiation due to destructive interference, and the directors form the forward directional pattern due to constructive interference. The purpose of such nanoantennas is not only to increase the rate of spontaneous emission and EQS fluorescence, but also to achieve a high directivity of photon emission for their effective detection.

In nanooptics, the first Yagi-Uda type antennas were presented in [132], where it was demonstrated that the non-directional radiation of CdSeTe/ZnS quantum dots was converted into directional one.

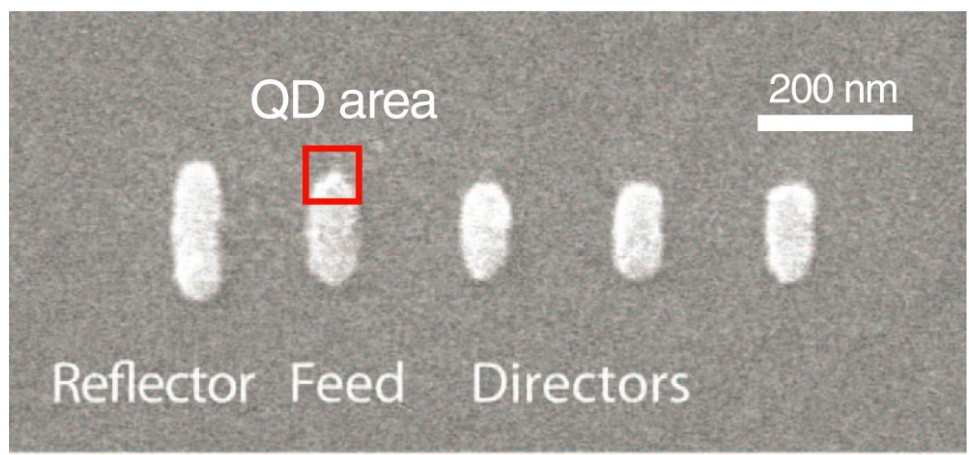

Fig. 19. An optical nanoantenna of the Yagi-Uda type based on gold nanoparticles. The red square shows the region where the emitting CdSeTe/ZnS quantum dots were deposited [132].

This was followed by a number of works on EQS radiation control using plasmonic and dielectric nanoantennas [133-135].

Recently, hybrid plasmon-dielectric Yagi-Uda nanoantennas have been proposed, which effectively use the advantages of both plasmon and dielectric nanoantennas [136,137]. In [136], it was proposed to use silver nanocylinders covered with a silicon shell as elements of the nanoantenna (Fig. 20a). This design makes it possible to achieve a radiation efficiency (49) close to 1 in a wide frequency range, which is significantly higher than the efficiency of fully dielectric nanoantennas [3] (see Fig. 20b).

In [137], a design of a hybrid plasmon-dielectric antenna based on silver nanospheres covered with a silicon shell was proposed (Fig.21a).

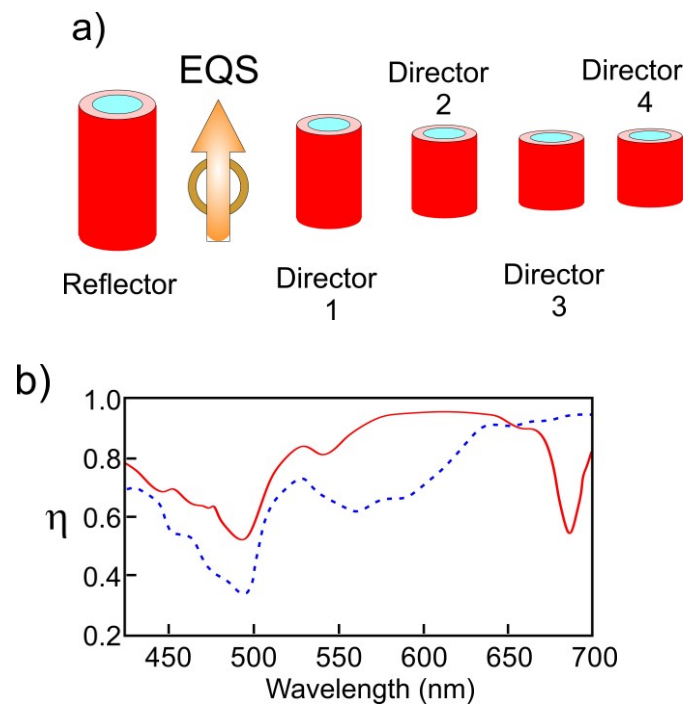

Fig. 20. a) Geometry of a hybrid plasmon-dielectric antenna [136]; b) radiation efficiency η of such a nanoantenna (red curve, [136]) in comparison with that of an all-dielectric nanoantenna (blue dashed curve, [3]).

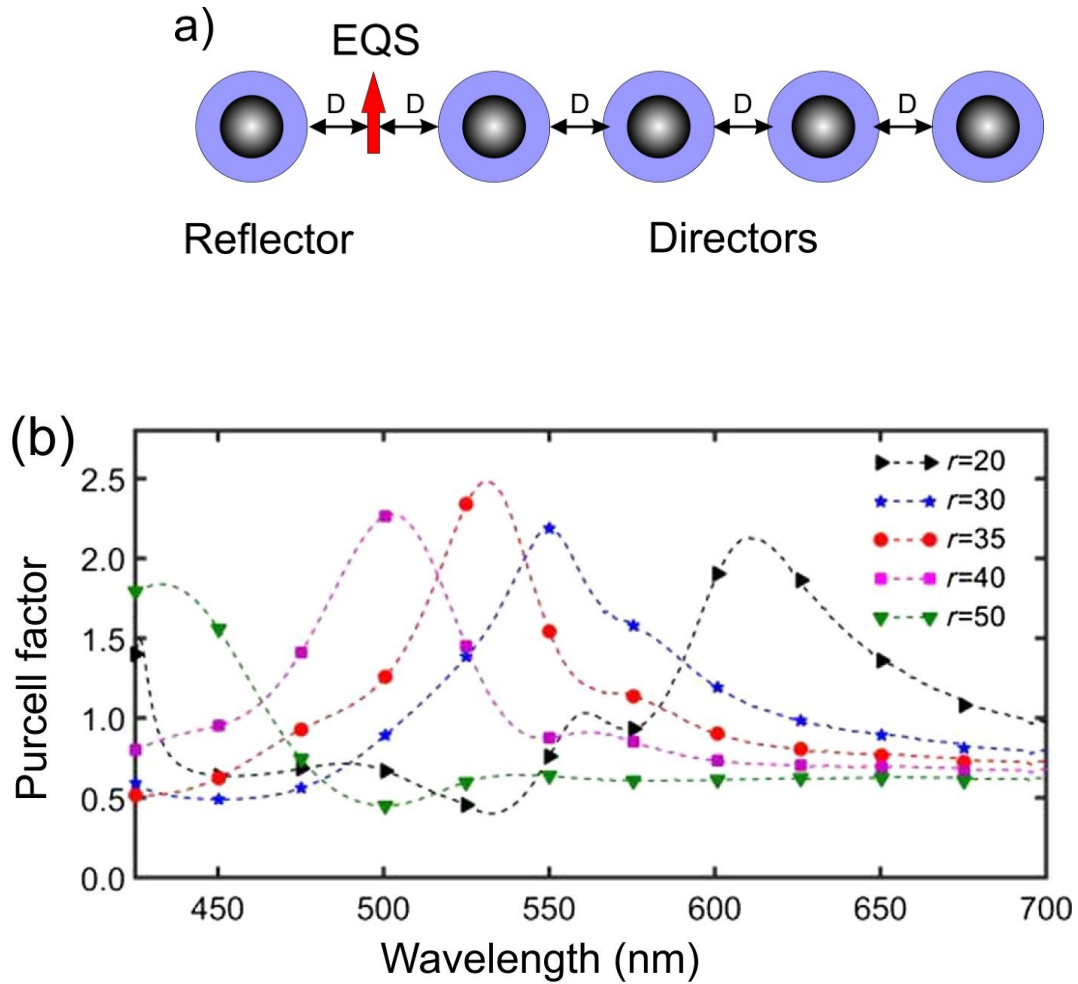

Fig. 21. a) Geometry of a hybrid plasmon-dielectric antenna based on layered spherical nanoparticles; b) the Purcell factor for such an antenna for different radii of the core at a fixed distance between the elements of the nanoantenna $D = 70$ nm [137].

Quite recently, a very interesting experimental work has appeared on the implementation and study of a hybrid nanostructure consisting of a cubic gold nanocrystal with the edge of 100 nm, with dielectric polymer "wings" deposited on the opposite edges ($n = 1.5$) and doped with CdSe/CdS/Zn (core / shell / shell) colloidal quantum dots [138] (see Fig. 22a). For such a structure, a full range of spectroscopic studies was carried out, including measurements of photoluminescence with time resolution (Fig. 22b). Analyzing the decay times of quantum dots in the presence of an Au nanocube and without it, one can estimate the Purcell factor, that turns out to be of the order of 30 [138].

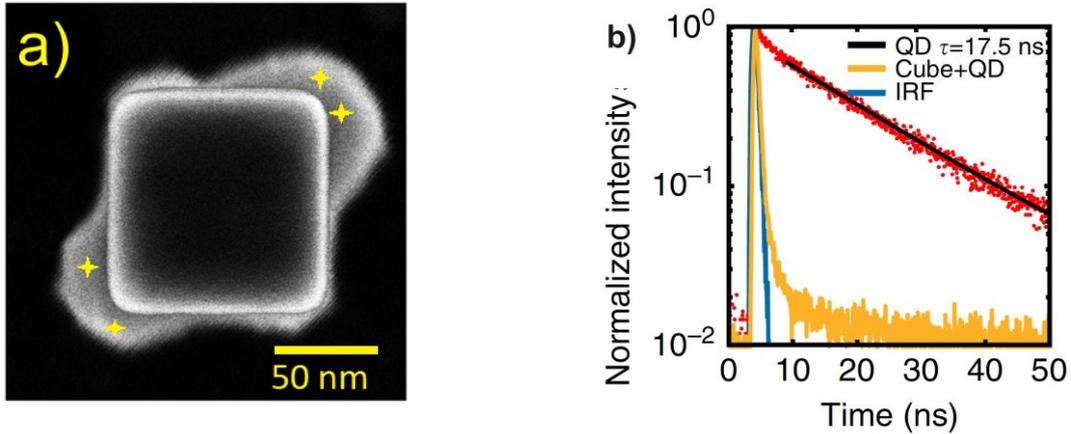

Fig. 22. a) A SEM-image of a single-crystal gold nanocube with an edge of 100 nm, with a polymer deposited on its opposite sides and doped with CdSe/CdS/Zn (core/shell/ shell) colloidal quantum dots (yellow stars); (b) time-resolved measurements of the photoluminescence of quantum dots in a bulk polymer (red dots, $\tau$ = 17.5 ns) and quantum dots in the polymer "wings" on the gold nanocube (yellow dots, $\tau$ = 0.63 ns) [138].

In the same work, the autocorrelation function of emitted photons $g^{(2)}(0)$ ~0.35 was investigated, showing that there is emission of single photons. Interestingly, the emission of photons stopped if the polarization of the exciting field became perpendicular to the direction of deposition of the polymer "wings" with quantum dots (see Fig. 22). The authors of [138] showed that, in this case, the plasmonic enhancement of the fields near the quantum dots did not occur and they were not excited (see (54)).
More detailed information on EQS radiation control using hybrid nanoantennas can be found in [6].

In [139], the authors went further and proposed to create optical interconnectors for the next generation of computers based on receiving and transmitting nanoantennas.

By studying single molecules using scanning aperture microscopes (SNOM), the problem of EQS radiation near a nanohole in a metal film is of great importance (Fig. 23). In [140-142], such a problem was solved both analytically - for a hole in a perfectly conducting film, and numerically - for a hole in a nanofilm made of a real metal. In [142], a design and materials were proposed allowing to increase significantly the rate of spontaneous emission of a photon into the tip of a scanning microscope for the purpose of its subsequent detection.

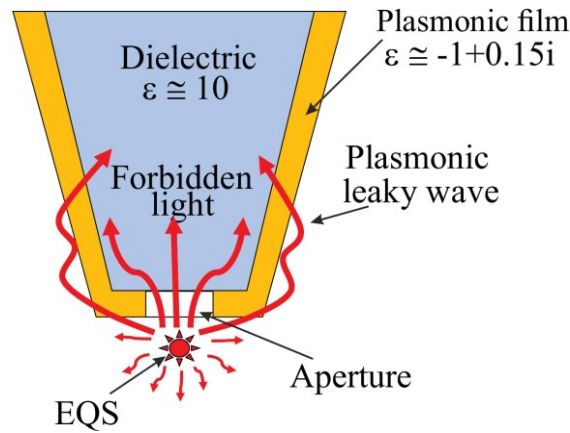

Fig. 23. Geometry of the problem of EQS radiation in the presence of a nanohole in a scanning microscope (SNOM) tip.

Above, examples of control of radiation of EQS with allowed electric dipole (*ED*) transitions were considered. However, for many applications, forbidden magnetic dipole (*MD*) and electric quadrupole (*EQ*) transitions are also of interest. The influence of a planar nanocavity and a dimer of plasmonic nanoparticles on the radiation of EQS with *EQ* transitions was considered theoretically in [68, 143]. In particular, it was shown in [143] that a dimer (nanoantenna) of plasmonic nanoparticles can increase significantly the rate of forbidden quadrupole transitions in comparison both with free space and with a single spherical plasmonic nanoparticle.

In [144], the system of full control of both allowed and forbidden transitions of the $OsO_3$ molecule was studied theoretically. This system is based on a symmetric patch nanoantenna (see Fig. 24).

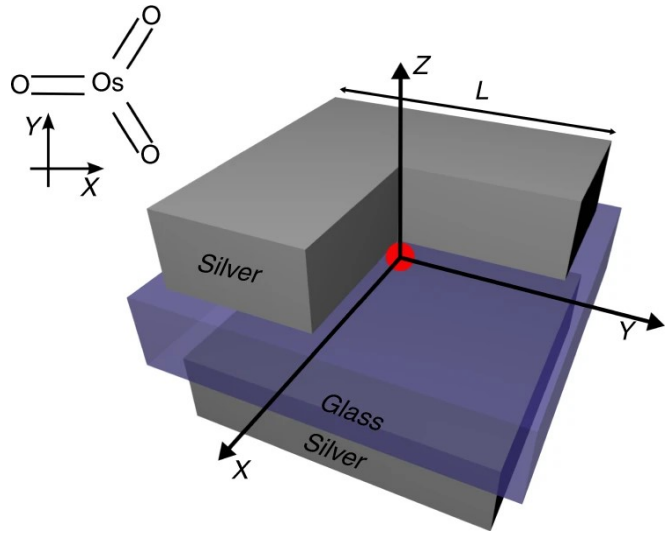

Fig.24. A patch nanoantenna to control allowed and forbidden transitions in $OsO_3$. The nanoantenna consists of two silver cuboids 50 nm thick and edge *L*. The cuboids are separated by a dielectric spacer ($\varepsilon = 2.25$) 30 nm thick and side $L + 20$ nm. The emitting $OsO_3$ molecule is located in the center of the antenna [144].

By changing the length *L* of the silver nanoparticles edge, it is possible to control the interference of allowed and forbidden transitions, from significant amplification (more than two orders of magnitude for the case of *MD* transitions and more than four orders for the case of *EQ* transitions) to complete suppression of radiation. In this paper, several issues are considered, conditioning the rapid experimental implementation of such a system for controlling the EQS radiation.

Above, the control of radiation processes was presented using metals having a negative permittivity and therefore plasmonic properties, that is, allowing the existence of surface and localized oscillations. However, the negative permittivity can be caused not only by plasmon oscillations of conduction electrons in metals and their alloys. For longer wavelengths, especially in the very important mid-IR band of the spectrum (3–20 μm), there are alternative materials where the permittivity is negative. These alternatives include polar dielectrics in the reststralhen band, where phonon-polaritons exist, [145], and heavily doped semiconductors with high carrier mobility [146]. In such materials, internal losses are significantly (by an order of magnitude or more) lower than in noble metals in the visible and near-IR bands of the spectrum, and therefore, they can be used to provide higher rates of spontaneous emission. A comparison of the effect of nanoparticles made of plasmonic materials (Ag, Na) and polar dielectrics (SiC) on the radiative rate of spontaneous emission was carried out in [117-119].

Figure 25 shows the decay rates of the EQS spontaneous emission near a cluster of two nanospheres made of Na and SiC [119] in the ultrastrong coupling regime (see Section 2.3).

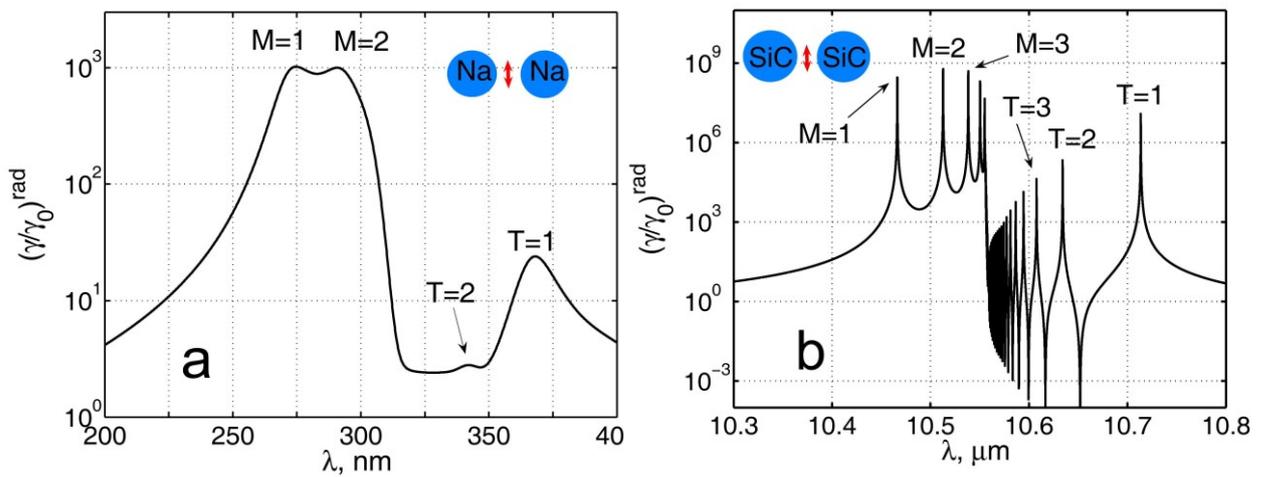

Fig. 25. Relative radiative decay rate of emission from EQS located in the gap between two nanospheres as a function of the radiation wavelength [119]. (a) Na nanospheres. (b) SiC nanospheres. In both cases, the radius of the nanoparticles is $R_0 = 50$ nm and the distance between their centers is $R_{12} = 105$ nm. The orientation and location of the EQS is shown by arrows. The letters $T$ and $M$ denote ordinary **T**ransversal oscillations of a cluster of two nanospheres and bound modes arising in the regime of ultrastrong coupling (see Section 2.3) between the nanospheres ($M$ stands for the word "molecule").

It can be seen from these figures that the rates of EQS spontaneous emission near the SiC nanoantenna are significantly higher than for the Na nanoantenna. For a nanoantenna made of Ag nanoparticles, the situation is similar.

On the other hand, the high Q-factor of the phonon-polariton oscillations limits the operation spectral band of the nanoantennas, and in each case the choice of material depends on the task and resources available in the laboratory [147-149].

## 4. Control of EQS Radiation Using Dielectric Nanoparticles and Nanostructures

Despite the interesting results of plasmon nanostructures use for control of the EQS radiation, the development of the plasmon direction in general has a number of significant difficulties associated with both significant losses (if we talk about the visible and near-IR ranges) and with insufficient compatibility of the process of manufacturing plasmon nanostructures with modern technologies of production of electronic micro- and nanocircuits. The latter are based on silicon technology, that does not conform to the use of plasmonic materials such as Au and Ag.

On the other hand, modern silicon technologies make it possible to fabricate silicon nanostructures that practically have no losses in the near-IR range [14,150,151]. The high permittivity of silicon ($\varepsilon \approx 16$) gives localization and enhancement of electric fields that are quite comparable with the case of plasmon nanostructures.

This circumstance makes it very promising to use dielectric nanostructures for controlling EQS radiation and to create new nano-optical devices combined with electronic nanodevices on this basis [152-154].

As in the case of plasmonic nanostructures, the influence of dielectric spherical nanoparticles on the EQS radiation is studied better. Thus, in [42-44,74,75,82], the radiation of an EQS located near or inside a dielectric microsphere with a high refractive index was investigated. In these works, all the features of nanoparticles with high refractive indices were found and it was shown that they can be effectively used to control the EQS radiation and, in particular, that with the help of dielectric nanoparticles the rate of spontaneous emission can be increased by more than 500 times (Fig. 26 )

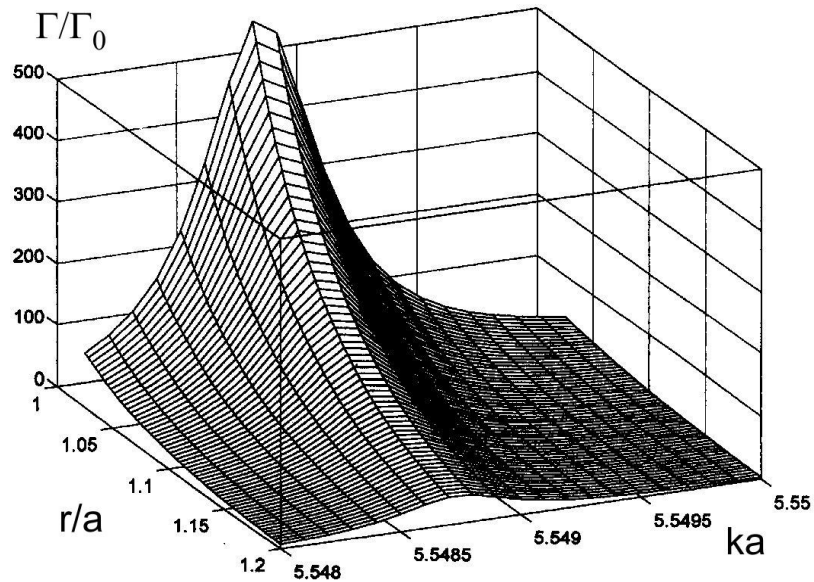

Figure: 26. Relative decay rate of spontaneous emission (Purcell factor) of the EQS located near a dielectric microsphere with $\varepsilon = 6$ (diamond) depending on the position *r/a* and on the size parameter *ka* (radial orientation of the EQS dipole moment) [42].

The conclusions of the theory were experimentally verified in [155], where the decay rate of the $Eu^{3+}$ complex (TTA) embedded in polystyrene spheres ($n = 1.59$) at a weight concentration of 5% was studied. The results of measuring of $Eu^{3+}$ ions decay rate at the wavelength of $\lambda = 615$ nm, normalized to the decay rate in bulk polystyrene, are shown in Fig. 27.

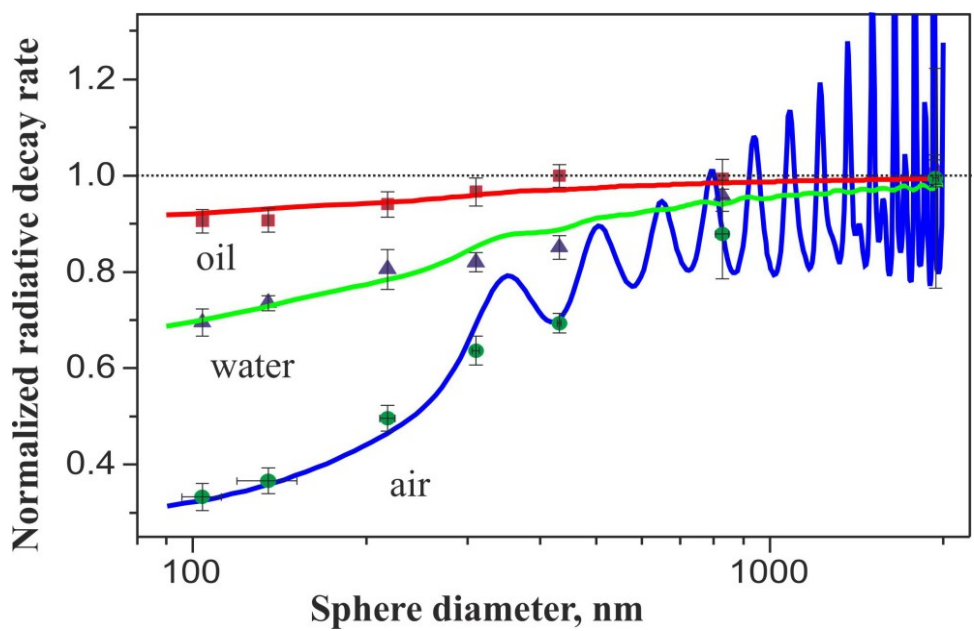

Fig. 27. Results of experimental measurements of $Eu^{3+}$ complex (TTA) decay rate in polystyrene nanospheres placed in air, water, and immersion oil [155]. Solid curves – theory [43].

Similar results were obtained in [156], where the photoluminescence of NV centers was measured in spheroidal nanodiamonds with the sizes ranging from 300 to 1000 nm (see Fig. 28). It can be seen from this figure that in the case of the resonance between the nanoparticle and the NV center, the decay rate increases nearly twofold.

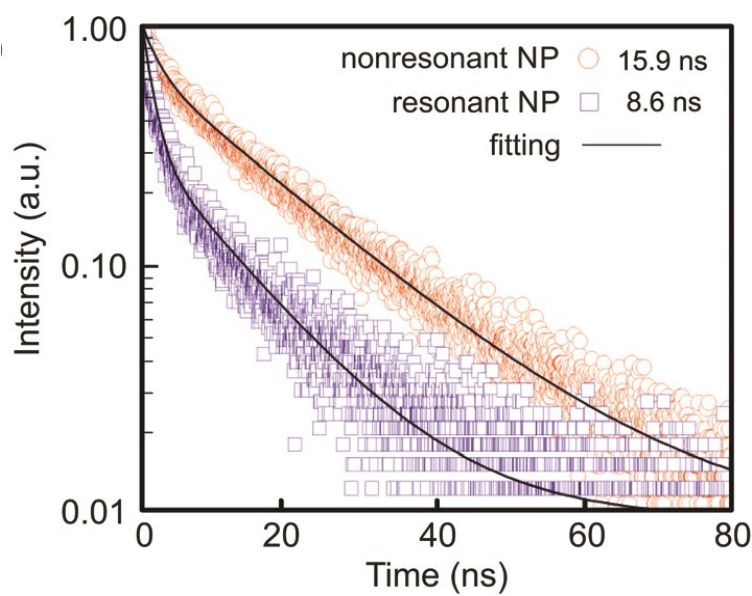

Fig. 28. Time-resolved photoluminescence measurements for two cases of overlap of Mie-type resonances with the spectrum of NV centers [156].

In [157], it was shown that in the presence of a single cylindrical Si nanoparticle the quantum efficiency of radiation and the Purcell factor of the EQS can be significantly increased if it is placed over a silver nanofilm. In such a system, a complex system of high-quality plasmon-dielectric resonances is formed, allowing an efficient control of the EQS radiation. In particular, here it is shown theoretically that at a high (> 90%) efficiency of total (into plasmons and photons) radiation and radiation into photons (> 70%), the Purcell factor reaches 2600!

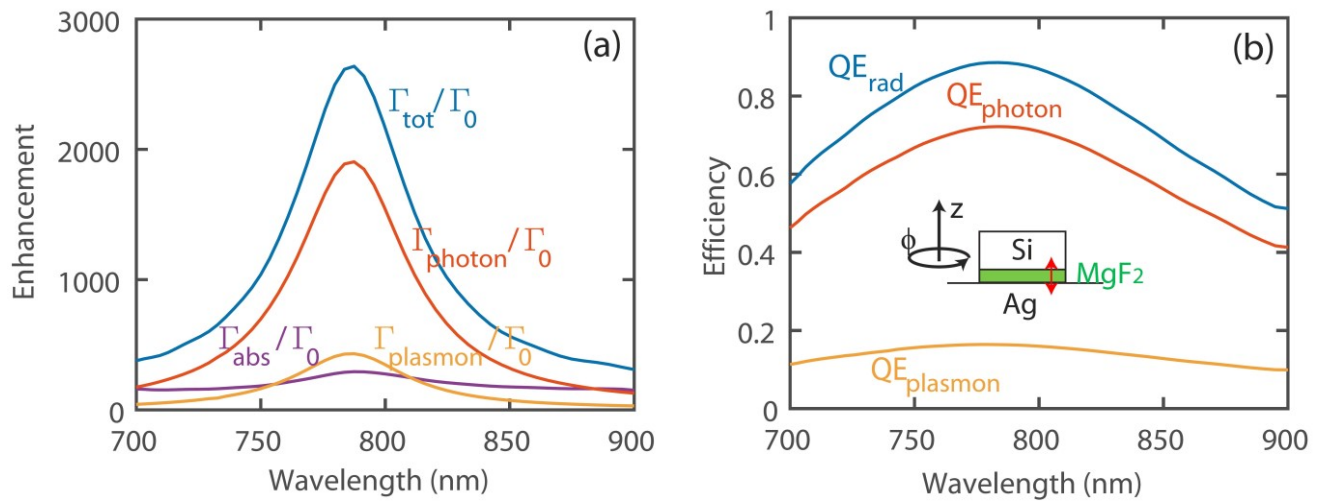

Fig. 29. Enhancement of spontaneous emission from an EQS by a Si nanocylinder at optical frequencies. The nanocylinder (inset in Fig. 29b) is separated from the silver film by a 2 nm layer of $MgF_2$ ($n \sim 1.375$). a) Purcell factors of total, radiative, and nonradiative decay rates and also decay rates into plasmons; b) radiative efficiencies η into different channels. The inset in Fig.29b shows the geometry of the structure of a Si nanocylinder with the radius of 72 nm and the height of 100 nm above the silver film. The emitting z-oriented EQS is located at the distance of 60 nm from the axis and in the center of the 2-nm $MgF_2$ layer (red arrow) [157].

In [158], to enhance spontaneous EQS transitions, clusters of silicon ($\varepsilon = 16$) nanospheres with the radius of $r = 70$ nm and the period of $a = 200$ nm were considered, and the EQS was placed in the center of the cluster. By approximating the spheres with point dipoles, the authors showed theoretically and experimentally (transferring the measurements to the microwave range) that the spontaneous emission decay rate can be increased by almost 70 times for the transverse orientation of the dipole (Fig. 30, Fig. 31). No enhancement effect was found for the longitudinal orientation. This result differs from the case of a plasmon antenna, where the situation is exactly the opposite: longitudinal oscillations are enhanced, and transverse ones are inhibited [117-119].

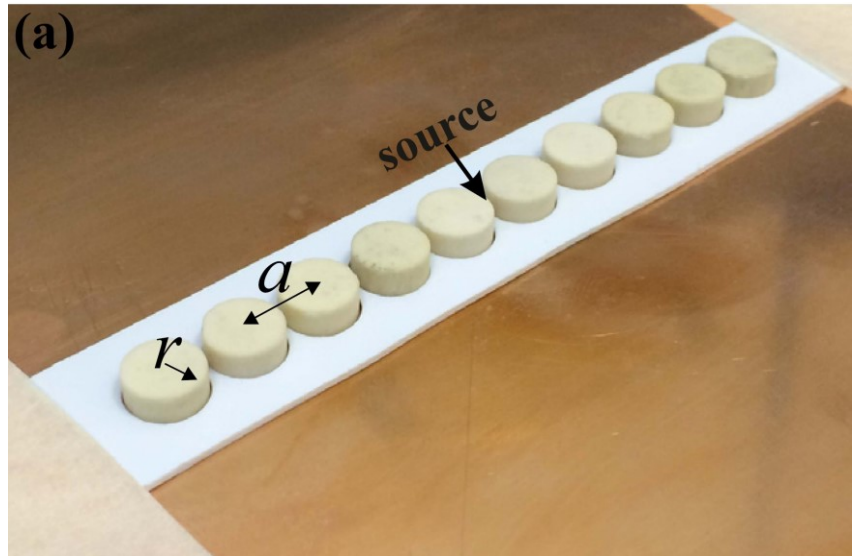

Fig. 30. Geometry of a microwave sample of disks of the radius $r = 4$ mm, the height $H = 4$ mm, and the period $a = 5$ mm for experimental verification of the theory ($\varepsilon = 16$) [158].

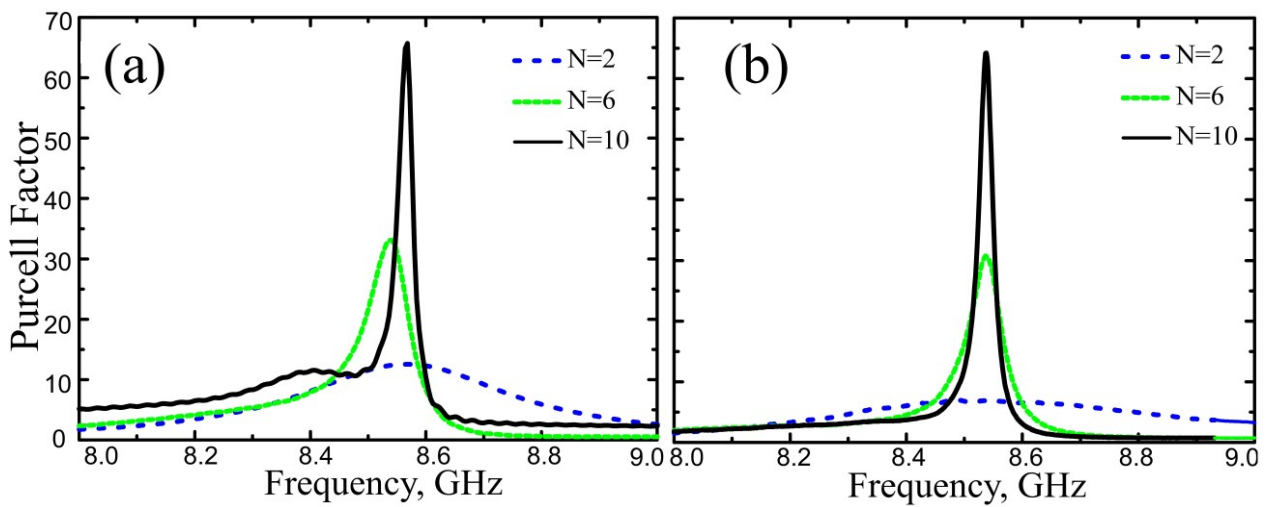

Fig. 31. Purcell factor for a microwave system shown in Fig.30. a) Theory, b) microwave experiment [158]. $N$ is the number of disks in the sample.

In [159], basing on the expansion in local spherical harmonics, the spontaneous decay rates averaged over the directions of the dipole moment (see (19)) were calculated for a cluster of 25 TiO$_2$ nanospheres with $\varepsilon = 7.35$. The calculation results for fixed geometry, different EQS positions, and different frequencies are shown in Fig. 32. It can be seen from this figure that for periods small compared to the wavelength, the change is small if normalized to the decay rate in the medium (compare with Fig. 27), but as the wavelength decreases, lattice resonances appear in the system, and the spontaneous radiation increases by two orders of magnitude when normalized to the vacuum value and by an order of magnitude if the influence of the sphere is normalized to the decay rate in a continuous medium with the sphere's permittivity.

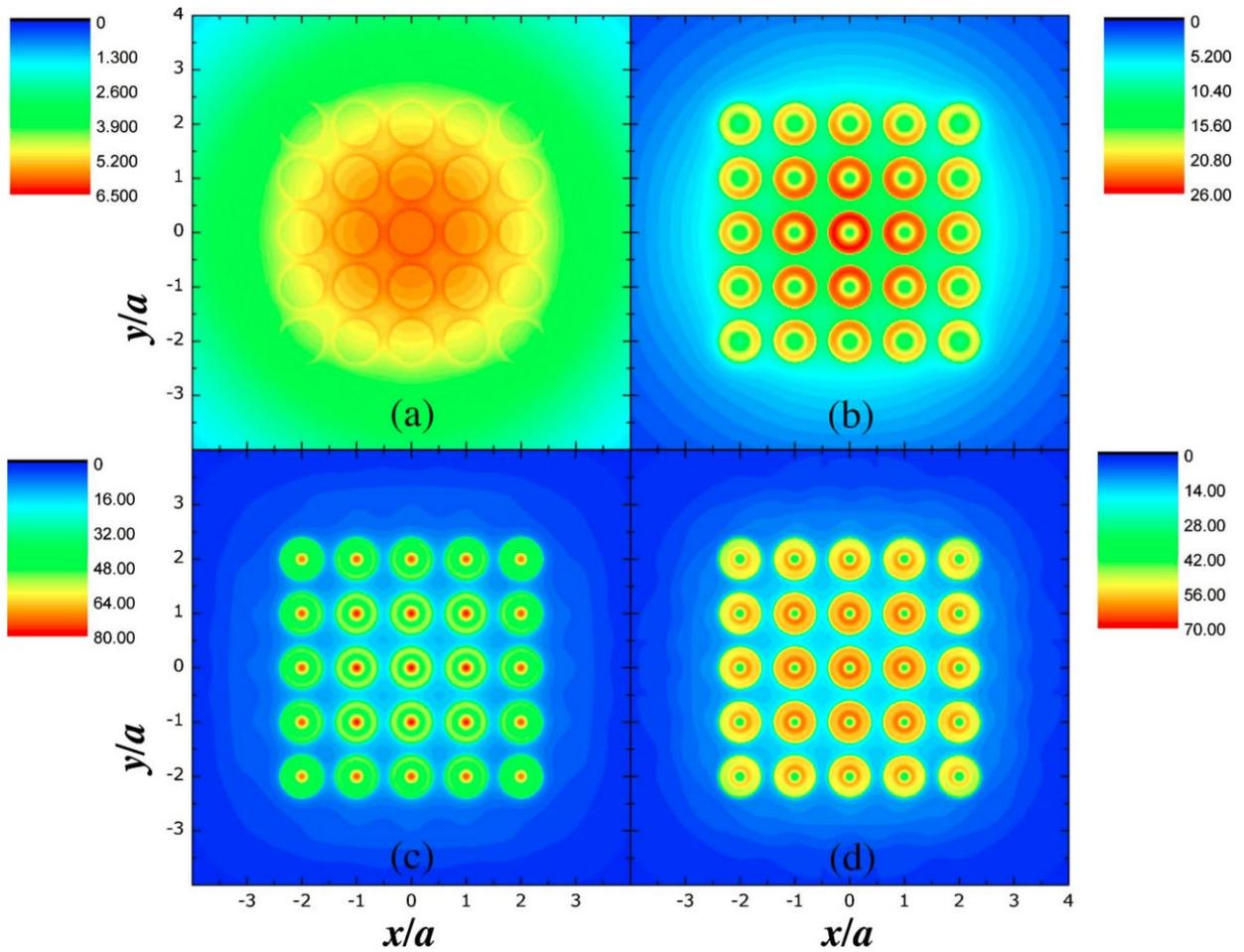

Fig. 32. Dependence on the position in the lattice of the spontaneous decay rate averaged over the EQS orientations (see (19)), normalized to the vacuum value for a lattice with a period $a$ of 25 TiO$_2$ nanospheres of the radius $0.4a$ and the permittivity of ε =7.35 for different frequencies. a) $\omega a/c$=1.5; b) $\omega a/c$=2.5; c) $\omega a/c$=3.85; d) $\omega a/c$=5 [159].

The influence of single dielectric spheres on the emission of forbidden (*MD* and *EQ*) transitions was studied in [74,75], where it was shown that nanoobjects with a size parameter $ka \ll 1$ affect such transitions much more strongly than allowed (*ED*) transitions. In particular, for quadrupole *EQ* transitions, the Purcell factor $F_P^{EQ}$ can be estimated as

$$F_P^{EQ} \sim \frac{F_P^{ED}}{(ka)^2} \gg 1, \qquad (55)$$

due to the fact that forbidden transitions "feel" not the field itself, but its gradients (see (25)). Figure 33 shows the relative decay rates of spontaneous emission (Purcell factors) for quadrupole transitions (Figure 33a) and their comparison with the decay rates of dipole transitions (Figure 33b) near the dielectric nanosphere.

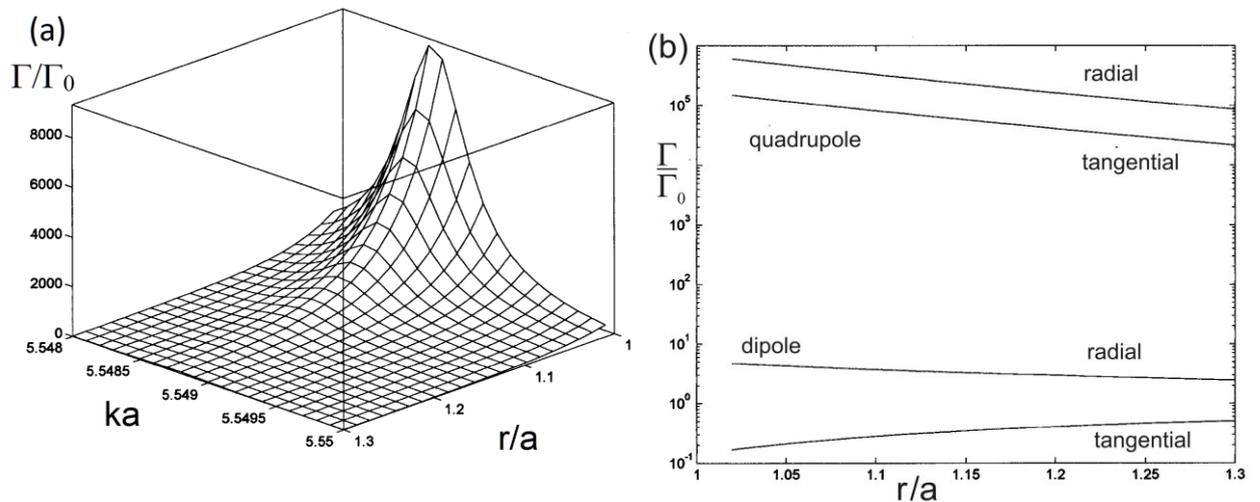

Fig. 33. Relative decay rates of spontaneous emission (Purcell factors) for quadrupole transitions (a) and their comparison with the decay rates of dipole transitions (b) near a dielectric nanosphere with ε=6 и ka=0.01.

It is indeed seen from these figures that forbidden transitions are affected by nanoobjects more substantially than allowed ones.

In [160], to control the emission of *ED* and *MD* transitions, a more complex system than a sphere was proposed. This system consists of a hollow silicon nanocylinder (Fig. 34).

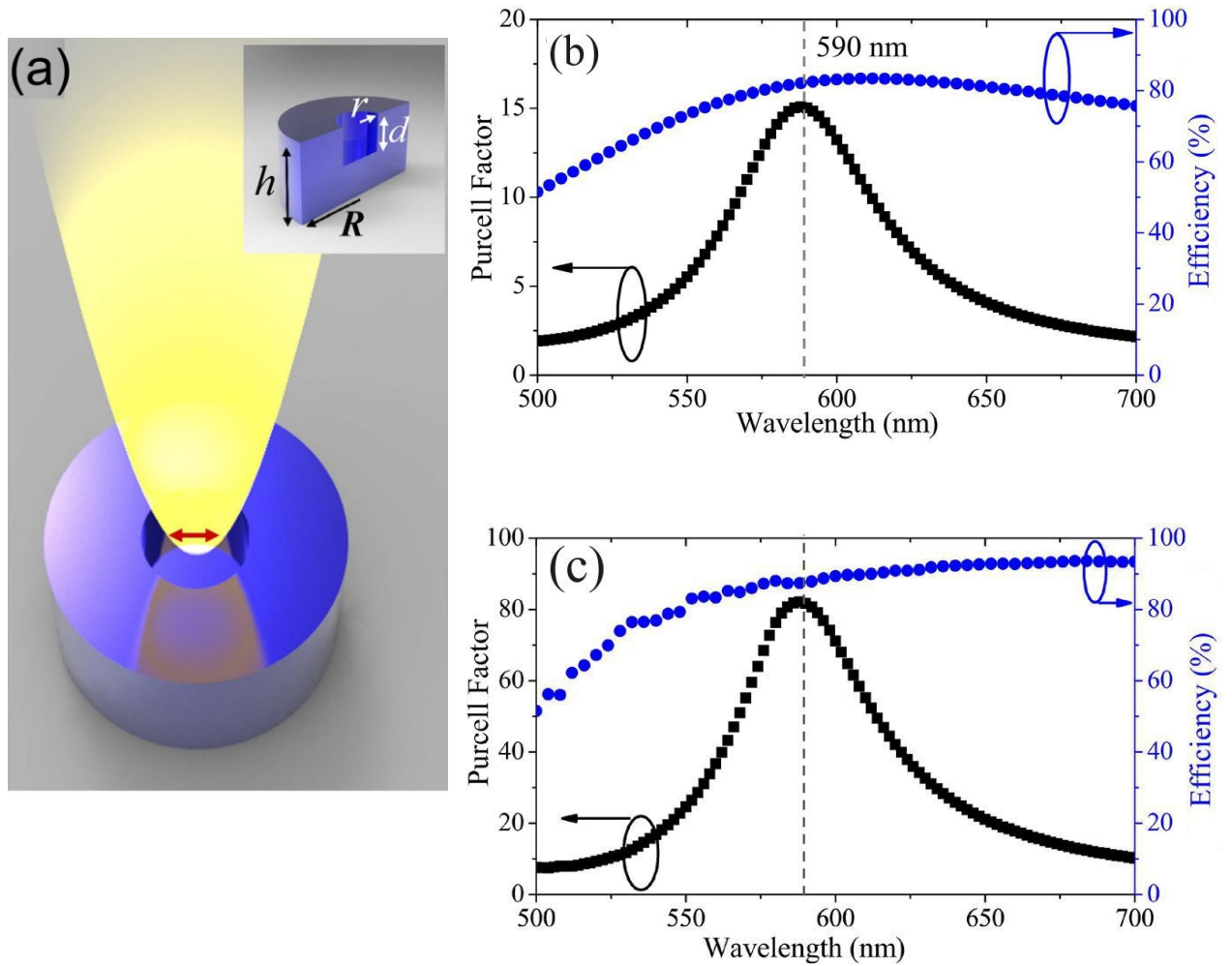

Fig. 34. a) Geometry of a dielectric nanostructure in the form of a silicon cylinder with a void to control the radiation of forbidden transitions; b), c) Purcell factors and radiation efficiencies for *ED* and *MD* transitions, respectively. b) $R$ = 110 nm, $h$ = 86 nm, $d$ = 19 nm, $r$ = 25 nm; c) $R$ = 110 nm, $h$ = 80nm, $d$ = 30 nm, $r$ = 25 nm [160].

Theoretical calculations showed [160], that such a system can enhance spontaneous *ED* transitions by a factor of 15 at a radiation efficiency of 80% and spontaneous *MD* transitions by a factor of 80 at a radiation efficiency of 85%.

An experimental study of the dielectric nanostructure influence on the emission of both allowed (*ED*) and forbidden (*MD*) transitions was carried out in [161]. In this work, they studied the effect of a dimer made of silicon cuboids 300 × 300 × 90 nm³ on the photoluminescence of a 30 nm $Gd_2O_3$ layer doped with $Eu^{3+}$ ions. The geometry of the experiment is shown in Fig. 35a.

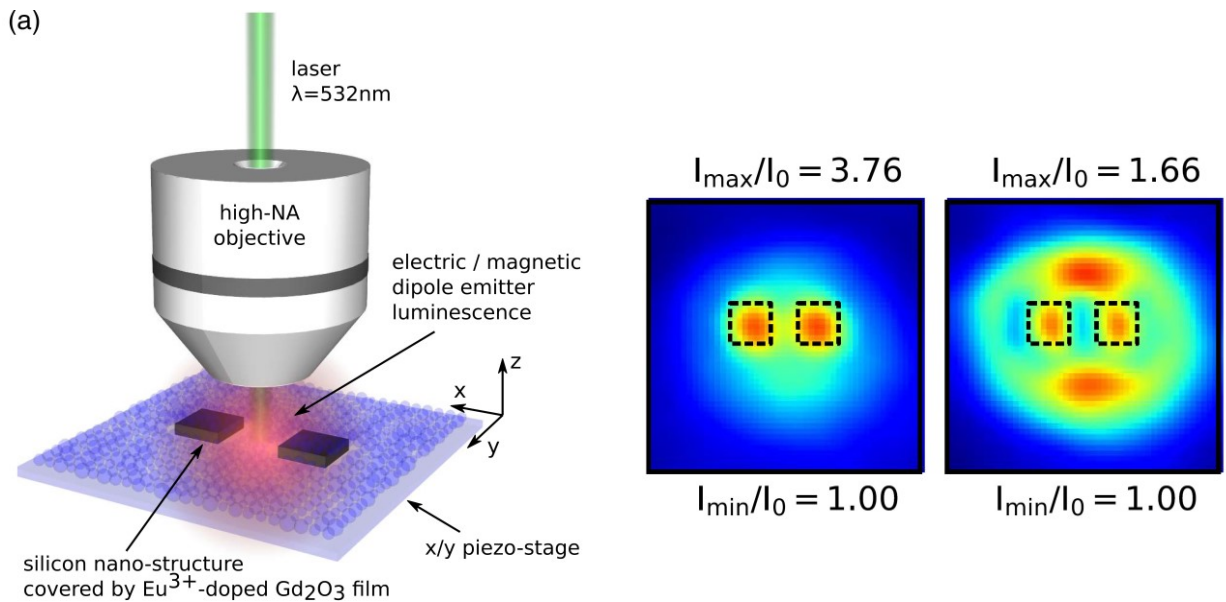

Fig. 35. a) The geometry of the experiment; b) typical measured distributions of local photoluminescence (left - *MD* transitions, right - *ED* transitions). The slit width between the cuboids is 200 nm [161].

Magnetic (*MD*) or electrical (*ED*) transitions in $Eu^{3+}$ occur at wavelengths λ = 590 ± 10 nm and λ = 610 ± 10 nm, respectively. In the same frequency ranges, photoluminescence was measured, and it turned out to be proportional to the orientation-averaged Purcell factor (19) at these wavelengths. Interestingly, the enhancement of *MD* transitions was found to be more significant than the enhancement of *ED* transitions. This is consistent with the general results [74,75,66]. The agreement between theory and experiment for *MD* transitions is not bad, but for *ED* transitions, probably the model should be refined.

So far, the main attention was paid to works devoted to the enhancement of radiation processes and the increase in the Purcell factor. However, as already mentioned in Section 2, an equally important characteristic of the influence of dielectric nanoparticles on the EQS radiation is the resulting radiation pattern.

In the domain of dielectric nanoantennas, the control of EQS radiation becomes more promising, since in such systems it is possible to superpose the spectral positions of *ED* and *MD* Mie resonances, and as a result even a single nanoparticle can lead to directional EQS radiation due to complete suppression of backward radiation (Huygens element) [4,162, 163]. In a Si nanoparticle with the radius of 65 nm, this effect is achieved, since in a certain frequency range ($\lambda = 570$ nm) it is possible to provide induced in the particle, electric and magnetic dipole moments with the same amplitude and phase. In this case, the EQS radiation will be directed from the dipole to the side where the nanoparticle is located. At $\lambda = 490$ nm, the electric and magnetic dipole moments induced in the same particle have the same amplitudes, but the phases are shifted by 1.3 radians, and so the radiation direction becomes reverse in the system under consideration. Even greater directivity can be achieved with a Yagi-Uda Si nanosphere antenna [4,162,163]. The use of a dielectric nanoantenna in the form of a sphere with a void makes it possible to extract radiation from diamond nanocrystals with NV centers efficiently and directionally [164].

## 5. Influence of Nanoparticles and Metamaterials on EQS Radiation

In the previous sections, methods of controlling the EQS radiation using nanostructures from traditional materials - metals and dielectrics, and also from their combinations - hybrid systems - were considered. However, at present, nanotechnology makes it possible to synthesize materials with fundamentally different properties. We are talking about nanostructures consisting of artificial metaatoms with dimensions much smaller than the wavelength specific to the problem under consideration.

## 5.1. Nanoparticles and Metamaterials with a Negative Refractive Index

The history of metamaterials begins with a paper in Physics Uspekhi [165], where V.G. Veselago showed that, unlike metals, a substance (at that moment hypothetical) with simultaneously negative permittivities and permeabilities allows the propagation of waves with a negative refractive index, that is, with oppositely directed phase and group velocities of waves.

The rapid development of this direction began with works [166,167], where metamaterials with a negative refractive index were realized in the microwave frequency band. Later, the metamaterials were realized in the IR and visible bands (for more details see, for example, [10,168,169]). Now, such materials are called NIM (Negative Index Materials) or DNG (double negative) materials. As an example of implementation, Fig. 36 shows the design of a metamaterial with a negative refractive index in the visible frequency range [170].

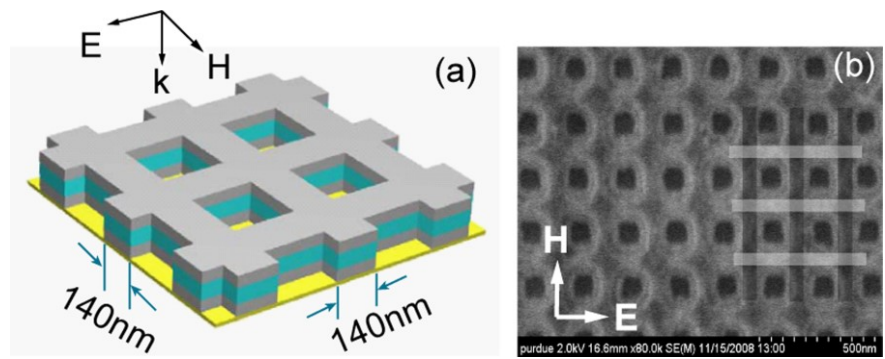

Fig. 36. (a) Schematic and (b) SEM images of a metamaterial formed by perforated layers of $Al_2O_3$ (45 nm, middle) and Ag (43 nm, top and bottom) and having a negative refractive index at $\lambda = 580$ nm [170].

In [171], the influence of a metamaterial sphere with arbitrary negative permittivities and permeabilities on the decay rate of an EQS with allowed (*ED*) and forbidden (*MD*) transitions was studied in detail. In this work, it was shown that such nanoparticles allow control of spontaneous emission in much wider ranges than dielectric or plasmonic nanoparticles. In particular, using DNG metamaterials, huge values of the Purcell factor are possible already in particles with the size parameter $ka \approx 2$, which can be actually manufactured (see Fig. 37).

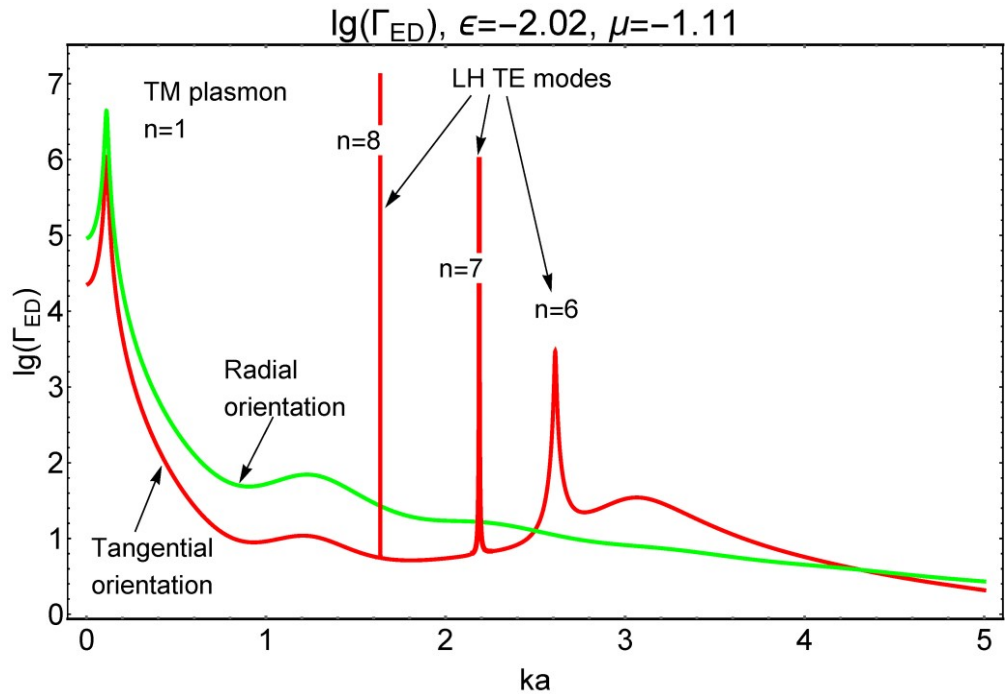

Fig. 37. Dependence of the rate of *ED* spontaneous emission of an EQS located near a spherical nanoparticle with $\varepsilon = -2.02$, $\mu = -1.11$ on the size parameter $ka$ [171].

This effect is associated with a completely new type of mode - LH (Left-Handed) modes, appearing in DNG nanoparticles along with modes similar to plasmon and whispering gallery modes. These superhigh quality modes are surface modes and arise as a result of a phase transition from conventional plasmon modes. Fig. 38 shows the decay rates of EQS placed near usual and DNG nanoparticle and the related field distribution of the whispering gallery and LH modes.

It can be seen from this figure that the huge Purcell factor for spheres with $\varepsilon = -15$, $\mu = -1.1$, $n = -4$ ($F_P \sim 10^{10}$) is associated with the surface character of the LH modes. For a dielectric sphere with a higher refractive index $n = 7$, the modes have a volume character, leading to a decrease in the Purcell factor by several orders of magnitude ($F_P \sim 10^5$) in comparison with the LH modes.

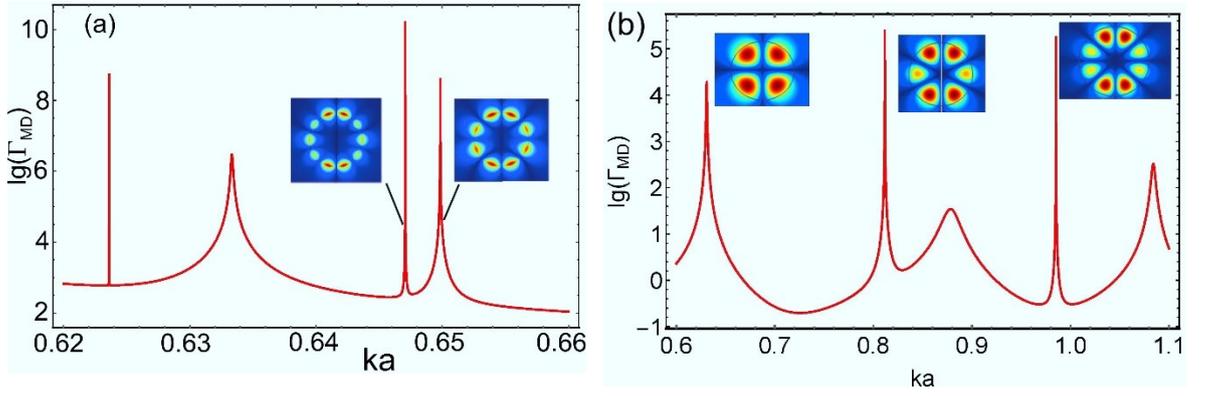

Fig. 38. The relative decay rate of spontaneous emission of a radially oriented magnetic dipole as a function of the size parameter and the distribution of the electric field intensity for TE modes for a sphere made of a metamaterial with $\varepsilon = -15$, $\mu = -1.1$ (a) and for a dielectric sphere with $\varepsilon = 50$, $\mu = 1$ (b).

An interesting example is metamaterials with a refractive index close to zero - ENZ (Epsilon Near Zero) materials [172]. Their influence on the rate of spontaneous emission was considered in [173] theoretically for ordinary materials in the frequency range where their permittivity is close or equal to zero. SiC and ITO (Indium Tin Oxide) were taken as such materials, having zero permittivity at $\lambda$ = 10.3 μm and 1.223 μm, respectively. By calculating the imaginary part of the Green's function (13) for the ENZ half-space, the total and radiative decay rates were found, and it was shown that when the distances between EQS and the surface are less than 475 nm (for SiC) and 129 nm (for ITO), non-radiative decays dominate, while at larger distances, the radiation goes through the radiation channel.

In [174], spontaneous emission was considered near a half-space filled with random metal-dielectric (Au-polysterene) nanocomposite. At a certain filling factor *f*, such a nanocomposite undergoes a percolation transition, and the composite turns from dielectric into conductor. At the point of the percolation transition, the permittivity is equal to zero, and the composite becomes an ENZ material. As in [173], in the region of the ENZ material, there is a huge increase in the total decay rate - by 5-6 orders of magnitude, depending on the orientation of the dipole.

It may not be quite correct to use the local effective values of permittivities to calculate the decay rate in the presence of ENZ metamaterials, since in the region of zero permittivity the wavelength in the substance becomes very large and it is necessary to take into account nonlocal effects [175] (see also the experimental work [33]).

In any case, significant changes in the rate of spontaneous emission occur for ENZ materials, and on the basis of this fact, it was proposed in [176] to use the strong nonlinear dependence of the properties of the ENZ metamaterial on the external field to control the rate of spontaneous emission.

5.2. Chiral Nanoparticles and Metamaterials

Chiral metamaterials, where polarization depends on both electric and magnetic fields, are also directly related to DNG. The material equations of such media can be expressed in the form [177,178]:

$$\mathbf{D} = \varepsilon(\mathbf{E} + \chi \nabla \times \mathbf{E}/k_0), \quad \mathbf{B} = \mu(\mathbf{H} + \chi \nabla \times \mathbf{E}/k_0), \tag{56}$$

where $\chi$ is the dimensionless chirality parameter. In bi-isotropic chiral metamaterials, where $\varepsilon$, $\mu$ and $\chi$ are scalar quantities, the refractive indices $n_R$, $n_L$ for right-hand (RCP) and left-hand (LCP) polarized waves have different values:

$$k_L = k_0 \frac{\sqrt{\varepsilon\mu}}{1 - \chi\sqrt{\varepsilon\mu}} = k_0 n_L; \quad k_R = k_0 \frac{\sqrt{\varepsilon\mu}}{1 + \chi\sqrt{\varepsilon\mu}} = k_0 n_R \tag{57}$$

It can be seen from expressions (57) that at a sufficiently large $\chi\sqrt{\varepsilon\mu}$, one of the refractive indices ($n_R$ or $n_L$) becomes negative necessarily. In [179], it was proposed to use this fact for the realization of media with a negative refractive index. Chiral media can be implemented in a number of ways [180-183]. Moreover, chiral properties are inherent in all living things. In particular, DNA has a helical structure. Fig. 39 shows Parapoxvirus, which is actually a chiral nanoparticle.

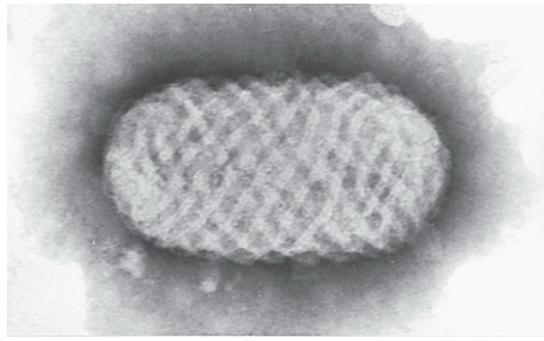

Fig. 39. Electron micrograph of Parapoxvirus with distinct spiral structures on the surface. The approximate size of the virus is 160 nm in diameter and 260 nm in length. Adapted from [184].

In view of the above, it is extremely interesting and important to know how chiral nanostructures affect the EQS radiation. In this case, under EQS it is natural to understand not only systems with simple electrical or magnetic transitions, but also chiral EQS, which can have both electric and magnetic dipole moments of the transition (see Fig. 40).

The quantum and classical theories of radiation of arbitrary EQS in the presence of arbitrary chiral metamaterials were developed in the [69,70,72,73,185-188]. Since an arbitrary bi-isotropic metamaterial is described by 3 independent parameters ($\varepsilon$, $\mu$, and $\chi$), the effect of nanoparticles made from such materials on the radiation of EQS becomes more effective than the effect of simple dielectric or plasmonic nanoparticles.

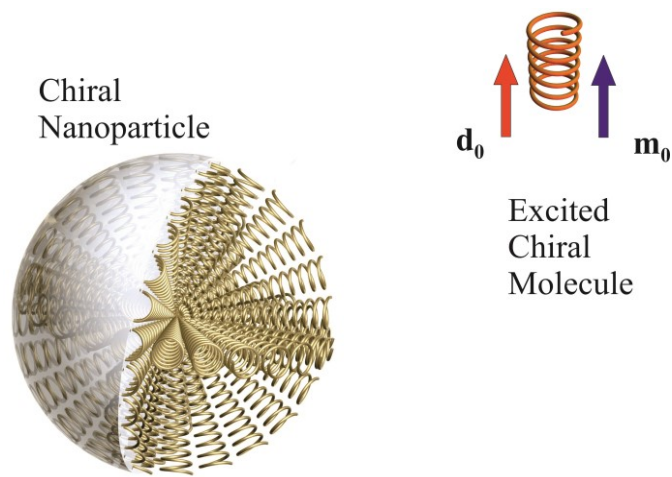

Fig. 40. Geometry of the problem of the emission of a chiral molecule near a chiral nanoparticle. The chiral molecule possesses both electric and magnetic dipole moments (red and blue arrows).

In particular, in [70] it was shown within the framework of the quasi-static approximation, that the decay rates of right-handed and left-handed enantiomers can increase and decrease in the presence of the same chiral nanoparticle significantly. It was shown in [70] that the decay rates can increase significantly if the chiral-plasmon resonance condition is met:

$$(\varepsilon+2)(\mu+2)=4\varepsilon\mu\chi^2. \tag{58}$$

Fig. 41 shows the decay rates of the right and left enantiomers of the molecule, averaged over its orientations, at different values of the permittivity and permeability.

From Fig. 41, it is seen that at $\mu \approx -2$ and $\varepsilon \approx -0.5$ the decay rate of right-handed molecules is close to zero, and at $\mu \approx -2$ and $\varepsilon \approx +0.5$, the decay rate of left-handed molecules vanishes. The vanishing of the decay rates is associated with destructive interference of the radiation of electric and magnetic dipoles and, in the general case, occurs at

$$\mu=-\frac{2d_0}{d_0+2\chi m_0}; \varepsilon=-\frac{2m_0}{m_0+2\chi d_0}. \tag{59}$$

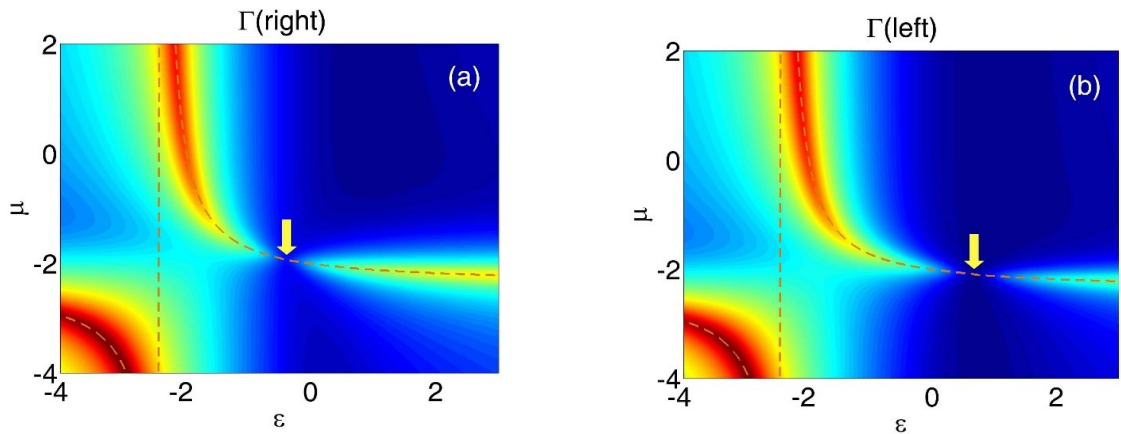

Fig. 41. Averaged over orientations, the radiative decay rate of spontaneous emission of chiral molecules near a chiral nanoparticle with chirality $\chi = 0.2$ as a function of ε and μ. a) The case of the right molecule $m_0 / d_0 = + 0.1$ b) The case of the left molecule $m_0 / d_0 = - 0.1$. The imaginary parts of the permeabilities are $\varepsilon'' = \mu'' = 0.1$. The dotted line corresponds to the chiral-plasmon resonance (58). The yellow arrow shows the minimum value on the curve.

The complete electrodynamic theory of emission of chiral molecules in the presence of chiral nanoparticles of arbitrary sizes was constructed in [69,72,185,186]. Fig.42 shows the dependence of the decay rate of chiral molecules of different orientations in the presence of a dielectric chiral sphere on the chirality parameter of the latter.

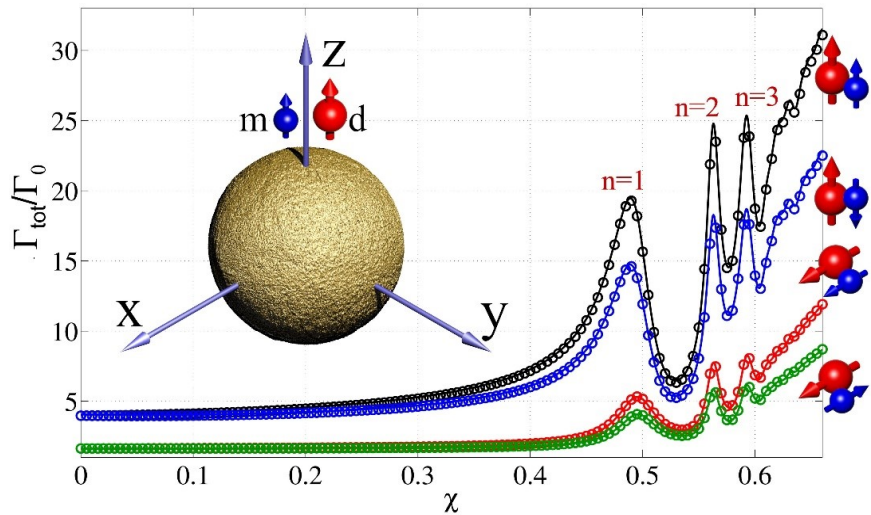

Fig. 42. Total decay rate of chiral molecules located near a chiral dielectric sphere with $\varepsilon = 2 + 0.04i$, radius $a = 70$ nm at $\lambda = 570$ nm, depending on the dimensionless chirality parameter $\chi$. Solid line — analytical solution [72], circles — numerical solution [185].

An even greater enhancement of the transition rates can be achieved by using a chiral sphere with simultaneously negative permittivities and permeabilities for control (Fig. 43). It can be seen from this figure that as the value of the chiral parameter approaches the critical one $\left(\chi\sqrt{\varepsilon\mu}=1\right)$, when one of the refractive indices tends to infinity and then changes the sign (see (57)), the quality of the chiral sphere modes becomes better, since the corresponding wavelength in the sphere tends to zero, and the decay rate can increase indefinitely, without taking into account losses and the effects of strong interaction, of course.

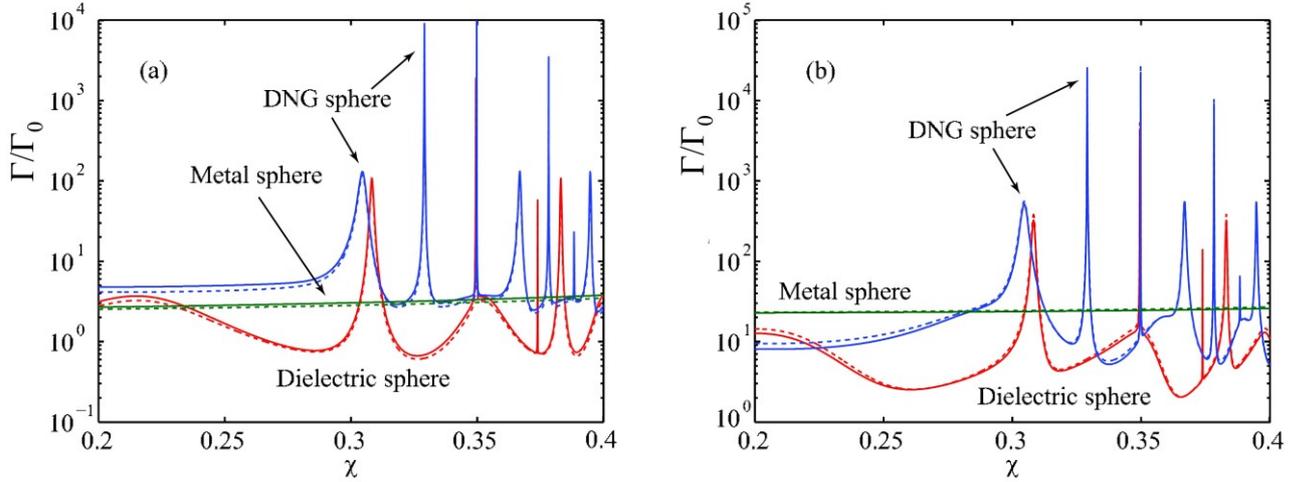

Fig. 43. Radiative decay rate of spontaneous emission of a chiral molecule located near a chiral spherical nanoparticle ($k_0 a = 1$) as a function of the chiral parameter $\chi$. The parameters of the dielectric, metallic, and DNG spheres are ($\varepsilon = 4$, $\mu = 1$), ($\varepsilon = -4$, $\mu = 1$), and ($\varepsilon = -4$, $\mu = -1.11$), respectively. (a) $\mathbf{d}_0 = d_0 \mathbf{e}_x$, parallel to the particle surface. (b) $\mathbf{d}_0 = d_0 \mathbf{e}_z$, perpendicular to the particle surface. The solid line corresponds to $\mathbf{m}_0 = m_0 \mathbf{e}_x$, the dotted line corresponds to $\mathbf{m}_0 = m_0 \mathbf{e}_z$ ($m_0 = 0.1 d_0$) [72].

The use for control chiral nanoparticles with more complex geometry (an ellipsoid, a cluster of two nanoparticles, or an asymmetric nanoparticle with a chiral core and a non-chiral shell) was considered in [69,73,87,88,186] (see Fig. 44). In [187,188], EQS radiation and its focusing by a chiral layer with a negative refractive index were studied.

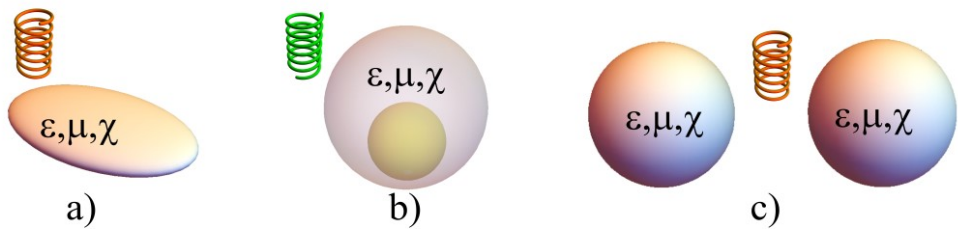

Fig. 44. Geometries of more complex chiral nanostructures for controlling the radiation of chiral molecules: a) a triaxial ellipsoid [87]; b) an asymmetric nanoparticle with a dielectric core and a chiral shell [186]; c) a chiral nanoantenna [88].

Additional parameters in these geometries lead to an even greater variety of ways to control the radiation of the EQS. In particular, the rates of spontaneous emission in a chiral nanoantenna can be significantly higher than the rates for an isolated nanoparticle due to the enhancement of local fields in the gap. The asymmetry between the emission of right-handed and left-handed enantiomers can also be even greater than in the case of a single chiral nanoparticle (see Fig. 45).

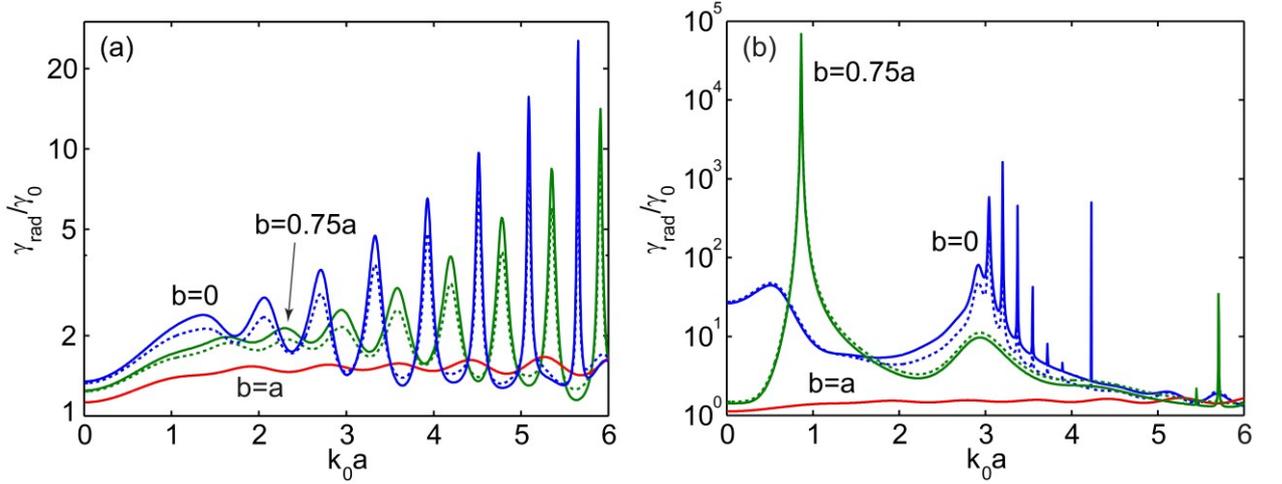

Fig. 45. Radiative decay rate of spontaneous emission of a chiral molecule located near the surface of a spherical particle with a dielectric core of the radius $b$ $(\varepsilon_2 = 2, \mu_2 = 1)$ and a concentric spherical shell of the radius $a > b$, depending on the size parameter $k_0 a$. (a) A chiral dielectric shell $(\varepsilon_1 = 3, \mu_1 = 1, \chi_1 = 0.1)$; (b) a shell of chiral DNG metamaterial $(\varepsilon_1 = -3, \mu_1 = -1, \chi_1 = 0.1)$. The solid line corresponds to the right-hand molecule $(\mathbf{m}_0 = 0.1\mathbf{d}_0)$ and the dotted line corresponds to the left-hand molecule $(\mathbf{m}_0 = -0.1\mathbf{d}_0)$ $(\mathbf{d}_0 = (1,1,1))$ [186].

### 5.3. Hyperbolic Nanoparticles and Metamaterials

Until now, we have considered isotropic and bi-isotropic media, that is, media where wave propagation does not depend on the direction of the propagation. However, since both in nature and during nanofabrication anisotropic metamaterials emerge very often, even almost always, it is important to know how they affect the EQS radiation.

In the most general (linear) case, materials of this kind are described by the material equations [189,190]:

$$\mathbf{D} = \hat{\varepsilon}\mathbf{E} + \hat{\xi}\mathbf{H}, \quad \mathbf{B} = \hat{\mu}\mathbf{H} + \hat{\eta}\mathbf{E}, \tag{60}$$

where the material parameters $\hat{\varepsilon}, \hat{\mu}, \hat{\xi}, \hat{\eta}$ are arbitrary tensors already. Eliminating the electric or magnetic field from the Maxwell's equations, we obtain the equations that determine the propagation of plane waves [190]:

$$\left(\left(\tilde{\mathbf{k}} + \xi\right)\hat{\mu}^{-1}\left(\tilde{\mathbf{k}} - \mathbf{\eta}\right) + k_0^2\hat{\varepsilon}\right)\mathbf{E} = \hat{N}\mathbf{E} = 0,$$
$$\left(\left(\tilde{\mathbf{k}} - \mathbf{\eta}\right)\hat{\varepsilon}^{-1}\left(\tilde{\mathbf{k}} + \xi\right) + k_0^2\hat{\mu}\right)\mathbf{H} = \hat{M}\mathbf{H} = 0, \tag{61}$$
$$\tilde{\mathbf{k}}_{ij} = \varepsilon_{ijl}k_l$$

Dispersion equations are obtained from conditions $\det(\hat{N}) = 0$ or $\det(\hat{M}) = 0$. It can be shown that the equifrequency surfaces of the fourth order in $k_x, k_y, k_z$ correspond to the dispersion equations (61). Such materials can possess very complicated and interesting topological properties of the equifrequency surfaces [190], and their study with regard to the EQS radiation control just begins and is mainly limited to hyperbolic metamaterials (HMM), where the permittivity tensor contains components of the opposite signs on the diagonal:

$$\hat{\varepsilon} = diag\left(\varepsilon_\perp, \varepsilon_\perp, \varepsilon_\parallel\right), \varepsilon_\perp \varepsilon_\parallel < 0 \tag{62}$$

Usually, such metamaterials consist of metal-dielectric layers perpendicular to the z axis, or nanowires parallel to the z axis. The effective permittivity of a layered HMM can be estimated from the relations:

$$\varepsilon_\perp = \frac{a\varepsilon_a + b\varepsilon_b}{a+b}; \frac{1}{\varepsilon_\parallel} = \frac{1}{a+b}\left(\frac{a}{\varepsilon_a} + \frac{b}{\varepsilon_b}\right), \tag{63}$$

where $a, b, \varepsilon_a, \varepsilon_b$ are the layer thicknesses and their permittivities, respectively. HMM production technology is well developed [191]. Fig. 46 shows an example of an HMM that consists of 5 nm thick single crystal layers! Moreover, several naturally occurring materials also exhibit hyperbolic anisotropy [192]. An anisotropic permittivity tensor is also characteristic of cold magnetized plasma [193, 194]. More details about the properties of the HMM one can find in [195].

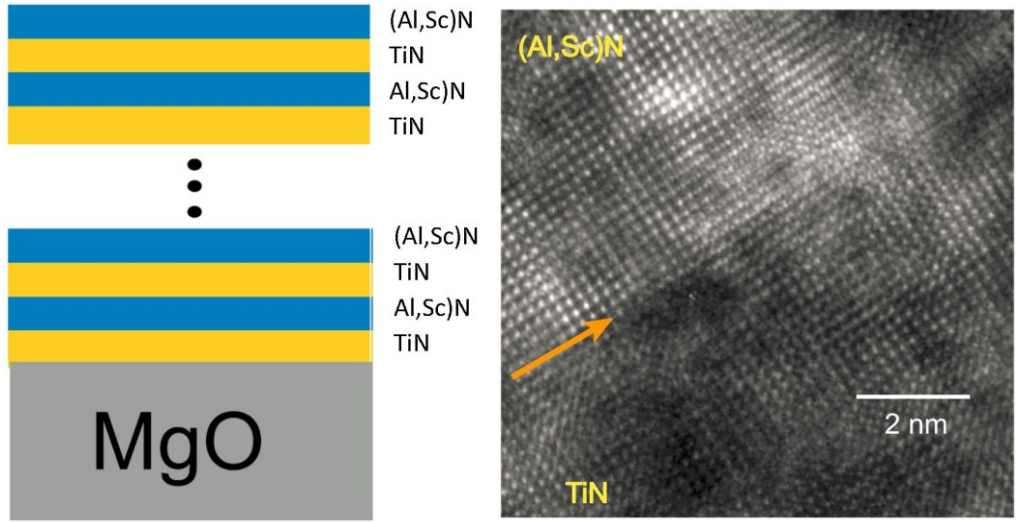

Fig. 46. A hyperbolic metamaterial formed by 5 nm thick TiN-Al$_{0.72}$Sc$_{0.28}$N layers. a) Geometry of the sample, b) HRTEM image of the interface between the layers. Adapted from [191].

The equifrequency surface for HMM with the permittivity tensor (62) consists of two branches, a sphere, and a hyperboloid:

$$\frac{1}{\varepsilon_\perp}\left(k_x^2 + k_y^2 + k_z^2\right) = k_0^2 \text{(TE waves)}$$
$$\frac{1}{\varepsilon_\parallel}\left(k_x^2 + k_y^2 + k_z^2\right) + \frac{1}{\varepsilon_\perp}k_z^2 = k_0^2 \text{(TM waves)} \quad (64)$$

Fig. 47 shows the isofrequency surfaces of isotropic and some bi-anisotropic materials.

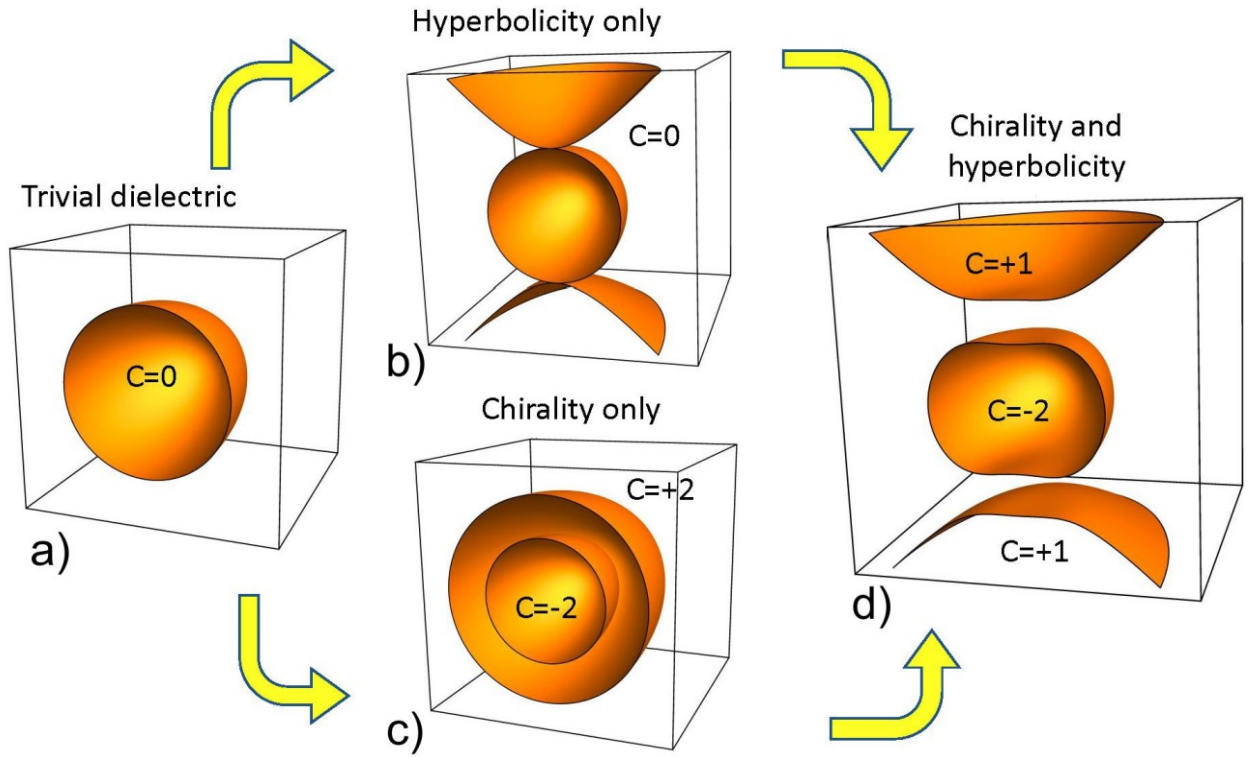

Fig. 47. Equifrequency surfaces of a) an ordinary dielectric, $\varepsilon > 0$, a sphere, the topological Chern number is zero; b) an anisotropic metamaterial with a sign-changing permittivity tensor, $\varepsilon_\perp > 0, \varepsilon_\parallel < 0$, the topological Chern number is zero; c) a chiral material (57), $\varepsilon_\perp = \varepsilon_\parallel > 0, \chi \neq 0$, the Chern numbers are ±2 and d) of a chiral-hyperbolic metamaterial $\varepsilon_\perp > 0, \varepsilon_\parallel < 0, \chi \neq 0$, see (61), the Chern numbers are +1, +1, -2.

It can be seen from Fig.47b that the equifrequency surfaces of the HMM allow the propagation of waves with $k > k_0 = \omega/c$ (high-k -mode) and therefore have an infinite volume in *k*-space and, therefore, an infinite density of states. Since the rate of spontaneous emission is directly proportional to the density of photon states (see (1)), it can be expected that the decay rate will increase infinitely in the presence of HMM.

To verify this directly, one can, as usual (see (9)), investigate the imaginary part of the electric Green's function for an anisotropic material [193,194,196]:

$$\hat{G}(R) = \hat{\mathbf{A}} G_o(R) + \hat{\mathbf{B}} G_e(R); G_{o,e}(R) = \exp(ik_0 R_{o,e}) / R_{o,e};$$
$$R_o = \sqrt{\varepsilon_\perp (\rho^2 + z^2)}; R_e = \sqrt{\varepsilon_\parallel \rho^2 + \varepsilon_\perp z^2}$$
(65)

where $G_o(R)$ and $G_e(R)$ are the scalar Green's functions for ordinary and extraordinary waves. Specific expressions for nonsingular tensors $\hat{\mathbf{A}}, \hat{\mathbf{B}}$ can be found in [194,196].

Figure 48 shows the spatial distribution of the imaginary part of the Green's function, which determines the Purcell factor, in the vicinity of the location of the dipole source.

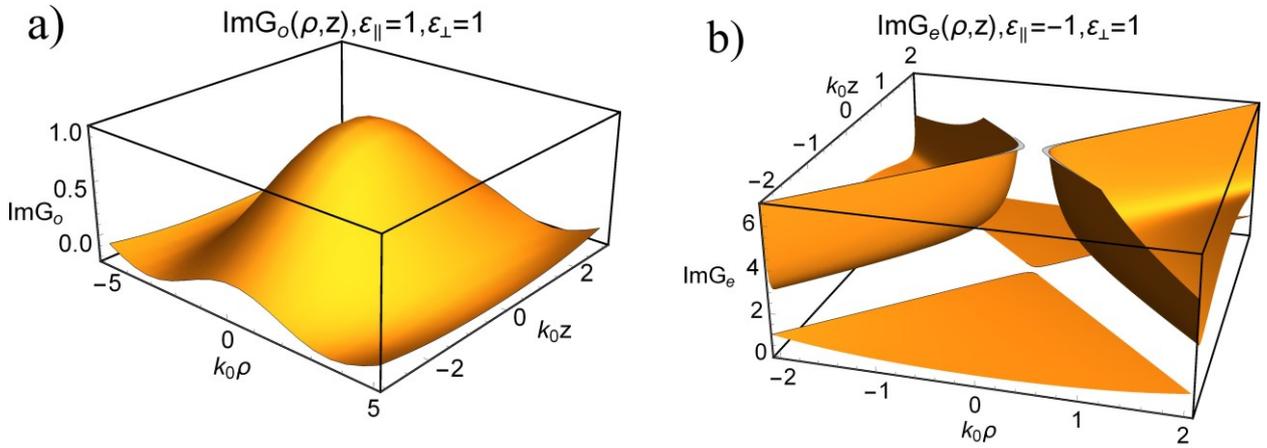

Fig. 48. The spatial distributions of Im$G$ ($\rho$, $z$) near the point of location of the dipole source in vacuum (a) and in a hyperbolic medium with $\varepsilon_\perp = 1, \varepsilon_\parallel = -1$ (b).

It can be seen from these figures that the imaginary part of the Green's function in vacuum is nonsingular near the dipole point and its value is 1 at this point, that is, the Purcell factor is 1. For a HMM, the Green's function has a significantly singular character, it goes to infinity on the line $\varepsilon_\perp z^2 + \varepsilon_\parallel \rho^2 = 0$. In the region $\varepsilon_\perp z^2 + \varepsilon_\parallel \rho^2 > 0$, its behavior is like the behavior of the imaginary part of the Green's function in vacuum (Fig. 48a), but in the region $\varepsilon_\perp z^2 + \varepsilon_\parallel \rho^2 < 0$ it grows infinitely, leading to an infinite Purcell factor.

Of course, in real situations, the Purcell factor does not go to infinity, since the geometry of the sample or the physics of the problem do not allow the appearance of infinitely large vectors of high-k modes.

In [197,198], the emission of Rh800 dye molecules near the HMM surface was studied theoretically and experimentally. In this work, the maximum value of the wave vector of high-k modes was associated with the periodicity of the metamaterial layers and the distance from the EQS to the HMM surface. Despite these limitations, it was shown experimentally [197] that the total decay rate near the HMM exceeds the decay rate near the metal surface, where fluorescence is quenched due to losses in the metal (Fig. 49). This is a clear confirmation of the high density of states in the HMM. Note that a similar enhancement occurs when the EQS approaches the surface of the dielectric, since in this case, too, the high-frequency components of the near fields are converted into waves that propagate in the dielectric (forbidden light [199,200]). However, in the case of a dielectric, the enhancement factor is limited by its refractive index, while for HMM the enhancement can be much greater.

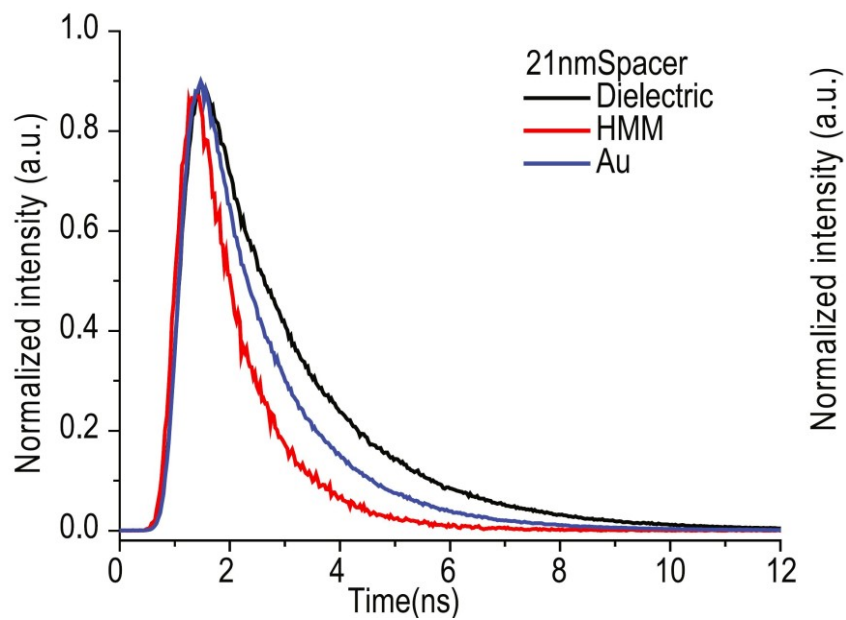

Fig. 49. Fluorescence intensity of Rh800 dye molecules ($\lambda_{peak}$ = 715 nm) in an epoxy matrix ($n$ = 1.58) located at 21 nm from the dielectric, metal (Au), and a HMM, consisting of 16 $Al_2O_3$ and Au layers 19 nm thick [197].

In theoretical works [201-203], a study was carried out of the effect of the EQS size and the HMM actual structure on the total radiation rate, and the presence of restrictions on the maximum value of the wave vectors of high-k modes and the Purcell factor was confirmed quantitatively, due to the finiteness of the size of the EQS, and also the thickness of the HMM layers and their sequence in the case of multiperiodic HMMs [204]. The emission of CdSe/ZnS quantum dots on the surface of an aperiodic HMM Thue-Morse metamaterial (Au-SiO$_2$) was studied in [205], where the authors showed theoretically and experimentally that the total decay rate in an aperiodic multilayer structure can be 10–15% higher than the total decay rate decay in the presence of a periodic HMM with the same metal filling parameter.

In [206], the influence of a planar HMM waveguide (Fig. 50a) on the spontaneous emission of molecules of various dyes (Rh590, Rh610, Rh640, LD700) was investigated experimentally. The waveguide was filled with such molecules to study all regimes of its operation, from hyperbolic to elliptical, including the ENZ regime. Time-resolved photoluminescence measurements showed that the total decay rate increased by a factor of approximately 50, while the enhancement in the radiative decay rate (into the substrate + into the superstrate) turned out to be much smaller (about 18% of the total decay rate). However, the authors hope that the radiation captured by the waveguide can also be used in optical nanodevices based on silicon photonics.

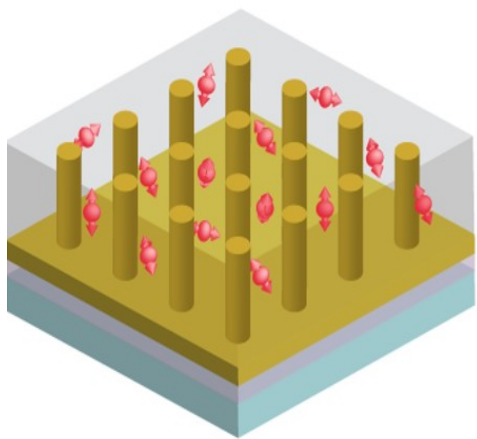
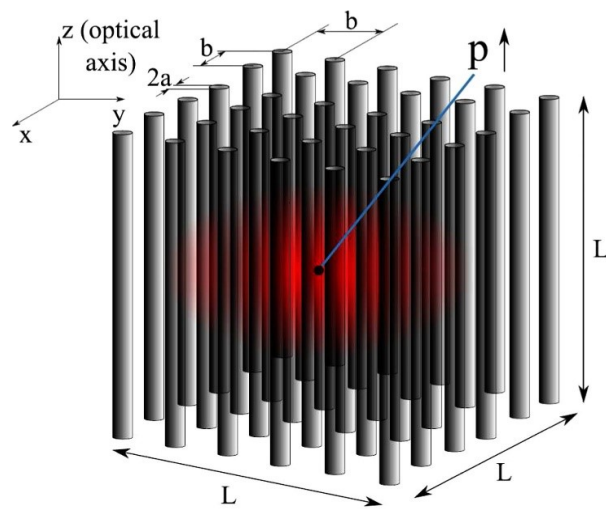

Figure: 50. a) A HMM waveguide of gold nanorods ($d = 38$ nm, $H = 150$ nm, period $P = 80$ nm), the space between the nanorods is filled with one of 4 dyes (R590, R610, R640, LD700) in a PMMA matrix [206]; b) a finite lattice of infinite (in z direction) LiTaO$_3$ nanowires with the radius $a = 32$ nm, the period $b = 200$ nm [207].

An experimental study of the influence of nonlocal effects (spatial dispersion effects), which become significant near the region where the dielectric constant of extraordinary waves vanishes (ENZ metamaterial), on spontaneous emission was carried out in [33] using the example of a metamaterial made of gold nanorods similar to the metamaterial in Fig. 50a, but with the diameter $d = 50$ nm, the height $H = 250$ nm, and the period $P = 100$ nm. In this work, it is shown that in the ENZ region ($\lambda = 575$ nm) the enhancement of the total decay rate by more than three orders of magnitude predicted by the local theory is not confirmed experimentally.

In a nonlocal theory that takes into account spatial dispersion, the permittivity along the optical axis takes the form [33]:

$$\varepsilon_\parallel(k_z) = \pi \left(\frac{d}{2P}\right)^2 \left(k_z^2 \frac{c^2}{\omega^2} - \left(n_z^1\right)^2\right) \frac{\varepsilon_{Au} - \varepsilon_{host}}{\varepsilon_{host} - \left(n_\infty^1\right)^2}, \qquad (66)$$

where $n_z^1$, $n_\infty^1$ are the effective refractive indices of waves propagating through the Au and an perfect conductor cylinder, respectively. Calculations of the rate of spontaneous emission with allowance for (66) are already in good agreement with experiment. It is interesting that the nonlocal theory, where a new type of waves - longitudinal waves - appears due to spatial dispersion, leads to the fact that a gap appears in the hyperbolic equifrequency surface (Fig.47b), its topology changes, and it becomes similar to equifrequency surfaces of chiral hyperbolic metamaterials, where, longitudinal waves do not appear, but the waves are split into right- and left-handed polarized (Fig. 47d).

The effect of a hyperbolic metamaterial in the form of a finite array of parallel LiTaO nanowires on the spontaneous emission of an EQS was studied theoretically in [207] (the geometry is shown in Fig.50b). Fig. 51 shows the total Purcell factor and its radiative part depending on the parameters of the system.

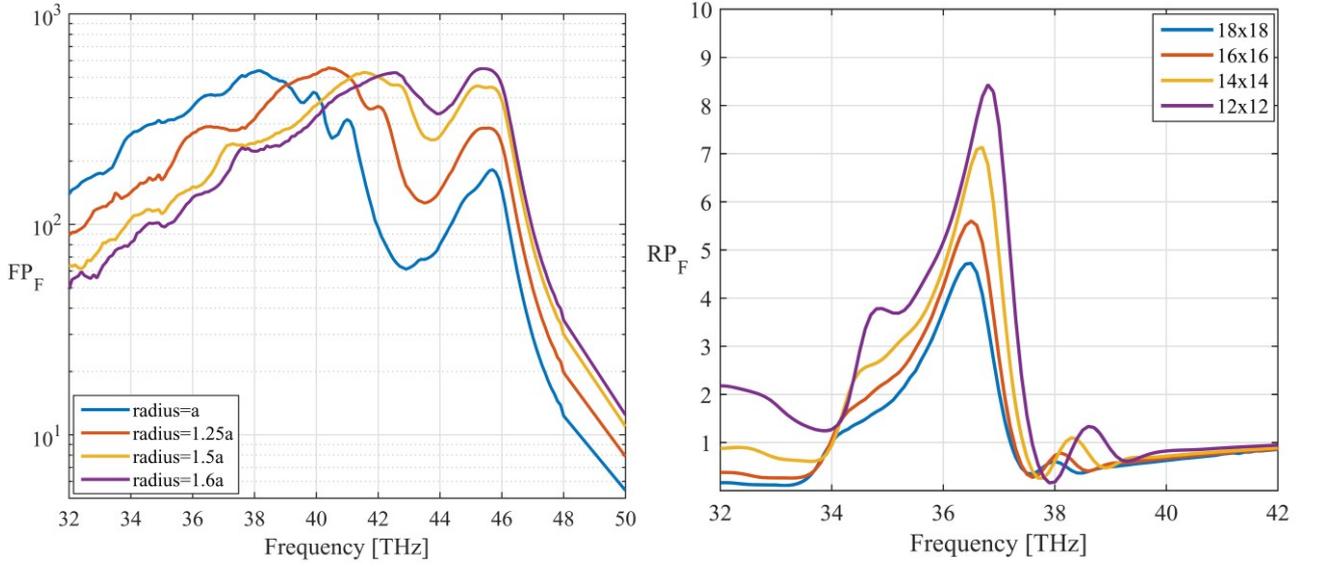

Fig. 51. Total (FP$_F$) and radiative (RP$_F$) Purcell factors for the HMM nanowire lattice of finite size. a) The initial radius $a = 32$nm, the period $b = 200$ nm, the lattice 18x18; b) the fixed radius $a = 32$, the period $b = 200$ nm, various lattice dimensions. EQS is in the center of the system (see Fig. 50b), and its dipole moment is parallel to the nanowires [207].

These plots show clearly that the total rate of spontaneous emission increases by more than two orders of magnitude, but its radiative part is much less influenced, and the Purcell radiation factor decreases with increasing array dimension rapidly and tends to zero in the case of an infinite array, since all radiation is absorbed in the wires. Often, such a significant difference between the total and the radiative Purcell factors is not considered, and it is assumed implicitly that the radiative Purcell factor is in order of magnitude equal to the total Purcell factor. Fundamental constraints on the radiative Purcell factor in the presence of an arbitrary uniaxial HMM layer without losses were found in [208] for allowed transitions and in [209] for forbidden ones. In these works, it is shown strictly that the relative radiative decay rate of spontaneous emission into the half-space where the EQS is located cannot be more than 2, that is $F_{P,rad}^{\uparrow} \leq 2$ for any HMM. The decay rate into the space behind the HMM layer depends substantially on the material located behind the HMM. If this is the same medium as the medium where the EQS is situated, then the Purcell factor for radiation behind HMM layer cannot be more than 1/2. If a dielectric with a sufficiently high refractive index is located behind the metamaterial layer, the radiative Purcell factor can increase by several orders of magnitude, and the maximum enhancement is achieved not for HMM, but for an isotropic metal with specially selected parameters [208,209]. The constraints on the radiative decay rate found in these works should serve as a starting point for planning any experiments with layered HMMs. The limitations on the total decay rate associated with losses in the metal were analyzed in [210], where the authors concluded that significant (by orders of magnitude) increase of spontaneous emission decay rate is impossible too. This result agrees with the results of [197] and does not fully agree with the experimental data [33, 206].

An interesting example of metamaterials are nanostructures, which, due to the simultaneous existence of electrical and magnetic resonances, have an effective impedance $Z = \sqrt{\mu_{eff}/\varepsilon_{eff}} = 1$ equal to the impedance of free space. In this case, as it is known, there is no reflection at normal incidence, and in such a metamaterial the radiation can be absorbed completely. In [211] such a metamaterial was realized experimentally (Fig. 52), and its influence on the emission rate of CdSe/ZnS quantum dots in a polystyrene nanofilm was investigated. The results of the study show that with the optimal geometry providing the minimum reflection coefficient, the total Purcell factor reaches $F_P \approx 5$, while the photoluminescence also increases by a factor of 3.6. This suggests that metamaterials of this kind are also very promising for controlling the emission of the EQS. Note that a similar system (see Fig. 36) is a negative refractive index metamaterial (NIM).

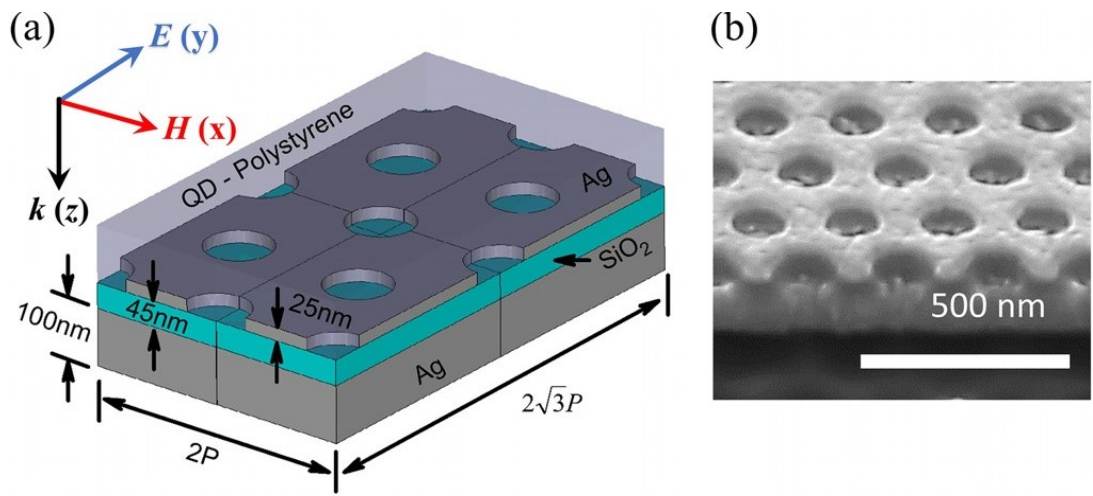

Fig. 52. a) Geometry of the sample with CdSe/ZnS quantum dots in a polystyrene nanofilm on a metamaterial with impedance equal to the impedance of free space; b) SEM image of the experimental sample [211].

## 6. Control of Radiation of Nanoscale Sources of Waves of Non-Electromagnetic nature

The previous sections discussed the effect of the environment on the emission of electromagnetic waves. However, when waves of any other nature are emitted, the environment will have an effect like that in the electromagnetic case.

### 6.1. Acoustic Wave Emitters

The case of acoustic waves is the closest to the electromagnetic case. There are so many loudspeakers, and their sound control has reached perfection, although their size is usually much smaller than the wavelength. However, there are still few works on the influence of the environment on the emission of acoustic waves in the aspect in which is relevant for electromagnetic waves [212-214], although this is very important for the creation of effective small-sized sources of low-frequency sound.

In [212], the change in the linewidth (decay rate) of a Chinese gong, excited by the impact of a wooden ball in the presence of a rigid reflective surface (Fig. 53a), was investigated and a good agreement between theory and experiment was found (Fig. 53b, compare with Fig. 8).

A more interesting study of the Purcell effect in acoustics was carried out in [214]. In this work, they considered the effect of a nanometadisk of 10 cm in diameter made of an acoustic metamaterial with a low effective speed of sound (analogous to a large refractive index in optics) on the radiation of a monopole sound source of 1 mm in diameter. The geometry of the experiment and the nanometadisk is shown in Fig. 54.

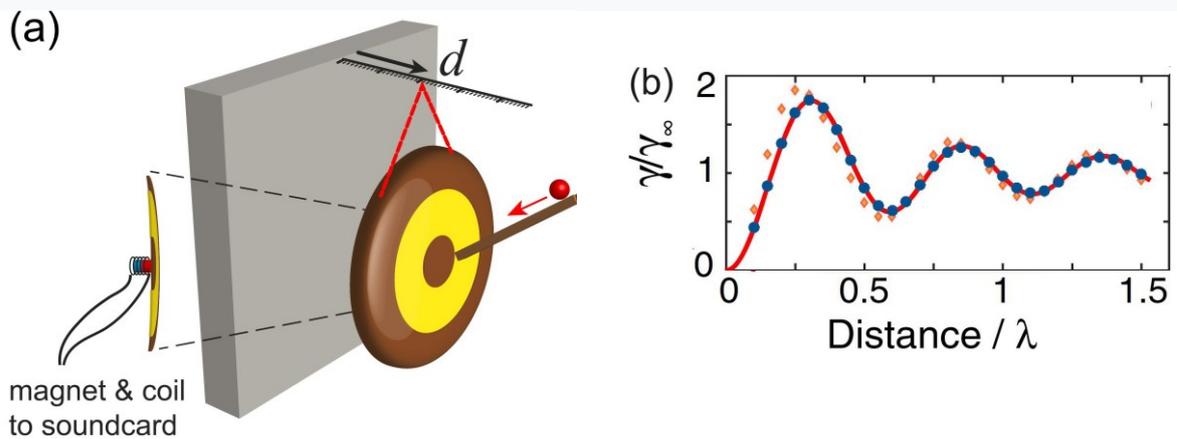

Fig. 53. A schematic of an experiment to measure the linewidth of the sound of a Chinese gong made of brass with the radius of 10 cm and the thickness of 0.5 mm, depending on the distance to the wall. b) Linewidth of the 306 Hz mode versus the distance of the gong from the wall. Orange circles - individual measurements, blue dots - averaging over 5 measurements, solid curve – theory [212].

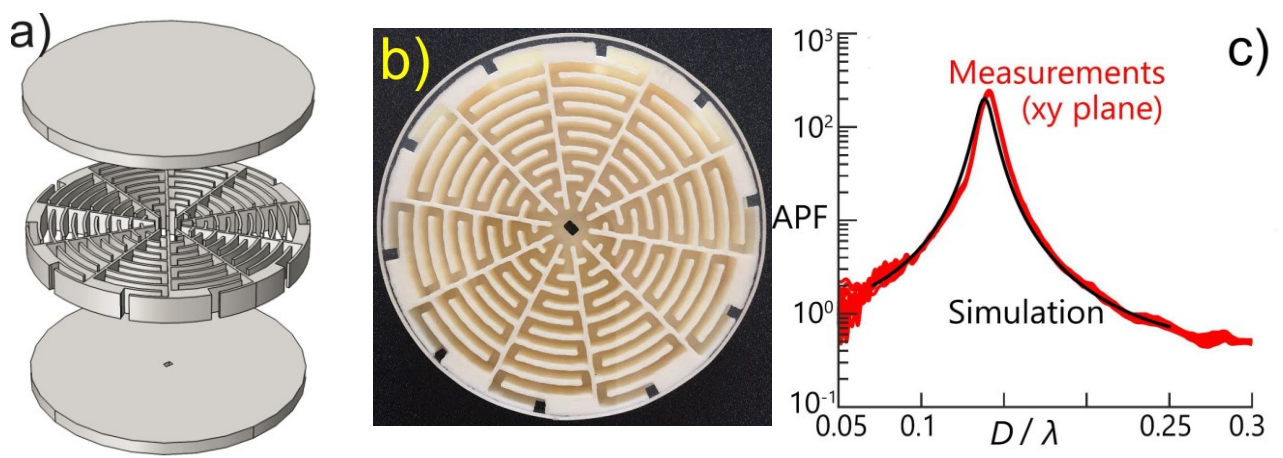

Fig. 54. a) Geometry of the experimental sample, where the main element is a 10 cm diameter nanometadisk made of plastic with zigzag nanochannels, excited by a monopole source of 1 mm in diameter from below (b); c) the dependence of the acoustic Purcell factor (APF) on the wavelength (red - measurements in the plane of the disk at a distance of 15 cm from the center, black - calculation through the imaginary part of the Green's function) [214].

As can be seen from Fig.54c, the agreement between theory and experiment is good, and that is not surprising, since losses in the system are small, and acoustic waves are scalar. A study of a similar system with more complicated geometry was carried out in [213], where high values of the Purcell factor were also predicted.

For acoustic waves, as well as for electromagnetic ones, there are media (acoustic crystals or metamaterials) where their refractive index is negative or has other anomalies (see, for example, [215]). In this case, one can also expect the manifestation of a strong Purcell effect, as it was for electromagnetic waves.

## 6.2. Elastic Wave Emitters

The influence of the environment on the radiation of elastic waves is much more complicated than the electromagnetic or acoustic cases, since the elastic waves have three independent components (one longitudinal and two transverse), and even isotropic medium elastic wave equations are described by three independent parameters (density $\rho$ and Lamé parameters $\lambda$, $\mu$):

$$\mu \nabla \times \nabla \times \mathbf{u} - (\lambda + 2\mu)\nabla(\nabla \cdot \mathbf{u}) - \rho\omega^2\mathbf{u} = 0, \qquad (67)$$

where **u** is the displacement field.

Equifrequency surfaces corresponding to (67) have the form of three concentric spheres:

$$\left(\mathbf{k}^2 - \frac{\rho}{\mu}\omega^2\right)^2 \left(\mathbf{k}^2 - \frac{\rho}{\lambda + 2\mu}\omega^2\right) = 0, \qquad (68)$$

where two spheres coincide with each other, but describe different polarizations, distinguishing them from the externally similar equifrequency surfaces of chiral metamaterials (Fig. 47c).

The boundary conditions in the case of the elastic waves are also more complex and include the continuity of the displacement vector and the normal component of the stress tensor, that is, in the general case of the elastic waves, there are 6 boundary conditions instead of 4 in the electromagnetic case. Nevertheless, the problem of radiation of a point elastic source in the presence of a gold microsphere was solved analytically in [216], where the resonant modes of such a system were found and the Purcell factor was calculated, which in the case of elastic waves is described by an expression similar to the electromagnetic case [216]:

$$\frac{\Gamma}{\Gamma_0} = 1 + \frac{6\pi}{|\mathbf{f}|^2} \frac{\operatorname{Im}\left\{\mathbf{f}^*\ddot{\mathbf{G}}^R(\mathbf{r},\mathbf{r})\mathbf{f}\right\}}{\frac{k_{l1}}{2(\lambda_1+2\mu_1)} + \frac{k_{s1}}{\mu_1}}, \qquad (69)$$

where $\ddot{\mathbf{G}}^R(\mathbf{r},\mathbf{r})$ is the scattered part of the Green's function of the elastic source at the point r of its location, $k_{l1} = \omega\sqrt{\rho_1/(\lambda_1+2\mu_1)}, k_{s1} = \omega\sqrt{\rho_1/\mu_1}$ are the wave vectors of longitudinal and transverse waves in the surrounding space, $\mathbf{f}$ is the amplitude of the force of a linear elastic source. Figure 55 shows the dependence of the Purcell factor for the emission of elastic waves by a point source, calculated using the exact expression (69).

The results [216] can be very useful for describing both existing MEMS and NEMS based on conventional materials, and radiation in the presence of more complex elastic metamaterials, with the properties very different from common materials and having the effective densities and stiffness coefficients able to take negative values [217, 218]. The resulting effects will be very interesting, as it was when the DNG metamaterials were created in the electrodynamic case.

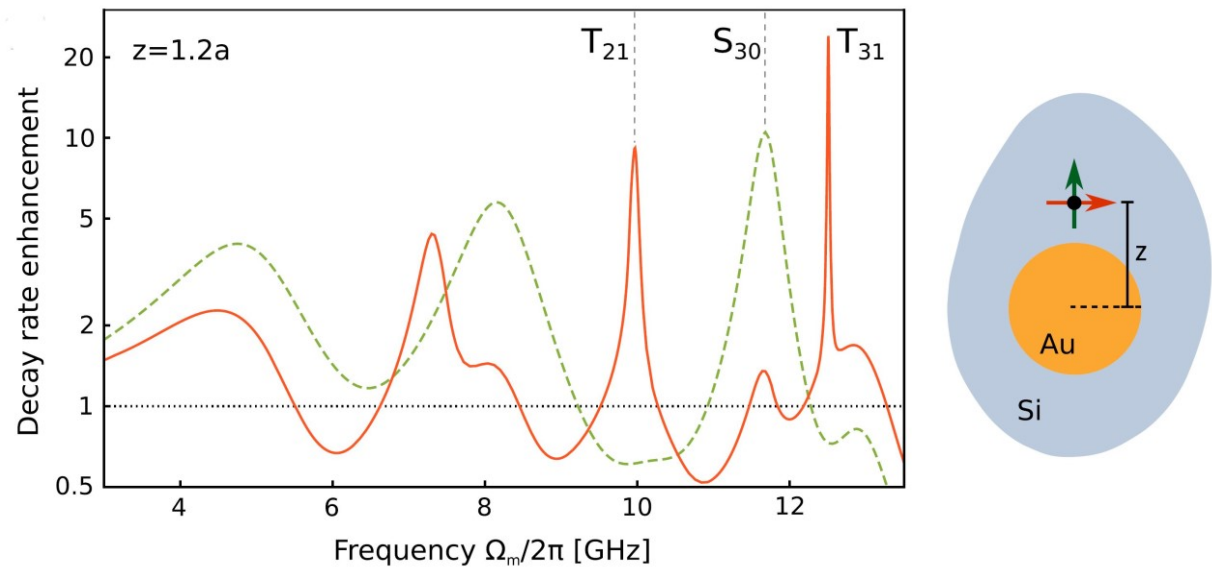

Fig. 55. Dependence of the relative decay rate of the spontaneous emission for a linearly polarized elastic source near a gold nanosphere with the radius of 100 nm located in a silicon matrix at 120 nm from the center of the nanosphere. Red and green colors show the results for different polarizations of the force **f** in the source [216].

## 7. Conclusion

Mainly, the fundamental aspects of radiation control methods using nanoscale structures and metamaterials were discussed in this review , but the results of their study are related to real life directly. At present, the methods of radiation control have already found or are finding practical application in a number of areas.

First of all, metamaterials and nanostructures are widely used to control the radiation of microwave antennas both to reduce their mass-dimensional characteristics and to increase the stability of their operation in modern systems [220, 221]. This is an important area of application of the Purcell effect and metamaterials - for example, dozens, if not hundreds, of antennas are installed on modern aircraft, and all of them should be lightweight and not experience strong interaction.

On the other hand, the above approaches for reducing the size of woofers are used actively in practice in music and other applications to increase their efficiency and power. Metamaterials are also used to suppress low-frequency sounds in manufacturing and defense products (although they are not always called that).

In NEMS and MEMS [222,223], nanoscale sources of elastic oscillations are the main element, and, as it was shown in Section 6, the emission of elastic waves can also be effectively controlled.

The dependence of the emission rate of quantum dots and dye molecules on the nanoenvironment is widely used in sensors and detectors of various types. Thus, in [29,224,225], the enhancement of fluorescence near plasmonic nanoparticles was proposed to be used both for visualization of cells marked with dyes and for operative decoding of the DNA structure [29]. The special structure of the nano-needle of aperture scanning microscopes increases the emission rate of detected molecules significantly, allowing to use SNOM much more efficiently [142].

The radical change in the decay rates of spontaneous emission during the metal-insulator phase transition can be used to control such phase transitions [226]. In [227], it was proposed to use the Purcell effect to characterize fluid flows near nanoscale objects.

The dependence of the emission of enantiomers of chiral molecules on the chiral environment allows one to create efficient systems for the separation of racemic mixtures without the use of additional chemical agents, significantly improving the quality of manufactured drugs [70, 228, 229].

The proposed methods [230] make it possible to create artificial fluorescent markers more efficient and brighter in all wavelength ranges than available ones.

In the new age of information technologies, the problem of creating a new element base for computers and other information processing systems is of particular importance. The size and power consumption of such systems should decrease, and their speed should increase. It is believed that the solution to the problem of the element base of information processing systems lies in the field of nanooptics and quantum optics. To create a nano-optical element base of information systems, nanoscale light sources and the ways to effective control of their radiation are needed.

There are all prerequisites for the development of nano-optical element base:

1) Modern nanotechnologies, and in particular CMOS and SOI technologies, make it possible to manufacture almost arbitrary nanostructures and metamaterials [150-152,231-233].

2) At present, as a rule, experimenters can effectively position emitters in a layer with nanometer precision vertically, however, methods of truly three-dimensional positioning of the EQS (vertically and horizontally) with nanometer precision are also being developed [234], since the precise positioning of nano-emitters in the required the location of the nanostructure is very important.

Significant progress towards the formation of the element base of nano-optics has already been achieved in the field of creating nanolasers [235-237] and incoherent nanoscale light sources with electrical excitation [32,238]. The precise selection of conditions for efficient extraction of light from nanoscale sources is of crucial importance for development of such devices.

Several types of efficient sources of single photons (see, for example, [239-243]) and even entangled photon pairs [244] have been developed using the optimization of the Purcell factor as an important element of quantum nano-optics required for the creation of quantum computers. The Purcell effect can also be used to control (suppress) the radiation of systems of several qubits [245].

Light emitting diode (LED) light sources have become a part of our daily life. It is impossible to increase the efficiency of extraction of their spontaneous emission without using the methods of optimization of the Purcell factor [246,247,248].

A new type of optoelectronic devices based on the electronic structure of two-dimensional materials is being actively developed. This type of nanodevices, by analogy with electronics, is designated by a general term "valleytronics" (see, for example, [249-251]). The methods for controlling the chiral Purcell factor [252], presented in this review also make it possible to control the state of electrons in different energy valleys effectively.

Thus, currently the effect of the environment on the rate of EQS emission predicted more than 70 years ago, is being dynamically studied, its potential is far from being fully realized, and the activity of the research in this direction will keep on growing.

**Acknowledgments:**

The reported study was funded by RFBR, project number 19-12-50157.